\documentclass[twocolumn,showpacs,showkeys,preprintnumbers,amsmath,amssymb,pre,floatfix]{revtex4}

\usepackage{dcolumn}
\usepackage{bm}
\usepackage{psfrag}
\usepackage{amsmath}
\usepackage{amssymb}
\usepackage{graphicx}
\usepackage[usenames,dvipsnames]{xcolor}
\graphicspath{{Figures/}}

\newcounter{resultcounter}[section]

\newtheorem{thm}[resultcounter]{Theorem}

\newtheorem{cor}{Corollary}

\def\proof{\noindent{\bf Proof.}\ \ }
\def\qed{\hfill $\Box$\medskip}

\begin{document}
\title{Noise-Assisted Quantum Exciton and Electron Transfer in Bio-Complexes\\ with Finite Donor and Acceptor Bandwidths}

\author{Alexander I. Nesterov}
   \email{nesterov@cencar.udg.mx}
\affiliation{Departamento de F{\'\i}sica, CUCEI, Universidad de Guadalajara,
Av. Revoluci\'on 1500, Guadalajara, CP 44420, Jalisco, M\'exico,}
\affiliation{the Center for Nonlinear Studies, Los Alamos National Laboratory,
Los Alamos, NM 87545, USA}
\author{Gennady P.  Berman}
 \email{gpb@lanl.gov}
\affiliation{Theoretical Division, T-4, MS-213, Los Alamos National Laboratory, Los Alamos, NM 87545, USA,}
\affiliation{New Mexico Consortium, Los Alamos, NM 87544, USA}
\author{Marco Merkli}
 \email{merkli@mun.ca}
\affiliation{Department of Mathematics and Statistics,  Memorial University of Newfoundland, St. John's, Newfoundland, Canada A1C 5S7,}
\affiliation{the Center for Nonlinear Studies, Los Alamos National Laboratory,
	Los Alamos, NM 87545, USA}
\author{Avadh Saxena}
\email{avadh@lanl.gov}
\affiliation{Theoretical Division and the Center for Nonlinear Studies, Los Alamos National Laboratory, Los Alamos, NM 87545, USA}
\begin{abstract}
We present an analytic and numerical study of noise-assisted quantum exciton (electron) transfer (ET) in a bio-complex, consisting of an electron donor and acceptor (a dimer), modeled by interacting continuous electron bands of finite widths.
The interaction with  the protein-solvent environment is modeled by a stationary stochastic 
process (noise) acting on all the donor and acceptor energy levels. We start with discrete energy 
levels for both bands. Then, by using a continuous {limit} for the electron spectra, we derive  
integro-differential equations for ET dynamics between  two bands. Finally, we derive from 
these equations rate-type differential equations for  ET dynamics. We formulate the conditions 
of validity of the rate-type equations.  We consider different regions of parameters 
characterizing the widths of the donor and acceptor bands and  the strength of the dimer-noise 
interaction. 
For a simplified model with a single energy level donor and a continuous acceptor band,  we derive a generalized analytic expression and provide  numerical simulations for the ET rate. They are consistent with Wigner-Weisskopf,  F\"orster-type, and Marcus-type expressions, in their corresponding regime of parameters. For a weak dimer-noise interaction, our approach leads to the Wigner-Weisskopf ET rate, or to the F\"orster-type ET rate, depending on other parameters. In the limit of strong dimer-noise interaction, our approach is non-perturbative in the dimer-noise interaction constant, and it recovers the Marcus-type ET rate. 
Our analytic  results are confirmed by numerical simulations. We demonstrate how our theoretical results are modified {when both the donor and the acceptor are described by finite bands}. We also show that, for a relatively wide acceptor band, the efficiency of the ET from donor to acceptor can be close to 100\% for a broad range of noise amplitudes, for both ``downhill" and ``uphill" ET, for sharp and flat redox potentials, and for reasonably short times. We discuss possible experimental implementations of our approach with application to bio-complexes. 
\end{abstract}


\preprint{LA-UR-18-31605}
\maketitle

\section{Introduction}

The exciton transfer in light harvesting complexes (LHCs) and the electron transfer in primary processes of charge separation in the reaction centers (RCs)  of photosynthetic bio-complexes such as plants, eukaryotic algae and cyanobacteria, take place on {a} very short time-scale, of the order of 1 - 5ps. It was recently discovered experimentally that these processes include quantum coherent effects, which should be taken into consideration even at room temperature. (For discussion see, for example, \cite{Len,ECR,IF,IFG,CWWC,PHFC}, and references therein.)  Quantum effects (including coherent ones) must be taken into account generally, even if they may not contribute significantly to the ET rate \cite{WD}. These results initiated  significant interest in modeling and creating adequate mathematical tools for describing quantum ET in these systems \cite{book1,RMKL,CFMB,BAF,MRSN,HDR,MBS}.

In the simplest theoretical approaches, the donor-acceptor complex (dimer) is modeled by two discrete electron energy levels interacting with each other and with {a} bosonic protein-solvent environment. For example, the dimer can consist of two excited electron energy levels of two neighboring chlorophyll molecules ($Chla$ and $Chlb$), interacting by the transitional dipole moments \cite{book1}. The main characteristic parameters of the model are: the difference of the donor and acceptor energy levels (redox potential), $\varepsilon$, the matrix element, $V$, of the  donor-acceptor interaction, the reconstruction energy, $\varepsilon_{rec}\propto \lambda^2$, which renormalizes the redox potential (here $\lambda$ is the interaction constant between the dimer and the environment), and the temperature, $T$.  

The bosonic protein-solvent environment is often modeled by a set of quantum linear  oscillators in equilibrium, at temperature, $T$.  The bosonic bath is usually characterized by its spectral density  (SD), $S(\omega)$, which is the Fourier transform of the bosonic correlation function of the environmental degrees of freedom, averaged over the equilibrium density matrix of the environment. The SD has  characteristic dependences on the frequency of bosonic modes, on temperature,  and on the cutoff frequency, $\omega_c$ \cite{book1}. Note that at sufficiently high temperature (such as room temperature),  the protein-solvent environment is often modeled by stationary stochastic processes, including $1/f$ noise \cite{MFL,DB,G,CBC,SPA} (see also below).

 In many situations (for example, in quantum computation and in some bio-complexes) the interaction constant, $\lambda$, is smaller than $\varepsilon$ and $V$. In this case,  standard perturbation theory, based on a power series expansion in $\lambda$, can be applied. (See, for example, \cite{MBR2}, and references therein.) In this situation the relaxation rate of the dimer (qubit), caused by interaction with the environment, is proportional to $\gamma_r \sim \lambda^2 S(\omega_{01})$, and the dephasing (decoherence) rate is proportional to $\gamma_d \sim \lambda^2 S(0)$ [when $S(0)\gg S(\omega_{01}$)], where $\omega_{01}$ is the dimer transition frequency. In quantum computation, this situation takes place at very low temperatures. For details see, for example, \cite{MBR2,BGA,GABS,NB1}, and references therein. 
 
 In biological systems, both situations emerge, when  $\lambda$ is small,  but also when $\lambda$ is relatively large. Indeed, one of the adequate and very powerful theoretical approaches for estimating the exciton transfer rates in bio-systems is based on the F\"orster theory (see, for example, \cite{Govorov,Andrews}, and references therein). In this case, the  dipole-dipole interaction (due to transitional dipole moments), $V$, between {the} donor and {the} acceptor is mediated by a relatively weak interaction through the electromagnetic field (see, for example,  Sec. VI in \cite{Andrews} and \cite{Milonni}). The interaction, $\lambda$, of the dimer with the protein-solvent environment is also relatively weak. So, standard perturbation theory {for small $\lambda$} can be applied \cite{Govorov,Andrews}. 

In other bio-complexes, the interaction of the dimer with the protein-solvent environment can be large, for instance for ``resonant" ET, when the reconstruction energy is large, $\varepsilon_{rec}\approx\varepsilon$. In this case, perturbation theory in $\lambda$ cannot be applied, and the Marcus theory, and its various modifications, is used \cite{MRSN,HDR,MBS}. 
   The main advantage of the Marcus approach, in application to bio-systems, is that even though both the interaction constant and the temperature are large (including room temperature), the ET rate has a simple analytic expression  \cite{MRSN,HDR,MBS}. The interaction constant, $\lambda$ (or the reconstruction energy, $\varepsilon_{rec}$), appears in this expression as a singular perturbation, in the denominator of a fraction. In this sense, the Marcus theory is non-perturbative in $\lambda$ (or, in the reconstruction energy). In 
   order to calculate the reconstruction energy, one needs to know the SD,  $S(\omega)$, of the {reservoir}. The conditions of applicability of the Marcus theory are complementary to the above mentioned case of the standard perturbation theory, in which small  $\lambda$ is assumed \cite{MBS}. (See also the text below.)

   Another well-known approach for calculating the ET rate from a discrete electron energy level (donor) into 
   the acceptor with infinite bandwidth and with finite electron density of states,    $\varrho_0$, was suggested by Wigner and Weisskopf in 1930 \cite{WW}. (See also \cite{SM}.) The model does not include a bosonic bath.  It was shown in \cite{WW}, that the ET  rate of the initially populated donor {is}  $\Gamma \approx 2\pi V^2 \varrho_0/\hbar $. In 
   this case, the irreversible dynamics {is not} caused by {a} thermal bosonic bath with continuous spectrum, but by the ``entropy factor" - a continuous electron energy spectrum of the acceptor band.

   In the present paper, we study analytically and numerically the ET between the donor and the acceptor which are modeled by continuous energy bands of finite  widths, centered at energies,  $E^{d}_0$ and $E^{a}_0$, respectively. In the corresponding discrete model, the interaction between the donor and acceptor energy levels  is provided by non-diagonal matrix elements, $V_{mn}$. Our results can be applied {both to the} exciton transfer {process} in LHCs and {to the} primary charge separation process in the RCs. The main difference is in the functional form of the corresponding matrix elements, $V_{mn}$. Namely, in the case of the exciton transfer, these matrix elements are related to the  interaction between transitional dipole moments of {the} donor and {the} acceptor. In the case of primary charge separation processes (as in the Marcus theory) the matrix elements, $V_{mn}$, are related to the direct Coulomb interaction, with possible effects of charge transfer.
   
    Instead of the thermal protein-solvent bosonic environment, we include in our model an external (classical) diagonal noise which interacts with both the donor and the acceptor electron energy levels. This approach is often used for modeling the protein-solvent environment at sufficiently high  temperatures, including room temperature, and when the characteristic time-scales are very short, so no thermal equilibrium is reached. Our approach provides: (i) {the derivation of a closed set of integro-differential equations for the reduced density matrix (averaged over noise), obtained  by taking a continuous limit of a model with discrete electron energy levels},  (ii) {the derivation of simplified rate-type differential equations describing the ET between interacting donor-acceptor bands with finite bandwidths}, and (iii) {the identification of the conditions of validity of the rate-type equations.} 

  The redox potential, the donor and acceptor bandwidths, and the intensity of noise can be varied in our theory. This allows us to consider the ET rates and efficiency for narrow or wide electron bands, for bands overlapping to different degrees,  and for various noise intensities. Our approach is quite general and can be applied to {a variety of} systems with noisy environment. 
  \smallskip

\noindent
Among our main results are the following.

1) We derive a closed set of the rate-type differential equations for interacting donor-acceptor bands of finite widths and for different intensities of {the} external noise.

2) When the donor consists of a single energy level,
we obtain an analytic generalized expression for the ET rate. It is consistent with Wigner-Weisskopf, F\"{o}rster-type, and Marcus-type results in their appropriate regimes. We compare our analytic and numerical results for a continuous acceptor band with the results for the acceptor band consisting of a large number of discrete energy levels. 

3)	We demonstrate that in the above case 2), there exists an important parameter in the ET rate which includes the electron bandwidths and the dimer-noise interaction constant and which characterizes the asymptotic ET rate for large times. 

4) We calculate analytically the {ET} dynamical rates for two interacting bands. We show the differences in the ET dynamics for two models,  when the donor consists of a single level and when it has a band of energies. 

5)	We find explicitly the conditions of applicability for our approach in all considered limits.

6)	We demonstrate that the efficiency of {populating the acceptor} can be close to 100 \%, in 
relatively short times, for both ``downhill" and ``uphill" redox potentials. \\

The paper is organized as follows. In Sec. II, a simplified model is considered in which the donor 
has a single electron energy level, and the acceptor is modeled by an electron band of finite 
width. Diagonal noise acts on both the donor and the acceptor. A set of simplified rate-type 
differential equations is derived. Analytic expressions for time-dependent and asymptotic ET 
rates are presented. In Sec. III, we compare the results of a discrete model, when  the donor has 
a single energy level and the acceptor includes many energy levels, with the solutions of the 
rate-type equations. In Sec. IV, we present the results for a generalized  model when both, 
donor and acceptor, are modeled by electron bands of finite widths. In the Conclusion, we 
summarize our results, and discuss possible experimental realizations. The Supplemental 
Material (SM) contains technical details.

\section{A simplified model}

\begin{figure}[tbh]
\scalebox{0.5}{\includegraphics{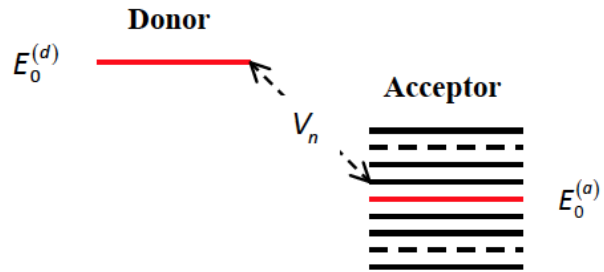}}
\caption{(Color online) Schematic of our simplified model consisting of the donor with a single electron energy level, and the acceptor with a discrete (nearly continuous) electron spectrum. The donor energy level and the center of the acceptor band are indicated by red color.
\label{S1a}}
\end{figure}

In this section, we consider a simplified model shown in Fig. \ref{S1a}. A single electron donor site, denoted by the  quantum state, $|d\rangle$,  interacts with an acceptor site with $2N_a+1$ $(N_a\gg 1)$  quasi-degenerate discrete energy levels.  

In the presence of a classical diagonal noise, the quantum dynamics of the ET can be described by the Hamiltonian \cite{G0},
\begin{align} \label{H1}
\mathcal H(t) = & (E_0^{(d)}+\lambda_d(t))|d\rangle\langle d|  +  \sum^{N_a}_{n=-N_a}  
(E_{n}+\lambda_a(t))|n\rangle\langle  n | \nonumber\\
& +\sum^{N_a}_{n=-N_a}\big ( V^\ast_{n}|n\rangle\langle  d| + V_{n}|d\rangle\langle n |\big),
\end{align}
where $E_0^{(d)}$ is the energy level of the donor, $E^{(a)}_0$ is the center of the acceptor band, $\lambda_d(t)$ and $\lambda_a(t)$ describe the noise acting on the donor 
and the acceptor.

We assume, for simplicity, that the noisy environment, $\xi(t)$, is the same for the donor and the acceptor (collective noise). This assumption is similar to that used in the Marcus theory where thermal protein-solvent environments are considered \cite{MRSN,HDR}. One can write $\lambda_d(t)=g_d\xi(t)$ and $\lambda_a(t)=g_a\xi(t)$, where $g_{d}$ and  $g_{a}$ are the interaction constants. We  consider  {a} stationary noise described by {a} random variable,  $\xi(t)$, with correlation function $\chi(t-t')=\langle \xi(t)\xi(t')\rangle$ and $\langle\xi(t)\rangle=0$. The averaging, $\langle ... \rangle$, is taken over a random process describing the noise.

We assume that the electron energy spectrum of the acceptor is sufficiently dense,  so that $E_n$ and $V_{n}$ can be considered  as continuous variables. Thus, in Eq. 
 (\ref{H1}) one can perform an integration instead of a summation, so that 
\begin{align}\label{ah1}
\mathcal H(t) = & (E_0^{(d)}+\lambda_d(t))|d\rangle\langle d|)  \nonumber \\
&+ \int (E +\lambda_a(t))|E\rangle\langle E|\varrho(E)dE  \nonumber \\
&+  \Big( \int V(E)|d\rangle\langle E|\varrho(E) dE + \rm h.c.\Big),
 \end{align}
where, $\varrho(E)=dn(E)/dE$, is the density of electron states of the acceptor, and $V_{n}$ is replaced by $V(E)$.

We consider a Gaussian  density of states of the acceptor, centered at the energy, $E^{(a)}_0$,
\begin{align}
\varrho(E) =   \varrho_0 e^{-\alpha (E-E_0^{(a)})^2},
\label{G1a}
\end{align}
where, $\varrho_0=dn(E)/dE|_{E^{(a)}_0}$, is the density of states at the center of the acceptor band.
Formally, the expression (\ref{G1a}) allows the existence of energy levels in the acceptor band with very large energies, $|E|\rightarrow\infty$, although with very small density of states. To simplify our consideration, we assume that all the $2N_a+1$ levels of the acceptor band are distributed inside the finite interval, $\delta_a$ (acceptor energy band),  $[E_0^{(a)} - \delta_a/2, E_0^{(a)} + \delta_a/2]$. Since the number of levels also equals $\int^{\delta_a/2}_{-\delta_a/2} \varrho(E+E_0^{(a)}) dE$, we have 
\begin{align}
	2N_a+1 = \varrho_0\sqrt{\pi/ \alpha}\, {\rm erf} (\sqrt{\alpha}  \delta_a/2),
	\label{BW1}
\end{align}
	where, ${\rm erf}(z)=(2/\sqrt{\pi})\int_{0}^{z}\exp({-x^2})dx$, is the error function \cite{abr}. We define the acceptor bandwidth by $\delta_a= 2\sqrt{\pi/ \alpha}$. Employing this expression in Eq. (\ref{BW1}), we obtain,
\begin{align}
	\varrho_0 = \frac{4N_a+2}{\delta_a \, {\rm erf}(\sqrt{\pi})}.
	\label{R1}
\end{align}
Using the result (\ref{R1}), Eq. (\ref{G1a}) can be written as,
\begin{align}
\varrho(E) =   \frac{4N_a+2}{\delta_a \, {\rm erf} (\sqrt{\pi}  )}\, e^{-4\pi (E-E_0^{(a)})^2/\delta_a^2}.
\label{G21}
\end{align}
Note, that the factor, ${\rm erf} (\sqrt{\pi})$, is very close to one: ${\rm erf} (\sqrt{\pi})\approx 
0.988$.

The expression (\ref{R1}) allows us to establish a formal relation between the number of levels, $(2N_a+1)$, in the acceptor of a discrete model (\ref{H1}) and the density of states, $\varrho_0$, at the center  of the acceptor band, in a continuum approach (\ref{ah1}). This relation is used below, in  numerical simulations, for comparison of these two approaches.\\

{\bf \em{Limit $N_a\rightarrow\infty$}.} It follows from (\ref{R1}), that when the acceptor bandwidth is finite, $\delta_a>0$, then the density of states satisfies $\varrho_0\rightarrow\infty$ when $N_a\rightarrow\infty$, which should be expected. In what follows, we will require that, in this limit, the {variance} of energy (\ref{H1}) (at $\lambda_{a,d}=0$) is finite. This requirement involves the initial state of the system. Suppose that initially only {the} donor is populated. {From  (\ref{H1}) the variance of energy in the donor state, $|d\rangle$, is}  
 \begin{align}
 	D_d\equiv\langle d|\mathcal H^2|d\rangle-\langle d|\mathcal H|d\rangle^2=\nonumber\\
 	\sum^{N_a}_{n=-N_a}|V_{n}|^2=(2N_a+1)|V|^2\equiv v^2{\rm erf}(\sqrt{\pi}),
 	\label{D}
 	\end{align}
 	where we have used, for simplicity: $V_{n}=V$, and 
 	$v^2=V^2{\varrho_0\sqrt{\pi/\alpha}}$ (see below). The requirement $D_d= \rm 
 	const$  means that, when $N_a\rightarrow\infty$, the matrix elements, 
 	$V_n\rightarrow 0$ (or $V\rightarrow 0$), the renormalized matrix element, 
 	$v={\rm const.}$

\begin{figure}[tbh]
\scalebox{0.425}{\includegraphics{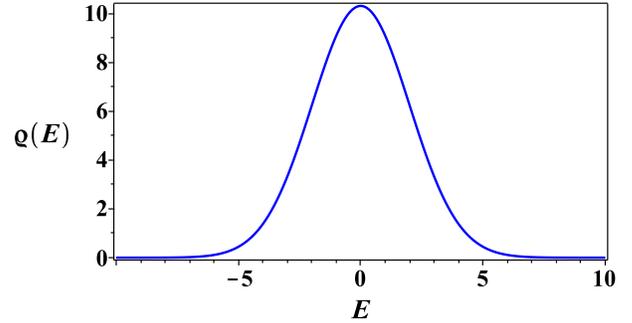}}
\caption{(Color online) Dependence of  the density of states, $\varrho(E)$, on the energy, $E$. Parameters: $N_a=25$, $\delta_a=10$.
\label{Gd1}}
\end{figure}

 In Fig. \ref{Gd1}, the density of states for the acceptor band, centered at $E_0^{(a)} =0$, is  depicted. As one can see, the density of states, $\varrho(E)$, very quickly goes to zero  outside of the 
 interval $(-\delta_a/2, \delta_a/2)$. This supports our choice of the relation between $\alpha$ and the  acceptor bandwidth: $\delta_a= 2\sqrt{\pi/ \alpha}$. 

We denote  the donor and acceptor populations (occupation probabilities) at time $t$, averaged over the noise, by $p_d(t)$ and $p_a(t)$. Suppose the donor has a single energy level and the acceptor has a continuous energy band. The total donor-acceptor state $\rho(t)$ has matrix elements we denote by $\rho_{dd}(t)$, $\rho_{dE}(t)$, $\rho_{Ed}(t)$ and $\rho_{EE'}(t)$.  We have $p_d(t)= \langle \rho_{dd}(t)\rangle $ and $\rho_a(t) =  \int\varrho(E) \langle \rho_{EE}(t)\rangle dE = 1-p_d(t)$. 

We obtain the following system of integro-differential equations for the 
average of the populations (for technical details, see SM)
\begin{align} \label{IB4a}
\frac{d}{dt} p_d(t) =&- \int_0^t { K_1}(t,t') p_d(t') dt' \nonumber \\
&+\int_0^t { K_2}(t,t') p_a(t')  dt' ,\\
\frac{d}{dt} p_a(t) =& \int_0^t { K_1}(t,t') p_d(t') dt' \nonumber \\
 & -\int_0^t {K_2}(t,t') p_a(t') dt'.
\label{IB4b}
\end{align}

The kernels, $K_{1,2}(t,t')$, are found to be,
\begin{align}\label{IK1}
  K_1(t,t') = & 2v^2\Phi(t,t') \cos (\varepsilon(t-t'))  \exp\Big ({-\frac{(t-t')^2}{4\alpha}}\Big) , \\
   K_2(t,t') =  &2v^2\Phi(t,t') \cos (\varepsilon(t-t'))  \exp\Big ({-\frac{t^2 +t'^2}{4\alpha}}\Big),
   \label{IK2}
\end{align}
where,  
\begin{align}\label{v}
\varepsilon= E_0^{(d)} - E_0^{(a)}, ~~v^2 = V^2 \varrho_0 \sqrt{\pi/\alpha},
\end{align}
 and  {$\Phi(t,t') = \langle e^{i\varphi(t)} e^{i\varphi(t')}\rangle$, $\varphi(t) =(g_d-g_a) \int_0^t \xi(s) ds$,} is the characteristic functional of the random process. {To arrive at Eqs. \eqref{IB4a} - \eqref{v}, we have modeled the noise by a random telegraph process which guarantees a certain splitting of correlations.}

 {\bf \em{Rate-type equations}.} One can show (for detail see SM.) that if
  \begin{align}
 \Big |\int_{0}^{t} (t-t') K_{1,2}(t,t') d t' \Big | \ll 1,~~t\in (0, \infty),
  \label{Ker}
  \end{align}
 then  the integro-differential equations (\ref{IB4a}) and (\ref{IB4b}) are well  approximated by the rate-type ordinary differential equations
\begin{align} \label{PN6ar}
\frac{d}{dt} p_d =&- {\mathfrak R}_1(t) p_d + {\mathfrak R}_2(t) p_a , \\
\frac{d}{dt} p_a = & \ \ {\mathfrak R}_1(t) p_d - {\mathfrak R}_2(t) p_a ,
\label{PN7r}
\end{align}
where $\mathfrak {R}_{1,2}(t) =\int_0^t K_{1,2}(t,t') d t'$. We call the functions, $\mathfrak {R}_{1,2}(t)$, the ``ET dynamical rates". The rate-type equations (\ref{PN6ar}) and (\ref{PN7r}) remind us of the Master equations (18) in \cite{Struve} (see also \cite{Forster2}). 

The solution of Eqs. (\ref{PN6ar}) and (\ref{PN7r}) can be written as,
\begin{align}
 p_d(t) = & e^{-f(t)}\big ( 1 + \int^t_0{\mathfrak R}_2(s) e^{f(s)} ds\big 
),\\
 p_a(t) =& 1 - p_d(t),
\label{C6b}
\end{align}
where, $f(t) = \int^t_0\big({\mathfrak R}_1(\tau)+ {\mathfrak R}_2(\tau)\big)d\tau$.

\subsection*{On the conditions of applicability of the rate-type equations}  

The strong conditions of validity of approximation leading to Eqs. 
(\ref{PN6ar}) and   
(\ref{PN7r}) can be written as (see SM for technical details),
\begin{align}\label{pa1}
 \frac{| \Delta \rho(t)|}{p_a(t)}  \ll 1, \quad {\rm and} \quad \frac{v}{p} \ll 1,
	\end{align}
where $p= \sqrt{\delta_a^2/4\pi + 2(D\sigma)^2}$,  and we set $\sigma^2 = 
\chi(0)$, and $D= g_d -g_a$. The perturbation, $\Delta \rho(t)$, is defined by Eq. (65) in SM.

As one can see, the left inequality in (\ref{pa1}) depends on time, and both conditions impose the limitations on the parameters $v$, $\varepsilon$, and $\delta_a$. The first inequality in Eq. (\ref{pa1}) is related to the derivation of Eqs. (\ref{IB4a}) and (\ref{IB4b}). Unfortunately, the explicit analytical form of this condition  cannot be obtained, and even its numerical analysis is rather complicated. The condition of applicability, presented by Eq.  (\ref{Ker}),  leads to the second inequality in Eq. (\ref{pa1}):  $v \ll p$. It allows us to replace the system of integro-differential equations,  (\ref{IB4a}) and (\ref{IB4b}), by the rate-type equations (\ref{PN6ar}) and (\ref{PN7r}).
 
\subsection*{Estimates of the ET rates}

To estimate the ET dynamical rates,
\begin{align}
{\mathfrak R}_{1,2}(t) = \int_0^t K_{1,2}(t,t') d t' ,
\end{align}
we use a Gaussian approximation to calculate the characteristic functional  
\cite{NB1,NBSS},
\begin{align}\label{a9a}
\Phi(t,t')=\big \langle e^{i\kappa(\tau)} \big\rangle = \exp\bigg(- \frac{D^2\sigma^2 (t-t')^2}{2}\bigg),
\end{align}
where $\kappa(\tau)=\kappa(t-t')=D\int_0^{(t-t')}\xi(\tau')d\tau'$,  $\sigma^2 = \chi(0)$, and $D= g_d - g_a$.

The Gaussian approximation is valid if the following conditions hold: 
\begin{align}
|D\sigma| \ll \frac{d}{dt}\ln \chi(t)\Big |_{t=0} \ll
\frac{3p^2}{2\varepsilon},
\label{Dsigma}
\end{align}
where 
\begin{align}
 p = \sqrt{1/\alpha +2(D\sigma)^2}=  \sqrt{\delta_a^2/4\pi +2(D\sigma)^2},
 \label{p}
\end{align}
{and we recall that $\chi(t-t')=\langle \xi(t)\xi(t')\rangle$.}

The parameter, $p$, includes contributions of the acceptor bandwidth and the intensity of noise, and  it plays an important role in characterizing   the ET rate in all cases considered below. 

After some computation, the ET dynamical rates become:
\begin{widetext}
\begin{align}
{\mathfrak R}_1(t)=& \frac{\sqrt{\pi}\,{v}^2}{p}\exp\bigg(-\frac{\varepsilon^2}{p^2}\bigg)\bigg({\rm erf}\Big(\frac{pt}{2} + i\frac{\varepsilon}{p} \Big )+{\rm erf}\Big(\frac{pt}{2} - i\frac{\varepsilon}{p} \Big )\bigg),
\label{IBR1}\\
{\mathfrak R}_2(t)=& \frac{\sqrt{\pi}\,{v}^2}{p}\Bigg\{\exp\Big( -\frac{t^2}{2\alpha}+\frac{(t/2\alpha- i\varepsilon)^2}{p^2}\Big)\bigg({\rm erf}\Big(\frac{pt}{2} - \frac{(t/2\alpha-
 i\varepsilon)}{p} \Big ) + {\rm erf}\Big(\frac{(t/2\alpha- i\varepsilon)}{p} \Big )\bigg)  +\rm  
 c.c.\Bigg\}.
\label{IBR1a}
\end{align}
\end{widetext}

From the properties of  the error function it follows that the asymptotic values of rates, $\Gamma_{1,2}=\lim_{t\rightarrow \infty}{\mathfrak R}_{1,2}(t)$, are:
\begin{align}
 \Gamma_1= &\frac{2\sqrt{\pi}{v}^2}{p} e^{-\varepsilon^2/p^2},
 \label{G1} \\
\Gamma_2 =&\left \{ \begin{array}{cc}
\Gamma_1,& \rm if \quad {\delta_a} =0,\\
0,& \rm if\quad {\delta_a} \neq 0,
\end{array}
\right.
\label{G2}
\end{align}
where $0\leq p<\infty$. 
Note, that, at a given value of $\varepsilon$, $\Gamma_1(p=0)=0$ and $\Gamma_1(p\rightarrow\infty)\rightarrow 0$.  
  The maximum of the asymptotic ET rate (\ref{G1})  corresponds to the ``resonant" condition, 
\begin{align}
p_{res}= \sqrt{2} \varepsilon,
\label{res}
\end{align}
and the maximum of the ET rate (\ref{G1}) is,
\begin{align}
\Gamma_1^{(max)}=\sqrt{{2\pi}\over{e}}{{v^2}\over{\varepsilon}}.
\label{max}
\end{align}
We find that the leading terms in Eqs. (\ref{IBR1}) and (\ref{IBR1a}), as $t\rightarrow \infty$, 
are: 
\begin{align}
{\mathfrak R}_1(t)= & \Gamma_1\Big (1+{\mathcal O\Big( \frac{e^{-pt}}{\sqrt{\pi}\,pt}\Big)} 
\Big),
\label{IBR3}\\
{\mathfrak R}_2(t)\sim & \Gamma_1\cos\Big (\frac{\varepsilon t}{\alpha p^2}\Big)\exp\Big( -\nu t^2\Big),
\label{IBR3a}
\end{align}
where,
\begin{align}
\nu =\frac{2p^2 - 1/\alpha}{4\alpha p^2}.
\label{nu1}
\end{align}

  \subsection*{\bf Note on the asymptotic limits} 
  The expression (\ref{G1}) for $\Gamma_1$ can be considered as a generalized  ET rate for a single-level  (``zero bandwidth") donor  and a finite bandwidth acceptor. As we show below, it reduces to (i) the Wigner-Weisskopf ET rate for an infinitely wide acceptor band, and for any intensity of dimer-noise interaction constant, (ii) the modified Wigner-Weisskopf (or F\"orster-type) ET rate for weak dimer-noise interaction and finite bandwidth, and (iii) the Marcus-type ET rate, for relatively strong dimer-noise interaction, when the regular perturbation approach cannot be used. 
  
  It is important to note, that in the case of a single-level donor and a finite band acceptor, the ET rate, (\ref{G1}), is finite even at $t\rightarrow\infty$.  This results in re-population of the donor-acceptor complex during the time-interval $0\leq t<\infty$.  As we show below, this situation changes for a finite bandwidth donor.
  On the other hand, for any finite bandwidth, $\delta_a \neq 0$, of the acceptor, the asymptotic rate satisfies $\Gamma_2 = 0$. This rate provides a re-population of the ET dynamics only during finite times. 

Our analytical results, (\ref{G1}) and (\ref{G2}), are confirmed by the results of our numerical simulations for a continuous model, (\ref{PN6ar}) and (\ref{PN7r}), presented in Figs. \ref{Fig20} - \ref{Fig3c}, for different values of parameters characterizing the bandwidth and the intensity of noise.  

In the numerical simulations, we measure all energy parameters in units of $\rm ps^{-1}$, and time is measured in $\rm ps$. The values of parameters in the energy units can be obtained by multiplying our values by $\hbar\approx 6.58\times 10^{-13} \rm meV$. For example, $\varepsilon=60 \rm ps^{-1}\approx 40\, \rm meV$.

\subsection*{Dependence of asymptotic ET rate on acceptor bandwidth and  strength of dimer-noise interaction}

We use  the expression (\ref{G1}) to describe the  ET rates of the system. In the rest of the paper, it is assumed that initially only {the} donor is populated. We will analyze two limits: the ``narrow" acceptor band, when  $\delta_a\ll \varepsilon$ (the donor and the acceptor band do not overlap) and  the ``wide" acceptor band, when $\delta_a\gg \varepsilon$ (the donor overlaps with the acceptor band). We also define the ``weak dimer-noise interaction" regime by $|D\sigma|\ll \delta_a/\sqrt{8 \pi}$, and the ``strong dimer-noise interaction" regime to be the opposite limit, when $ |D\sigma|\gg \delta_a/\sqrt{8 \pi}$.


\subsubsection{Weak dimer-noise interaction}

For weak dimer-noise interaction, one can neglect the contribution of $D\sigma$, and write {$p \approx \delta_a/2\sqrt{\pi}$}.  Substituting $p$ into Eq. (\ref{G1}), we obtain,
\begin{align}
 \Gamma_1= \frac{4\pi v^2}{\delta_a}\, e^{-4\pi \varepsilon^2/\delta_a^2}.
 \label{sw}
 \end{align}
 
{\bf \em F\"{o}rster-type ET rate.} If we take into account the relation, ${v}^2 ={V^2} \varrho_0 \sqrt{\pi/\alpha}$, then expression (\ref{sw}) becomes
 \begin{align}
 \Gamma_1= {2{\pi}{V}^2}{\varrho_0} e^{-4\pi\varepsilon^2/\delta_a^2}. 
 \label{Forster1}
 \end{align}
 The ET rate (\ref{Forster1}) can be interpreted as the Wigner-Weisskopf-type ET rate with the renormalized electron density of states, $\varrho_0\rightarrow \varrho_0\exp(-4\pi\varepsilon^2/\delta_a^2)$. 
 We will call it the F\"{o}rster-type ET rate. It differs from the Wigner-Weisskopf-type ET rate by the factor, $\exp(-4\pi\varepsilon^2/\delta_a^2)$, which characterizes the overlap  between the donor energy level and the acceptor band. 
 \begin{figure}[tbh]
\scalebox{0.425}{\includegraphics{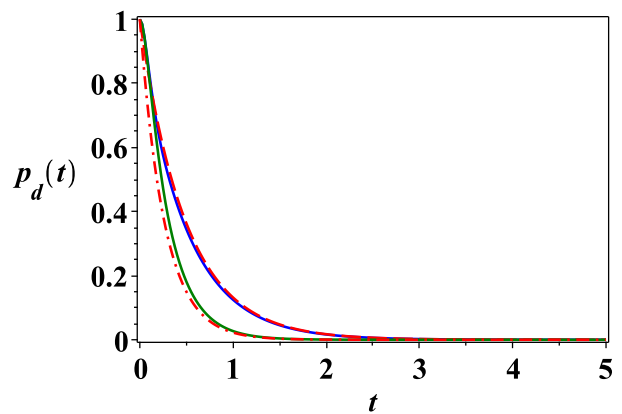}}
\caption{(Color online)  Dependence of $p_d(t)$  on time, for the F\"{o}rster-type ET rate: 
solutions of Eqs. (\ref{PN6ar}) and (\ref{PN7r}) (green and blue solid curves). Parameters: $ v=5$, $D\sigma = 0.01$,  $\delta_a = 50$; $\varepsilon = 10$ (green and red dash-dotted curves); 
$\varepsilon = 15$ (blue and red dashed curves). Plots presented by dashed and dash-dotted 
curves correspond to the rate given by Eq. (\ref{Forster1}).
\label{Fig20}}
\end{figure} 

In Fig. (\ref{Fig20}), we illustrate the F\"{o}rster-type ET dynamics for weak dimer-noise interaction and for a partial overlapping of the donor level with the acceptor band ($\varepsilon\ll\delta_a$). One can see a good agreement between the solutions of Eqs. (\ref{PN6ar}) and (\ref{PN7r}) (green and blue solid curves) with the plots presented by dashed and dash-dotted curves corresponding to the rate given by Eq. (\ref{Forster1}).\\

{\bf \em Wigner-Weisskopf-type ET rate.} For $\varepsilon \ll \delta_a/2\sqrt{\pi}$ (very wide band), the ET rate can be written as, 
 \begin{align}
 \Gamma_1= {2{\pi}{V}^2}{\varrho_0}=4\pi v^2/\delta_a,
 \label{Wigner}
 \end{align}
 which coincides with the Wigner-Weisskopf-type ET rate \cite{WW,SM}. 
 
 {\bf\em Relation to Heisenberg uncertainty principle.} Note, that the rate (\ref{Wigner}) has a simple connection with the Heisenberg uncertainty relation. Indeed, $\Gamma_1$ (\ref{Wigner}) can be written as:
 \begin{align}  
 \Gamma_1=4\pi(v/\delta_a)v<v, 
 \label{up}
 \end{align}
 where we have used the inequality, $v\ll\delta_a$, which means that  $\sqrt{D_d}$ in (\ref{D}), is significantly smaller than the acceptor bandwidth.   
 If we use for tunneling time in (\ref{Wigner}), $\Delta t_d\sim 1/\Gamma_1$, and for energy uncertainty from (\ref{D}), $\Delta E_d=\sqrt{D_d}\sim v$, we have from (\ref{up}) the Heisenberg uncertainty relation: $\Delta E_d\Delta t_d>1$. \\
 \begin{figure}[tbh]
\scalebox{0.425}{\includegraphics{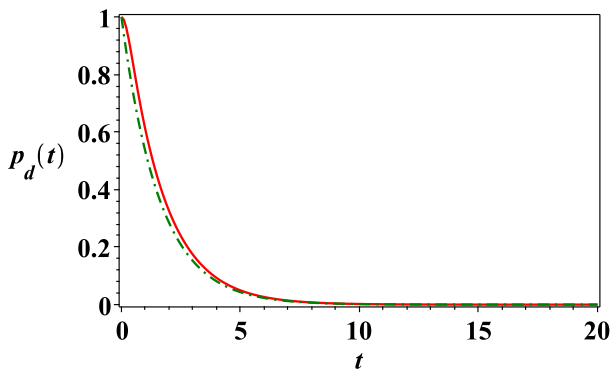}}
\caption{(Color online)  Dependence of $p_d(t)$  on time, for the Wigner-Weisskopf-type ET 
rate: solution of Eqs. (\ref{PN6ar}) and (\ref{PN7r}) (red curve); green dash-dotted curve 
corresponds to the asymptotic rate (\ref{Wigner}). Parameters: $v=1$, $\varepsilon = 0$, 
$D\sigma = 0.01$,  $\delta_a = 20$.
\label{Fig21}}
\end{figure} 

{\bf \em Fermi's golden rule.} The expression (\ref{Wigner}) means that the probability of population of the acceptor band is  $p_a(t)=1-\exp({-\Gamma_1 t})$. For $\Gamma_1t\ll 1$, we get  $p_a(t)\approx\Gamma_1 t$, which corresponds to the Fermi's golden rule \cite{Andrews}.

The dynamics of the donor population is illustrated, for weak dimer-noise interaction and wide band (the Wigner-Weisskopf limit), in Fig. \ref{Fig21}. The following parameters were chosen:  $v=1$, $\varepsilon = 0$, $D\sigma = 0.01$ and $\delta_a = 20$. In this case, noise does not make a contribution to the ET rate, and the population of the acceptor is determined by the ``entropy factor" -- the continuous electron energy spectrum of the acceptor band. 

The case of weak dimer-noise interaction and narrow band, which is close to a two-level system, will be discussed below, in Sec. III.

\subsubsection{Strong dimer-noise interaction}

In this case {we have $p\approx\sqrt2 D\sigma$} and for both, narrow and wide bands,  {(\ref{G1}) gives} the Marcus-type expression for the ET rate,
\begin{align}
\Gamma_1=&{v}^2\sqrt\frac{2\pi }{D^2\sigma^2}\exp\bigg(-\frac{\varepsilon^2}{2D^2\sigma^2}\bigg).
 \label{ns}
 \end{align}
 This result is similar to the Marcus ET rate for a donor-acceptor complex, modeled by a two-level system, having strong interaction with the environment, \cite{HDR,MBS}, even though for the latter one considers a thermal noise.
 
 As one can see from (\ref{ns}), the dimer-noise interaction, $D\sigma$,  represents, in the Marcus-type limit, a ``singular perturbation", and the ET rate (\ref{ns}) cannot be derived by using a regular perturbation theory in $D\sigma$.\\

\begin{figure}[tbh]
\scalebox{0.425}{\includegraphics{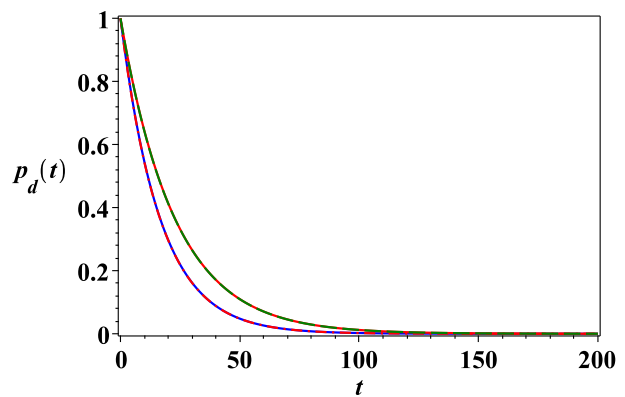}}
\caption{(Color online)  Marcus-type ET rate. Strong dimer-noise interaction and narrow band. Dependence of $p_d(t)$ on time, $t$ ($v=1$, $\varepsilon=25$, $\delta_a =5$). From the top to bottom, $D\sigma = 50$ and $D\sigma= 25$. Solid curves correspond to the solutions of Eqs. (\ref{PN6ar}) and  (\ref{PN7r}). Dashed curves correspond to the rate given by Eq. (\ref{ns}).
\label{Fig3a}}
\end{figure}

\begin{figure}[tbh]
\scalebox{0.425}{\includegraphics{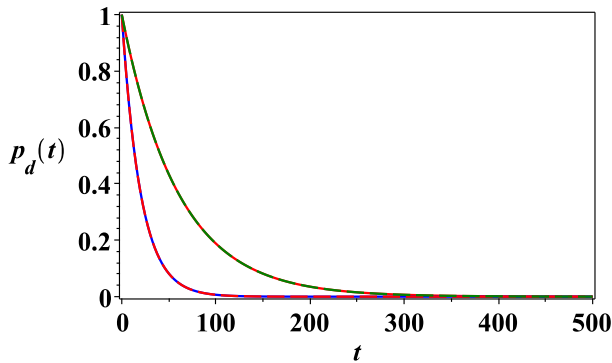}}
\caption{(Color online) Marcus-type ET rate. Strong dimer-noise interaction and relatively wide band. Dependence of $p_d(t)$ on time, $t$ ($v=1$, $\varepsilon=1$, $\delta_a =10$). From the top to bottom, $D\sigma = 150$ and $D\sigma=50$. Solid curves correspond to solutions of Eqs. (\ref{PN6ar}) and  (\ref{PN7r}). Dashed curves correspond to the rate given by Eq. (\ref{ns}).
\label{Fig3c}}
\end{figure}

Our analytic prediction (\ref{ns}) is confirmed by the numerical simulations of Eqs. (\ref{PN6ar}) and (\ref{PN7r}), presented in Fig. \ref{Fig3a} (for strong dimer-noise interaction and relatively narrow band) and in Fig. \ref{Fig3c} (for strong dimer-noise interaction and wide band). In Fig. \ref{Fig3a}, $p_d(t)$ is shown for different values of the dimer-noise interaction constant, $D\sigma$, and for a relatively narrow band, $\delta_a$. In Fig. \ref{Fig3c}, the donor population is given for a relatively wide acceptor band.  

\subsubsection*{Note on reconstruction energy}

Our approach, which models the protein-solvent environment by an external classical noise, instead of the thermal bath described by non-commuting quantum bosonic operators (such as in \cite{HDR}), leads to a zero reconstruction energy in the Marcus-type expression for the ET rate  (\ref{ns}). As shown in \cite{NBSS} (see Eq. (40)), one  way to introduce a reconstruction energy in our approach is to use classical noise with $\langle \xi(t)\rangle\equiv \bar{\xi}\not=0$. In this case, the renormalized redox potential becomes, $\varepsilon\rightarrow\varepsilon-\lambda$, where $\lambda=D\bar{\xi}$ is the ``reconstruction energy". Then, the ET rate in  (\ref{ns}) can be formally rewritten as \cite{NBSS},
\begin{align}
\Gamma=&{v}^2\sqrt\frac{2\pi}{D^2\sigma^2}\exp\bigg(-\frac{(\varepsilon-\lambda)^2}{2D^2\sigma^2}\bigg).
 \label{ns1}
 \end{align}
At the same time, the ``reconstruction energy" in (\ref{ns1}) differs from the reconstruction energy, $\varepsilon_{rec}$, in Marcus theory, where $\varepsilon_{rec}\propto\lambda^2$, and has a different meaning \cite{HDR,MBS}.

\section{Comparison of continuum and discrete models}

In this section, we compare the discrete system governed by the Hamiltonian (\ref{H1}) with the corresponding continuous model, described by the rate-type Eqs. (\ref{PN6ar}) and (\ref{PN7r}). We assume, for simplicity, that the amplitudes of transitions, $V_{n}$, are the same for all acceptor levels, $V_{n} =V$.

We consider the noisy environment to be described by a random telegraph process with the correlation function given by, 
\begin{align}
	\chi(t- t') = \sigma^2 e^{-2\gamma |t -t'|},
	\label{chi}
\end{align}
where $\sigma$ is the amplitude of noise, and $\gamma$ is the decay rate of correlations.
 Then, the equations of motion can be written as {follows} (for details see Ref. \cite{G0}). {It is convenient to introduce the subindex $0$ to denote the donor level, while $n=1,\ldots,2N_a+1$ denotes the $n$th acceptor level (this notation is different from the one previously used in the paper). Then,}
\begin{align}
\langle\dot\rho_{ {0}0}\rangle =&iV\sum_{n=1}^M(\langle\rho_{ {0}n}\rangle 
-\langle\rho_{n {0}}\rangle ),
\label{Teq1}\\
\langle\dot\rho_{nn'}\rangle =&-i\epsilon_{nn'}^{} \langle\rho_{nn'}\rangle +
iV\big(\langle \rho_{n {0}}\rangle  -\langle\rho_{ {0}n'}\rangle \big),
\nonumber  \\
\langle\dot\rho_{ {0}n}\rangle =&
i\epsilon_{n {0}}^{}\langle\rho_{ {0}n}\rangle +
iV\big(\langle\rho_{ {00}}\rangle 
-\sum_{n'= 1}^M\langle\rho_{n'n}\rangle \big)
-iD\sigma\langle{\rho}^\xi_{ {0}n}\rangle ,
 \nonumber \\
\langle{\dot\rho}^\xi_{ {00}}\rangle =&-2\gamma\,\langle{\rho}^\xi_{ {00}}\rangle +
iV\sum_{n=1}^M(\langle{\rho}^\xi_{ {0}n}\rangle -\langle{\rho}^\xi_{n {0}}\rangle ,
\nonumber \\
\langle{\dot\rho}^\xi_{nn'}\rangle =&(i\epsilon_{n'n}
-2\gamma)\langle{\rho}^\xi_{nn'}\rangle 
+iV\big(\langle {\rho}^\xi_{n {0}}\rangle - \langle{\rho}^\xi_{ {0}n'}\big),
\nonumber \\
\langle{\dot\rho}^\xi_{ {0}n}\rangle = &(i\epsilon_{n 
{0}}^{}-2\gamma)\langle{\rho}^\xi_{ 
{0}n}\rangle 
+ iV\big(\langle{\rho}^\xi_{ {00}}\rangle 
-\sum_{n'=1}^M\langle{\rho}^\xi_{n'n}\rangle \big)  
\nonumber \\
&-iD\sigma\langle\rho_{ {0}n}\rangle ,
\label{Teq2}
\end{align}
where $\epsilon_{n'n}=E^{(a)}_{n'}-E^{(a)}_n$, $\epsilon_{n0}=E^{(a)}_{n}-E^{(d)}_0$ 
and  {$\langle\rho^\xi_{n n'}\rangle = 
\langle\xi(t)\rho_{n,n'} (t)\rangle$.}
\begin{figure}[tbh]
\scalebox{0.42}{\includegraphics{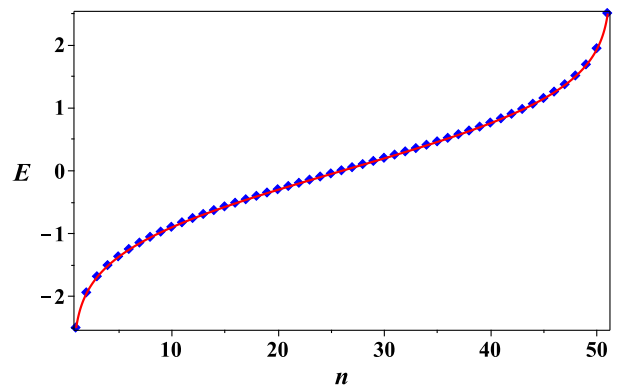}}
\caption{(Color online)  Energy distribution inside the acceptor band: discrete band (blue diamonds), continuous band (red solid curve). Parameters: $N_a = 25$, $\delta_a = 5$.
\label{Fig30}}
\end{figure}
Further, we assume that $M=2N_a+1$ acceptor levels are distributed inside the band, 
centered 
at the point $E^{(a)}_0=E^{(a)}_{N_a + 1}$, and according to the 
Gaussian distribution (\ref{G1a}). We find the following relation between the label $n$ of the energy levels and their energies, 
\begin{align}
\label{mq1}
	n= N_a+1 + \frac{N_a  {\rm erf }\Big (\displaystyle\frac{2\sqrt{\pi} (E^{(a)}_n - 
	E^{(a)}_{N_a + 1})}{\delta_a} \Big )}{\rm erf (\sqrt{\pi} )}. 
\end{align}
Solving this relation for $E\equiv E_n^{(a)}$ yields the curve plotted in Fig. \ref{Fig30}.

Employing Eq. (\ref{R1}), one can recast the renormalized amplitude of transition, $v^2 = V^2 
\varrho_0 \sqrt{\pi/\alpha}$, as 
\begin{align}
\label{mq2}
 v = V \sqrt{\frac{2N_a+1}{\rm erf (\sqrt{\pi })}}.  
\end{align}

According to the definitions in ({\ref{v}), and the expressions, (\ref{Dsigma}) and 
(\ref{chi}), the conditions of validity of the continuum approximation, {\eqref{pa1},} are 
given  by, 
\begin{align}\label{V1s}
&\frac{| \Delta \rho(t)|}{p_a(t)}  \ll 1, \quad {\rm and} \quad \frac{v}{p} \ll 1, \\
&|D\sigma| \ll 2\gamma \ll
\frac{3p^2}{3\varepsilon},
\label{V1aa}
\end{align}
where $\Delta \rho(t)$, is defined below  Eq. (51) in SM.

It is important to note, that in the region of parameters (\ref{V1aa}), the ET rate (\ref{G1}) does not depend on the decay rate, $\gamma$, of the noise correlation function (\ref{chi}).
In our numerical simulations, presented in Figs. \ref{Fig33} --  \ref{Fig35}  for selected parameters, the second condition in Eq. (\ref{V1s}) and the inequalities{, (\ref{V1aa})}, are satisfied for all cases, except for those presented in Figs. \ref{Fig34} and \ref{Fig3hr2}. The first condition in Eq. (\ref{V1s}) is analyzed in the  Appendix in SM.  

In Fig. \ref{Fig31}, we {plot} the population distribution, $\langle \rho_{ii}(t)\rangle$, inside the discrete acceptor band, 
for different times. As one can see, the evolution of $\langle \rho_{ii}(t)\rangle$ approaches, for large times, an ``equal distribution" \cite{G0} {(black diagonal crosses in Fig. \ref{Fig31}, for $t=1000$).  The energy level of the donor and of each acceptor is populated with equal probability, $1/(2N_a+2)$. With $N_a=25$, this gives
$\langle \rho_{ii}(t\rightarrow\infty)\rangle=1/(2N_a+2)\approx 0.019$.}

\begin{figure}[tbh]
\scalebox{0.42}{\includegraphics{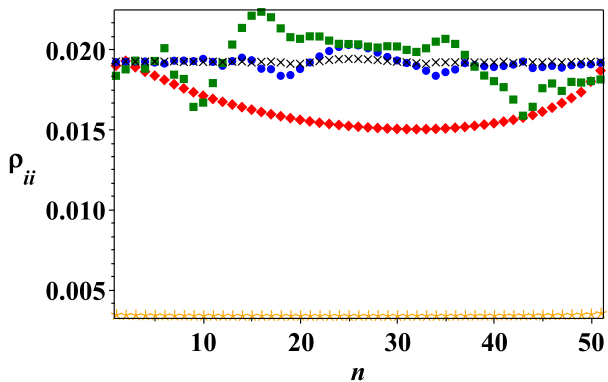}}
\caption{(Color online) Population distribution inside the discrete acceptor band, for different times, $t$: $t=1$ (orange asterisks), $t=10$ (red diamonds), $t=100$ (green boxes), $t=500$ (blue circles), $t=1000$  (black diagonal crosses). Parameters: $V=0.25$, $\varepsilon = 20$, $D\sigma = 40$,  $\gamma = 30$, $\delta_a = 5$, $N_a = 25$.
\label{Fig31}}
\end{figure}

In Figs. \ref{Fig33} - \ref{Fig34}, we compare the results of numerical 
simulations for discrete and continuous acceptor bands, for the Marcus-type ET rate, for different parameters, and for $N_a=50$. So, the total number of levels in the acceptor band is, $2N_a+1=101$. When both conditions of validity of continuous approximation, (\ref{V1s}) and 
(\ref{V1aa}), hold, one can observe a good agreement between {the} discrete and continuum models (Figs. \ref{Fig33} and \ref{Fig32}).  

In Fig. \ref{Fig34}, we compare the regime of the Marcus-type ET for discrete and continuous 
models, and for the choice of parameters when the first inequality in (\ref{V1s}) is violated, and 
the rate-type Eqs. (\ref{PN6ar}) and (\ref{PN7r}) cannot be used. {(For detail see Appendix C in 
SM .)}

\begin{figure}[tbh]
\scalebox{0.425}{\includegraphics{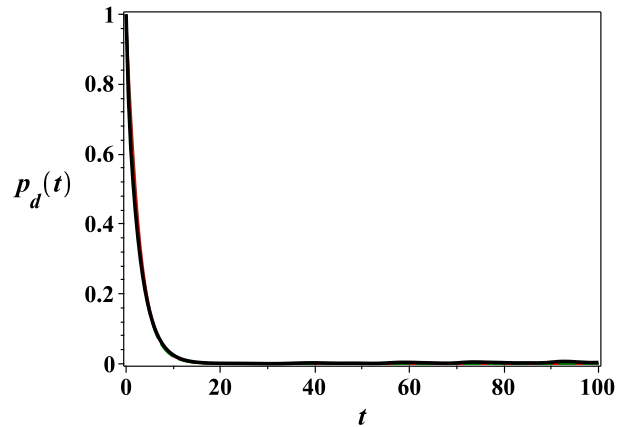}}
\caption{(Color online) The Marcus-type ET. Dependence of the
donor probability, $p_d(t)$,  on  time. Discrete band (blue curve), continuum band (red curve). Green dashed curve corresponds to an exponential decay with the rate given by Eq. (\ref{ns}). Parameters: $V=0.25$, $(v= 2.528)$, $\varepsilon = 10$, $D\sigma = 40$,  $\gamma = 60$, $\delta_a = 5$, $N_a=50$.
\label{Fig33}}
\end{figure}

\begin{figure}[tbh]
\scalebox{0.45}{\includegraphics{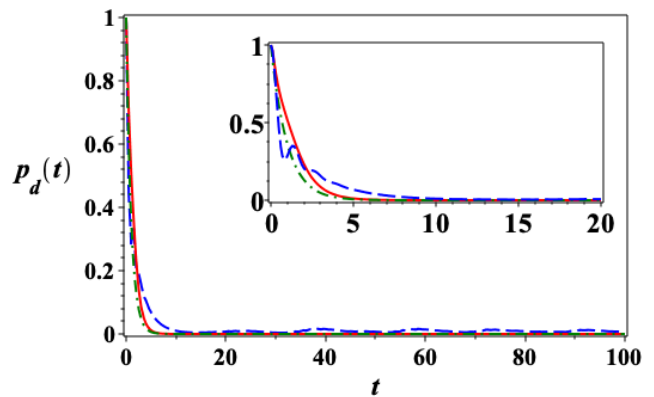}}
\caption{(Color online) The Marcus-type ET. Dependence of the 
donor probability, $p_d(t)$,  on  time. Discrete band (blue dashed curve), continuum band (red solid curve). Green dash-dotted curve corresponds to an exponential decay with the rate given by Eq. (\ref{ns}). The inset is a zoom of the main figure, to show the initial system behavior.  Parameters: $V= 0.2$, $(v= 2.02)$, $\varepsilon = 1$, $D\sigma = 10$,  $\gamma 
=30$, $\delta_a = 5$, $N_a=50$.
\label{Fig32}}
\end{figure}

\begin{figure}[tbh]
\scalebox{0.425}{\includegraphics{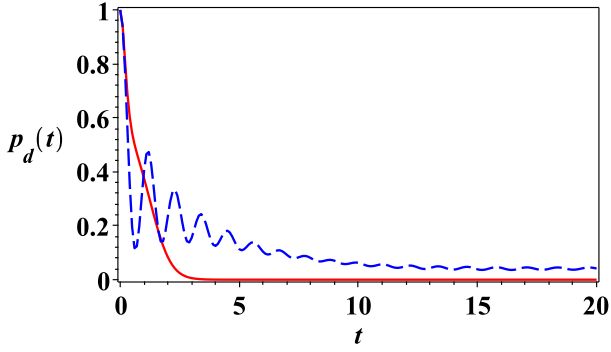}}
\caption{(Color online) Dependence of the donor probability, $p_d(t)$,  on  time, when the condition (\ref{V1s}) of applicability  of the continuum approach  is violated. Discrete band (blue dashed curve), continuous band (red solid curve).  Parameters: $ V=0.249$, $(v= 2.5)$, $\varepsilon = 0$, $D\sigma = 5$,  $\gamma =15$, $\delta_a = 5$, $N_a = 50$.
\label{Fig34}}
\end{figure}

\subsection*{\em  Intermediate ET dynamics for weak and strong dimer-noise interaction and narrow acceptor band}

\begin{figure}[tbh]
\scalebox{0.425}{\includegraphics{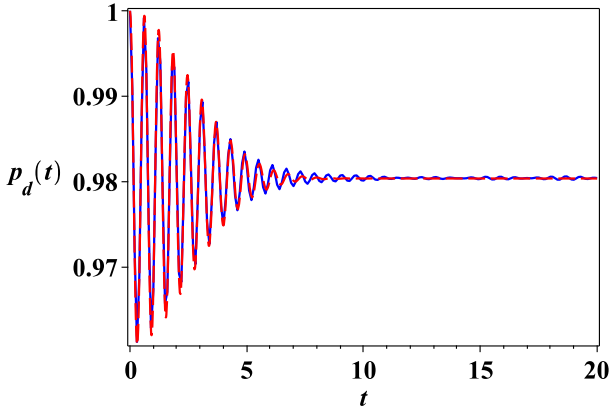}}
\caption{(Color online) The two-level system limit: narrow acceptor band and weak noise. 
Dependence of $p_d(t)$ on time. Parameters: $N_a=50$, $\varepsilon=10$, $\delta_a=2$, 
$D\sigma=0.005$, $\gamma =0.01$, $V =0.1$.  
Blue curve presents the solution of the discrete model, described by the 
system of Eqs. (\ref{Teq1}) and (\ref{Teq2}). Red dashed curve corresponds to the analytical 
expression (\ref{tls}). 
\label{Fig4e}} 
\end{figure}

For a continuous acceptor band, and for any (even a very small) asymptotic rate, $\Gamma_1$ \eqref{G1}, the probability of population of a single-level donor approaches zero as  $t\rightarrow\infty$: $p_d(t\rightarrow\infty)= 0$. However, our simulations show that there can be a two-scale ET dynamics of the donor-acceptor system. Indeed, suppose that the acceptor band is relatively narrow. Then, {on an {\it intermediate} time scale, the} dynamics {resembles} the dynamics of  {a} two-level system (with $\delta_a\approx 0$, see \eqref{tls}, and {after, on a longer time scale given by  $t_{slow}\sim 1/\Gamma_1$, a slow re-population of the acceptor band occurs}. 

In Fig. \ref{Fig4e}, we show the intermediate dynamics of the initially populated donor for {a} weak dimer-noise interaction constant ($D\sigma=0.01$) and for {a} relatively narrow acceptor band, $\delta_a<\varepsilon$ ($\delta_a=2$, $\varepsilon = 10$). In this case, the intermediate dynamics approaches the dynamics of the corresponding two-level ``donor-acceptor" system, in which noise is absent.  In the latter case, we found empirically that the solution of Eqs. (\ref{Teq1}) and (\ref{Teq2}) for $p_d(t)$ can be approximated as,
   \begin{align}
   \label{tls}
 p_d(t)=&1-\frac{4v^2}{\varepsilon^2+4v^2}\Big (1- e^{-\nu 
  t^2} 
  \cos(\Omega_Rt) \Big),
   \end{align}
where $\nu={\delta^2_a}/{16\pi}$ [see (\ref{nu1})] and $\Omega_R=\sqrt{\varepsilon^2+4v^2}$.
The amplitude of oscillations of $p_d(t)$ is found to be, $4{v^2}/({\varepsilon^2+4v^2})$.  As one can see from (\ref{tls}), the Rabi oscillations decay. For the parameters chosen in Fig. \ref{Fig4e}, $p_d^{min}\approx 0.98$, and the period of oscillations, $T=2\pi/\Omega_R\approx 0.616$. Figure \ref{Fig4e} shows a good agreement of analytical expression (\ref{tls}) with the numerical solution for the discrete band. The decay of the Rabi oscillations resulted from the finite width of the acceptor band. 

\begin{figure}[tbh]
\scalebox{0.4}{\includegraphics{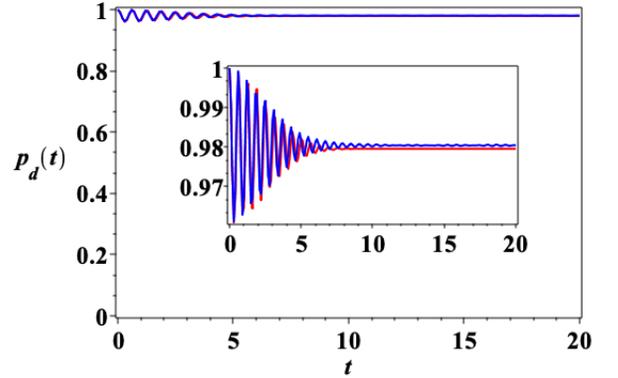}}
\caption{(Color online) The two-level system limit: narrow acceptor band and weak dimer-noise interaction. Dependence of $p_d(t)$ on time. Parameters: $N_a=50$, $\varepsilon=10$, $\delta_a=2$, $D\sigma=0.005$, $\gamma =0.01$, $V =0.1$ ($v=1.01$). Red curve corresponds to the results of rate-type Eqs. (\ref{PN6ar}) and (\ref{PN7r}), which describe the ET in the continuous acceptor band mode. The blue curve presents the solution of the discrete model described by Eqs. (\ref{Teq1}) and (\ref{Teq2}). Inset shows the zoom-in of the main figure.  
\label{Fig4g}}
\end{figure}

\begin{figure}[tbh]
\scalebox{0.415}{\includegraphics{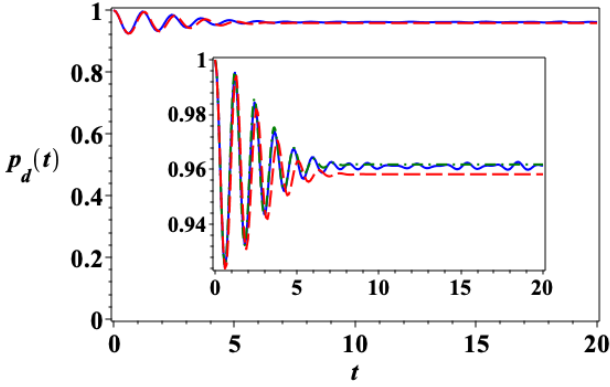}}
\caption{(Color online) The two-level system limit: narrow acceptor band and weak dimer-noise interaction. Dependence of $p_d(t)$ on time, $t$. Parameters: $N_a=25$, $\varepsilon=5$, $\delta_a=2$, $D\sigma=0.01$, $\gamma =0.025$, $V =0.1$ ($v=0.719$). 
The blue curve presents the solution of the discrete model described by Eqs. (\ref{Teq1}) and (\ref{Teq2}). The red dashed curve corresponds to the analytical expression (\ref{tls}).  The green dash-dotted curve corresponds to the results of Eqs. (\ref{PN6ar}) and (\ref{PN7r}) (continuous 
acceptor band). The inset shows the zoom-in of the main figure.
\label{Fig3hr1}}
\end{figure}

\begin{figure}[tbh]
\scalebox{0.285}{\includegraphics{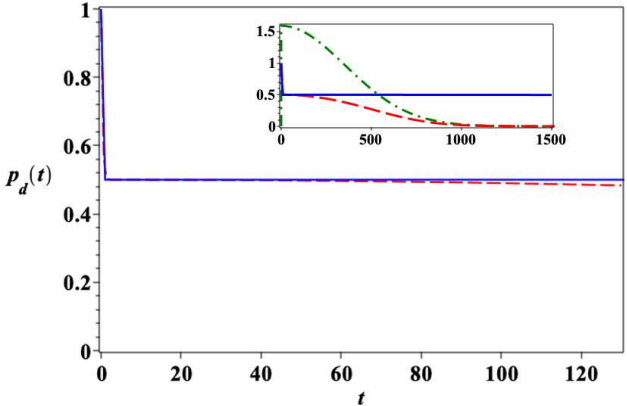}}
\caption{(Color online) The two time-scale behavior of the 
system for strong dimer-noise interaction and narrow band. Parameters: 
$N_a=50$, $\varepsilon=5$, $\delta_a=0.01$, $D\sigma=40$, 
$\gamma =60$, $V =0.5$ ($v=5.06$). Red dashed  curve 
corresponds to the solution of Eqs. (\ref{PN6ar}) and (\ref{PN7r}) (continuous band).  Blue curve represents the solution of the discrete model described by the system of Eqs. (\ref{Teq1}) and (\ref{Teq2}). The inset shows the asymptotic behavior of the continuum model (red dashed curve), of the discrete system (blue solid curve, $N_a=25$) and of the dynamical rate $\mathfrak R_2(t)$  (green dash-dotted curve).}
\label{Fig3hr2}
\end{figure}

In Figs.  \ref{Fig4g} and \ref{Fig3hr1}, we compare the results of {the} numerical simulations for discrete and continuum acceptor bands, for {a} narrow acceptor band and for {a} weak dimer-noise interaction, and when the conditions of applicability  of the continuum approximation, (\ref{V1s}) and (\ref{V1aa}), are satisfied. One can observe a good agreement between discrete and continuum models.  
\begin{figure}[tbh]
\scalebox{0.425}{\includegraphics{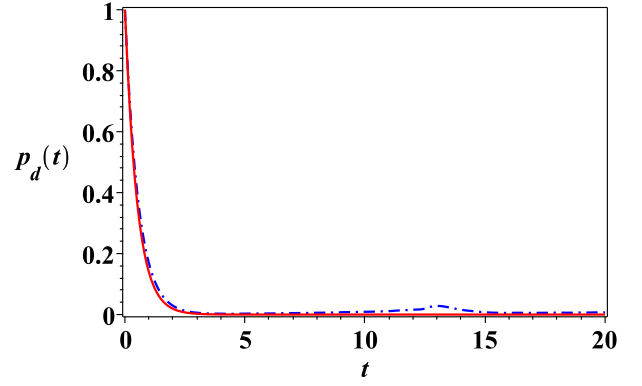}}
\caption{(Color online) The Marcus-type ET, for strong dimer-noise interaction and narrow band. Dependence of $p_d(t)$ on time. Parameters: $N_a=50$, $\varepsilon=150$, $\delta_a=100$, $D\sigma=150$, $\gamma 
=100$, $V=1.39$ ($v=14$). Blue dash-dotted line presents the solution of the discrete model described by the system of Eqs. (\ref{Teq1}) and (\ref{Teq2}). Red curve corresponds to the results of Eqs. (\ref{PN6ar}) and (\ref{PN7r}) (continuous band). 
\label{Fig3q}}
\end{figure}

As was mentioned above, for very large times, the probability of the donor population, $p_d(t\rightarrow\infty)=0$, and the acceptor becomes populated. 
For parameters chosen in Figs. \ref{Fig4e}, \ref{Fig4g} and \ref{Fig3hr1}, 
the characteristic timescale, at which the ET dynamics approaches its intermediate asymptotics [take $t\rightarrow\infty$ in \eqref{tls}]
\begin{align}\label{WN}
p_d \approx \frac{\varepsilon^2}{\varepsilon^2+4v^2},
\end{align}
can be estimated as: $t_{sat}=1/\sqrt{\nu}\approx 2/p\approx 4\sqrt{\pi}/\delta_a\approx 3.5$.   The intermediate asymptotics (\ref{WN}) and the saturation time, $t_{sat}$, are  in good agreement with numerical results for both discrete and continuum models.
 
 {In Fig. \ref{Fig3hr2}, we {show} the intermediate ET dynamics close to  the two-level system, for narrow acceptor band and for strong dimer-noise interaction. The  pure two-level system, with $\delta_a=0$, and all other  parameters as in Fig. \ref{Fig34}, experiences, for large enough times, the equal  distribution, $p_d(t)=p_a(t)=1/2$ \cite{G0}. As one can see from  Fig. \ref{Fig3hr2}, at  intermediate time {already}, $t\approx 100$, the equal distribution is approximately realized in a  continuum model. Because of finite width of the acceptor band, $\delta_a$, the probability,  $p_d(t)$, decays in both, discrete and continuous systems. However, an important 
 observation is that both curves start to diverge significantly after $t\approx 100$, due to  inapplicability of the rate-type equations. 
 Similar to Figs. \ref{Fig4e} - \ref{Fig3hr1}, 
the intermediate ET dynamics occurs, after which a slow exponential decay of the donor 
population takes place with a very small ET rate (see inset in Fig. \ref{Fig3hr2}).}

{\bf \em Dimers based on $Chla$ and $Chlb$ molecules in LHCs.}  In Fig. \ref{Fig3q}, we compare discrete and continuum models for a strong dimer-noise interaction and for a narrow band.  The  chosen parameters are: $\varepsilon=150ps^{-1}\approx 100meV$, $v=14ps^{-1}\approx 9.3meV$, $\delta_a=100ps^{-1}\approx 66.7meV$, $D\sigma=150ps^{-1}\approx 100meV$, which are close to the parameters of the donor-acceptor dimers realized by $Chla$ and $Chlb$ molecules in LHCs. (See, for example, \cite{Muh}, and references therein.) For {the} given choice of parameters, the conditions of validity of approximation, (\ref{V1s}) and (\ref{V1aa}), hold, and one can see a good agreement between the two models.

\subsection{Uphill ET $(\varepsilon<0)$}

Until now, we discussed the ``downhill ET" $(\varepsilon>0$) in the donor-acceptor 
photosynthetic complex. This situation in rather common--the energy of the donor 
band is positioned above the energy of the acceptor band. The question arises if it is 
possible to transfer the energy in photosynthetic complexes uphill (``uphill ET", 
$\varepsilon<0$), when the  energy of the donor band is positioned below the energy 
of the acceptor band (see Fig. \ref{Uph}). The answer is positive, and there exists significant experimental and  theoretical research in this field. (See, for example, 
\cite{SBK,MHU,S1,S2,LLC,AKZ,KND,TKK}, and references therein.) Even though many 
mechanisms  of  uphill ET are discussed  {in the literature,} a complete understanding has not yet 
been reached. 

\begin{figure}[tbh]
\scalebox{0.5}{\includegraphics{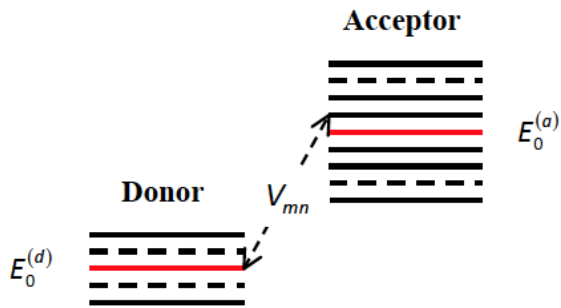}}
\caption{(Color online) Schematic of the uphill ET model consisting of the donor energy band positioned below the acceptor energy band.
\label{Uph}}
\end{figure}
In this sub-section, we discuss the uphill ET mechanism based on the `entropy 
factor''. Namely, we demonstrate that the uphill ET can be realized when the number 
of energy levels in the {higher} positioned acceptor band is larger than the number of 
energy levels in the lower positioned donor band,  and under some additional 
conditions, which can be easily satisfied. 
\begin{figure}[tbh]
	\scalebox{0.5}{\includegraphics{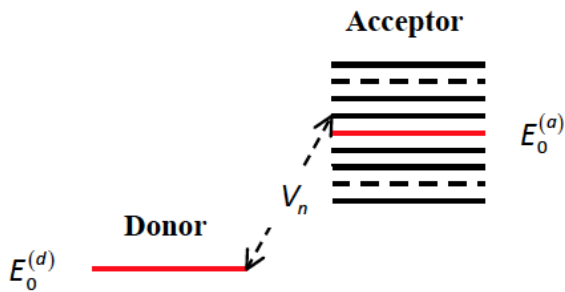}}
	\caption{(Color online) Schematic of uphill ET model consisting of the donor with a single electron energy level positioned below the acceptor energy band.
		\label{Uph1}}
\end{figure}

As was already mentioned, it was shown analytically and demonstrated numerically in \cite{G0} that {for} the asymptotic stationary solution in the discrete system of equations, described by the Hamiltonian (\ref{H1}), {there is} ``equal distribution"  of probabilities of all participating levels. Namely, for non-degenerate energy levels of the acceptor band ($\epsilon_{nn'}=E^{(a)}_n-E^{(a)}_{n'}\neq 0$) at large time $(t \rightarrow \infty)$, the probability of population of any energy level, $l$,  of donor and acceptor, is, 
\begin{align}
\label{mq3}
\langle \rho_{ll}(t \rightarrow \infty)\rangle = \frac{1}{2N_a +2}, 
\end{align}
where, as above, $2N_{a} +1$ is the total number of levels in the acceptor band. This result is independent of whether we consider   downhill ET or uphill ET.

{We also observe} independence of the ET on the sign of $\varepsilon$ in our continuum approach. Namely,  the above introduced dynamical rates, ${\mathfrak R}_1(t)$ and ${\mathfrak R}_2(t)$, do not depend on the sign of $\varepsilon$.  So, the results will be the same for the acceptor band located above or below the donor energy level, with $E_0^{(d)}-E_0^{(a)}=\pm|\varepsilon|$. This means that our results for the ET dynamics, in {the} continuum approach, will be {the same} for both {the} downhill and uphill ET dynamics.

Then, the asymptotic probability to populate an acceptor band is: 
\begin{align}
p_a(t \rightarrow \infty)= \frac{2N_a+1}{2N_a +2}, 
\end{align}
 and  $p_a(t \rightarrow \infty) \approx 1 $, when 
 $N_a\gg 1$.  In the continuum limit, $N_a\rightarrow\infty$, and $p_d(t\rightarrow\infty)=0$.
 
\begin{figure}[tbh]
\scalebox{0.425}{\includegraphics{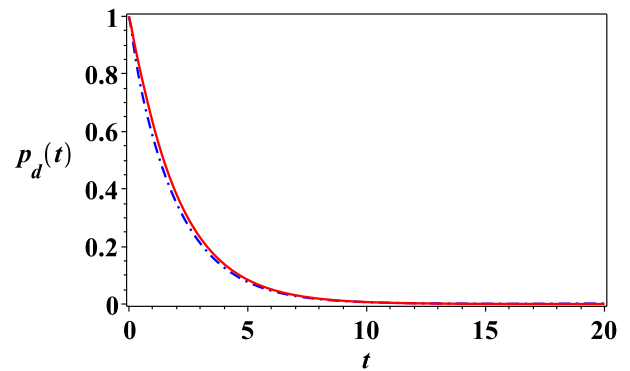}}
\caption{(Color online) Illustration of the uphill ET ($\varepsilon<0$), shown in Fig. \ref{Uph1}.  
Dependence of $p_d(t)$  on  time. Discrete band (blue dash-dotted curve), continuous band 
(red solid curve).  Parameters: $ V=0.25\, (v= 2.528)$, 
$\varepsilon = -20$, $D\sigma = 30$,  $\gamma =30$, $N_a = 50$, $\delta_a = 10$.
\label{Fig35}}
\end{figure}

 In Fig. \ref{Fig35}, we present the results of numerical 
simulations for the uphill ET ($\varepsilon<0$), for the case shown in Fig. \ref{Uph1}: a single-level donor and the acceptor with many quasi-degenerate levels, $2N_a+1$. As one case see, the results for a discrete model (blue dash-dotted curve) are in good agreement with the results of our continuum approach (red solid curve).

A significant difference between the ET dynamics of the discrete and  continuous models  occurs when both the donor and the acceptor have finite bandwidths. This case is discussed in the next section. 

\section{Finite electron bands of donor and acceptor}

\begin{figure}[tbh]
\scalebox{0.5}{\includegraphics{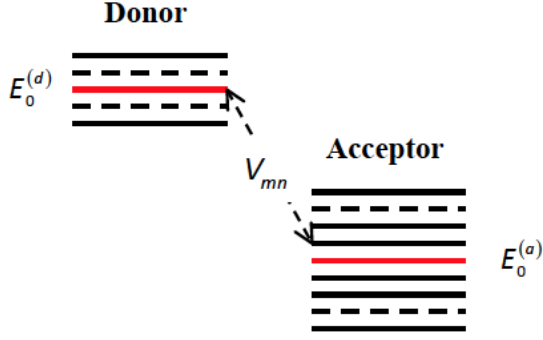}}
\caption{(Color online) Schematic of our model consisting of {a} donor and {an} acceptor with nearly continuous electron energy spectra.
\label{S1A}}
\end{figure}

In this Section, we consider both the donor and the acceptor sites with  $2N_d +1$ and $2N_a +1$ nearly degenerate discrete energy levels, respectively (See Fig. \ref{S1A}). 
The corresponding  Hamiltonian  can be written as:
\begin{widetext}
	\begin{align} \label{H1a}
\mathcal H(t) =  \sum^{N_d}_{m=-N_d}  (E_{m}+\lambda_d(t))|m\rangle\langle  m |  + \sum^{N_a}_{n=-N_a}  (E_{n}+\lambda_a(t))|n\rangle\langle  n | +\sum^{N_d}_{m=-N_d}\sum^{N_a}_{n=-N_a}\big (V_{mn}|m\rangle\langle n |+  V_{nm}
|n\rangle\langle  m| \big).
\end{align}
\end{widetext}
We assume that {the} electron energy spectra of {both the} donor and  {the} acceptor are sufficiently dense, so in Eq. (\ref{H1a}) one can perform {an} integration instead of {the} summation. We have, 
\begin{widetext}
	\begin{align}\label{Ah1}
\mathcal H(t) =  & \int (E_d +\lambda_d(t))|E_d\rangle\langle E_d|\varrho(E_d)dE_d  + \int (E_a +\lambda_a(t))|E\rangle\langle E_a|\varrho(E_a)dE_a \nonumber \\
& + \int\int  dE_d dE_a \varrho(E_d)\varrho(E_a) \Big(V(E_d,E_a)|E_d\rangle\langle E_a| + \rm 
h.c.\Big),
 \end{align}
\end{widetext}
where, $\varrho(E_{d})$, $\varrho(E_{a})$,  are densities of electron states in {the} donor and acceptor bands, {respectively}, and $V_{mn}\rightarrow V(E_d,E_a)$. Further, we assume that the amplitude of transition is a smoothly varying function of energy, so one can approximate $V(E_d,E_a)\approx V$ {is constant}.

For simplicity, we consider  Gaussian densities of states for {the} donor and acceptor bands centered at $E_0^{(a)}$ for the acceptor and at $E_0^{(d)}$ for the donor, 
\begin{align}\label{A_8}
\varrho(E_d) =   \varrho_d e^{-\alpha_1(E_d-E_0^{(d)})^2} , \\
\varrho(E_a) =   \varrho_a e^{-\alpha_2 (E_a-E_0^{(a)})^2} .
\end{align}
The corresponding bandwidths of the donor and the acceptor can be defined as in Sec. II. Namely, the
bandwidth of the donor is: $\delta_d=2\sqrt{\pi /\alpha_1}$,  and the bandwidth of the 
acceptor is: $\delta_a=2\sqrt{\pi/\alpha_2}$.

The {dynamics of the} donor-acceptor complex can be described by  the following rate-type system of ordinary 
differential equations (For detail see SM.). We obtain,
\begin{align} \label{Q1}
\frac{d}{dt}{\langle{\rho}}_{11}\rangle =&- {\mathfrak R}_1(t)\big\langle{\rho}_{11}\big\rangle + {\mathfrak R}_2(t)\big\langle{\rho}_{22}\big\rangle , \\
\frac{d}{dt}{\langle{\rho}}_{22}\rangle = & ~{\mathfrak R}_1(t)\big\langle{\rho}_{11}\big\rangle - {\mathfrak R}_2(t)\big\langle{\rho}_{22}\big\rangle ,
\label{Q2}
\end{align}
where
\begin{widetext}
\begin{align}
{\mathfrak R}_1(t)=& \frac{\sqrt{\pi}\,v^2}{p}\Bigg\{\exp\Big( -\frac{t^2}{2\alpha_1}+\frac{(t/2\alpha_1- i\varepsilon)^2}{p^2}\Big)\bigg({\rm erf}\Big(\frac{pt}{2} - \frac{(t/2\alpha_1-
 i\varepsilon)}{p} \Big ) + {\rm erf}\Big(\frac{(t/2\alpha_1- i\varepsilon)}{p} \Big )\bigg)  +\rm  c.c.\Bigg\},
\label{Q6}\\
{\mathfrak R}_2(t)=& \frac{\sqrt{\pi}\,v^2}{p}\Bigg\{\exp\Big( -\frac{t^2}{2\alpha_2}+\frac{(t/2\alpha_2- i\varepsilon)^2}{p^2}\Big)\bigg({\rm erf}\Big(\frac{pt}{2} - \frac{(t/2\alpha_2-
 i\varepsilon)}{p} \Big ) + {\rm erf}\Big(\frac{(t/2\alpha_2- i\varepsilon)}{p} \Big )\bigg)  +\rm  c.c.\Bigg\}.
\label{Q7}
\end{align}
\end{widetext}
Here we have set, $\varepsilon =E_0^{(d)}- E_0^{(a)}$, and
\begin{align}\label{V1}
v^2= &\frac{\pi\varrho_a \varrho_d |V|^2}{\sqrt{\alpha_{1}\alpha_{2}}} = \frac{1}{4}\varrho_a 
\varrho_d \delta_a \delta_d |V|^2
, \\
p=&\sqrt{ 1/\alpha_1+1/\alpha_2+2(D\sigma)^2} \nonumber \\
=  &\sqrt{\delta_d^2/4\pi + \delta_a^2/4\pi +2(D\sigma)^2}.
\label{V1a}
\end{align}
Note, that  {the} parameters, $v$ and $p$, generalize {those} defined in the previous sections {in the case $N_d=1$}. The difference with the previous sections is that in (\ref{V1}), for finite $\delta_{a,d}$, we require that 
$v^2\approx (2N_d+1)(2N_a+1)|V|^2=\rm const$, when both $N_{a,d}\rightarrow\infty$ 
and $V\rightarrow 0$. As in the previous sections, this is equivalent to the requirement of a finite dispersion for the initial population of the donor band. 

We find that the leading terms in Eqs. (\ref{Q6}) and (\ref{Q7}), as $t\rightarrow 
\infty$, are: 
\begin{align}
{\mathfrak R}_{1,2}(t)\sim & \Gamma\cos\Big (\frac{\varepsilon t}{\alpha_{1,2} 
p^2}\Big)\exp\Big( -\nu_{1,2} t^2\Big),
\label{TB1}
\end{align}
where
\begin{align}\label{TG1}
\Gamma =\frac{2\sqrt{\pi }v^2}{p}e^{-\varepsilon^2/p^2}, \\
\nu_{1,2} =\frac{2p^2 - 1/\alpha_{1,2}}{4\alpha_{1,2} p^2}.
\label{TG2}
\end{align}
Using the relationship between the bandwidths and the parameters $\alpha_{1,2}$,  $\alpha_1 = 4\pi/\delta^2_d$, $\alpha_2 = 4\pi/\delta^2_a$,  one can rewrite
Eqs. (\ref{TB1}), (\ref{TG2}) as,
\begin{align}\label{TB2}
{\mathfrak R}_{1}(t)\sim & \Gamma\cos\Big (\frac{\delta_d^2\varepsilon t}{4\pi p^2} 
\Big)\exp\Big( -\nu_{1} t^2\Big), \\
{\mathfrak R}_{2}(t)\sim & \Gamma\cos\Big (\frac{\delta_a^2\varepsilon t}{4\pi p^2} 
\Big)\exp\Big( -\nu_{2} t^2\Big), 
\label{TB3}
\end{align}
where,
\begin{align}\label{TG3}
\nu_{1} =\frac{\delta_d^2(8 \pi p^2 - \delta_d^2)}{64 \pi^2 p^2}, \quad
\nu_{2} =\frac{\delta_a^2(8 \pi p^2 - \delta_a^2)}{64 \pi^2 p^2}.
\end{align}\\

{\bf \em Conditions of applicability of the rate-type equations.} The conditions of validity of the approximation, leading to Eqs. (\ref{Q1}) and (\ref{Q2}), can be written as (see SM for technical details),
\begin{align}\label{QA1a}
 &|D\sqrt{\chi(0)}| \ll \frac{d}{dt}\ln \chi (t)\Big |_{t=0} \ll
\frac{3p^2}{2\varepsilon}, \\
&\frac{v}{p}\ll 1, \quad {\rm and 
}\quad  \frac{| \Delta \rho(t)|}{p_a(t)}  \ll 1.
 \label{QA1}
	\end{align} 
	They are similar to the conditions given by Eqs. (\ref{pa1}) and (\ref{Dsigma}). 
	However, here the parameters, $v$ and $p$,  are given by (\ref{V1}) and (\ref{V1a}), 
	and the perturbation, $\Delta \rho(t)$, is defined by Eq. (121) in SM.\\
	
Inserting $p_a(t) = 1 -p_d(t)$ into Eq. (\ref{Q1}), we obtain,
\begin{align} \label{Q8}
\frac{d}{dt}p_d =&-( {\mathfrak R}_1(t)+ {\mathfrak R}_2(t)) p_d+ {\mathfrak R}_2(t).
\end{align}
The solution is:
\begin{align}\label{Q9}
p_d(t) = & e^{-f(t)}\big ( p_d(0) + \int^t_0{\mathfrak R}_2(\tau) e^{f(\tau)} d\tau\big 
),
\end{align}
where $f(t) = \int^t_0\big({\mathfrak R}_1(\tau)+ {\mathfrak R}_2(\tau)\big)d\tau$. 

From here it follows, that if, for example, initially the donor was populated, $p_d(0) =1$ (assume that the donor band is populated homogeneously), then, as $t\rightarrow \infty$, the asymptotic population of donor becomes,
\begin{align}\label{Q10}
{\lim_{t\rightarrow\infty}}p_d(t) = & e^{-f(\infty)}\big ( 1 + \int^\infty_0{\mathfrak R}_2({\tau}) e^{f({\tau})} d{\tau}\big).
\end{align}

{\bf \em Peculiarities of the ET dynamics for finite donor and acceptor bandwidths.} When both {the} donor and acceptor bandwidths, $\delta_d$ and $\delta_a$, are finite, the ET dynamics is significantly different from the previous case of a simplified continuum model, with a 
single-level donor. This is mainly {caused by} the fact that both dynamical rates, $\mathfrak R_{1,2}(t)$, in (\ref{TB2}) and (\ref{TB3}) vanish, for the two-band model, at characteristic times, $t_{1,2}\sim 1/\sqrt{\nu_{1,2}}$, {respectively}. This was not the case for {the} simplified model, considered in Secs. II and III. [See Eqs. (\ref{IBR3}), (\ref{IBR3a}), and (\ref{nu1})]. Namely, the parameter, $\Gamma_1$ in (\ref{IBR3}) has {the} meaning of the asymptotic rate. However, a similar (by its form) parameter, $\Gamma$, in (\ref{IBR3a}) is a dynamical rate, which decays in time. This makes the ET dynamics of {the} simplified model and of the model with both finite donor and acceptor bands significantly different. 

The rigorous mathematical approach for the transition from two  discrete donor-acceptor energy bands to a continuum limit (including the intermediate dynamics), when  $N_{d,a}\gg 1$ and $\delta_{d,a}$ are finite, will be discussed in the future. So far, we understand well the situation where the dimer is in contact with a quantum heat reservoir (not classical noise as in the present manuscript)  in the limiting case when both bands are reduced to a single level, having arbitrary degeneracies $N_d$ and $N_a$.  In this situation we can apply the dynamical resonance theory \cite{MSB} and find the population dynamics for all times explicitly, in both the F\"orster and the Marcus regime. Taking this as a starting point, we plan to develop the resonance theory also for finite (small) donor and acceptor bandwidths. We expect to observe the emergence of two time scales, similar to \cite{MSBMulti}. On a shorter time scale, the dynamics shows rich behavior due to the fact that there are multiple quasi-stationary states (corresponding to fully stationary states when the bandwidths collapse to zero).  On a longer time scale the effect of the non-vanishing bandwidths becomes dominant and the dimer is driven to a final, unique stationary state (equilibrium).

Finally, the ``intermediate" ET dynamics, in the continuum two-band model, significantly depends on three parameters, given by Eqs. (\ref{TG1}) and (\ref{TG2}): $\Gamma$, and $\nu_{1,2}$. Indeed, according to Eqs. (\ref{TB2}) and (\ref{TB3}), when $\Gamma\gg\sqrt{\nu_{1,2}}$, the ET dynamics reveals itself on the time-scale, $\sim 1/\Gamma$. And the relaxation effects, related to time-scales, $\sim 1/\sqrt{\nu_{1,2}}$, are not important. In this case, at some additional conditions (see below), one can expect a high efficiency of the ET from donor band to acceptor band. However,  when $\Gamma\ll\sqrt{\nu_{1,2}}$, the ET dynamics is significantly suppressed, as the dynamical rates	(\ref{TB2}) and (\ref{TB3}) quickly decay. In this case, for the initially populated donor band, one cannot expect an efficient ET to the acceptor band. Below, we {plot} the ET dynamics for initial population of the donor band, and for different values of the parameters. \\ 

When $\delta_d\ll\delta_a $, for times $t\gg 1/\delta_a$, one can neglect the contribution of the term $ \int^t_0 {\mathfrak 
R}_2(\tau) e^{f(\tau)} d\tau$, and  (\ref{Q9}) can be simplified  as follows:
\begin{align}\label{Q11}
p_d(t) \approx e^{-\int^t_0{\mathfrak R}_1(\tau)  d\tau}.
\end{align}

For an arbitrary donor bandwidth, we did not succeed to obtain an analytical expression for $p_d(t)$. However,  for $\delta_d \ll \delta_a $ and $D\sigma \gtrsim \varepsilon/2$, we find that as $t\rightarrow \infty$, the donor population can be estimated as,
\begin{align}\label{Q12}
p_d\approx \exp\Big(-\frac{(2\pi)^{3/2} v^2}{p\delta_d }e^{-\varepsilon^2/p^2}\Big).
\end{align}
 This can be recast as follows:
\begin{align}
p_d = \exp\Big(-\frac{\sqrt{2}\pi \Gamma}{\delta_d }\Big),
\end{align}
where $\Gamma$ is defined by Eq. (\ref{TG1}).

Let us assume that both bandwidths are of the same order: $\delta_a \approx\delta_d $. Supposing that $p_d(0) =1$, we obtain from Eq. (\ref{Q9}) the following result:
\begin{align}
p_d(t) =\frac{1}{2}(1+e^{-2\int^t_0{\mathfrak R}_1(\tau)  d\tau} ).
\end{align}
For $\delta_a,\delta_d\ll \varepsilon$ ($\delta_a \approx\delta_d $) and $D\sigma \gtrsim \varepsilon/2$, we find that asymptotically, as $t\rightarrow \infty$, the donor population is given by,
\begin{align}\label{Q13}
p_d(\infty) \approx\frac{1}{2}\Big(1+e ^{-{2\sqrt{2}\pi 
\Gamma}/{\delta_d } }  \Big).
\end{align}

In Figs. \ref{Fig10z} and \ref{Fig10w}, the results predicted by Eqs. (\ref{Q12}), (\ref{Q13}) and {the}
numerical solution of Eq. (\ref{Q8}) are presented. Our numerical simulations show that the 
asymptotic formulas (\ref{Q12}) and (\ref{Q13}) yield a good agreement with the solution of Eq. 
(\ref{Q8}), when $D\sigma \gtrsim \varepsilon/2$.\\

{\bf\em Modified Wigner-Weisskopf, F\"{o}rster, and Marcus-type ET dynamics.} The 
parameters chosen in Fig. \ref{Fig10z} are:  $v=5$, $\delta_d =5$, 
$\delta_a =100$, $D\sigma =5$, $\gamma =10$,   
$\varepsilon=20$ (blue dashed curve) and
$\varepsilon=10$ (green dash-dotted curve). The lower green dash-dotted curve, represents the 
ET dynamics for the narrow donor band overlapped with the wide acceptor band, accompanied 
by a relatively weak dimer-noise interaction. We can say that this ET dynamics is 
of the Wigner-Weisskopf-type or of the F\"{o}rster-type. For $\varepsilon=100$ and $D\sigma 
=100$ (black curve), the donor and acceptor bands do not overlap and we have the 
Marcus-type ET dynamics.

{For the} green dash-dotted curve, {we chose} $\Gamma\approx 1.7$, 
$\nu_1\approx 1$, and $\nu_2\approx 211$. So, we have: $\nu_1<\Gamma\ll\nu_2$. In this 
case, the dynamical rate, $\mathfrak R_2(t)$, decays very fast (due to {the} relatively wide acceptor 
band), on the time-scale, $\sim 1/\sqrt{\nu_2}\approx 0.07$. The rate, $\Gamma$, provides 
the ET from donor band to acceptor band on the time-scale, $1/\Gamma\sim 0.6$. Finally, the 
dynamical rate $\mathfrak R_1(t)$, decays on the time-scale, $\sim 1/\sqrt{\nu_1}\approx 1$, 
and the ET dynamics saturates. In this case, the efficiency of the ET from donor to acceptor is, 
$p_a(t\rightarrow\infty)\rangle\approx 90\%$. (See Fig. \ref{Fig10z}, green 
dash-dotted curve.) 

On the other hand, for  parameters: $v=5$, $\varepsilon=100$, $\delta_d=5$, 
$\delta_a=100$, $D\sigma=100$ the donor and acceptor bands do  not overlap, and the dimer-noise interaction is strong (close to the resonant noise, $p=\sqrt{2}\varepsilon$). So, we 
can say that the upper black curve in Fig. \ref{Fig10z} corresponds to 
the Marcus-type ET dynamics.

In this case, $\Gamma\approx 0.337$, $\nu_1\approx 1$, and $\nu_2\approx 400$. So, we 
have: $\nu_1\lesssim\Gamma\ll\nu_2$. Similar to the previous case,  the dynamical rate, 
$\mathfrak R_2(t)$, decays very fast, on the time-scale $\sim 1/\sqrt{\nu_2}\approx 0.05$. 
Because $\nu_1\lesssim\Gamma$, the ET dynamics saturates, as in the previous case, at the 
time-scale, $\sim  1/\sqrt{\nu_1}\approx 1$. Because $\Gamma$ is smaller than in the 
previous case, the ET efficiency drops to $\approx 30\%$. (See Fig. \ref{Fig10z}, black curve.) 

 \begin{figure}[tbh]
\scalebox{0.41}{\includegraphics{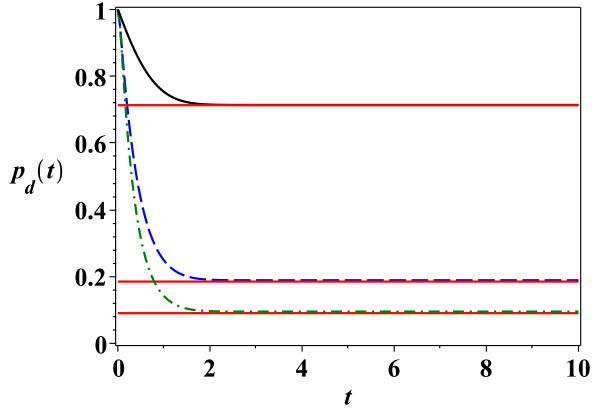}}
\caption{(Color online) Dependence of $p_d$ on time ($v=5$, $\delta_d =5$, 
$\delta_a =100$). Parameters:  $D\sigma =5$,   $\varepsilon=20$, $\gamma =10$ (blue 
dashed curve) and $\varepsilon=10$ (green dash-dotted curve); $D\sigma 
=100$,  $\varepsilon=100$, $\gamma =75$ (upper black solid curve). The 
results of the theoretical predictions, given by Eq. (\ref{Q12}), are depicted by red lines.
\label{Fig10z}}
\end{figure}

\begin{figure}[tbh]
\scalebox{0.425}{\includegraphics{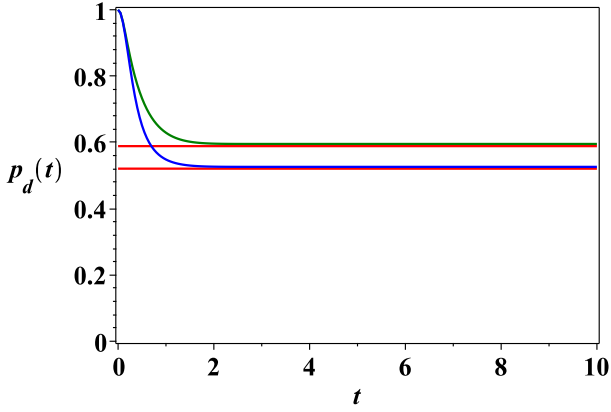}}
\caption{(Color online) Dependence of $p_d$ on time ($v=2$, $\varepsilon=2$, 
$\delta_d =5$,  $\delta_a =5$,  $\gamma = 10$).  From the top to bottom: 
$  D\sigma =10$ (green), $  D\sigma =5$  (blue). The results of the theoretical 
predictions, given by Eq. (\ref{Q13}), are depicted by horizontal red solid lines
\label{Fig10w}}
\end{figure}

In Fig. \ref{Fig10w}, the Marcus-type ET rate is plotted, $\Gamma\approx 
(\sqrt{2\pi}v^2/2D\sigma)\exp(-\varepsilon^2/2D^2\sigma^2)$.  In this case, the dimer-noise 
interaction constant is relatively large, $D\sigma\gg\delta_{d,a}/\sqrt{8\pi}$, for all values of 
$D\sigma$, presented in Fig. \ref{Fig10w}. Because the donor and acceptor bandwidths are 
equal, the efficiency is not high. For example, for $D\sigma=10$, the rate $\Gamma\approx 
0.86$, and $\nu_{1,2}\approx 0.25$. The efficiency in this case is approximately $40\%$. When 
the 
dimer-noise interaction decreases, $D\sigma=5$, we have: $\Gamma\approx 1.59$, and 
$\nu_{1,2}\approx 0.25$. In this case, the efficiency increases, and reaches approximately 
$48\%$. 
\begin{figure}[tbh]
\scalebox{0.425}{\includegraphics{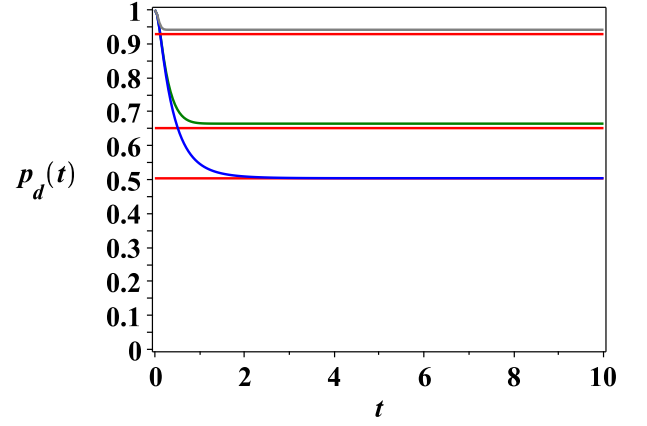}}
\caption{(Color online) Dependence of $p_d$ on time, ($\varepsilon=5$, 
$v=2.5$, $D\sigma =10$, $\gamma = 15$).  From the top to bottom: $ (\delta_d ,  \delta_a 
)=[(50,50), (10,10),(2.5,2.5)]$. The results of the theoretical 
predictions, given by Eq. (\ref{Q13}), are depicted by the horizontal red solid lines.
\label{Fig13a}}
\end{figure}

In Figs. \ref{Fig13a}  -- \ref{Fig12c}, the dependence of the donor population, $p_d(t)$, is shown for different values of parameters. For parameters chosen in Fig. \ref{Fig13a},  both bands have equal widths. For a relatively weak dimer-noise interaction,  the efficiency of population of the acceptor band is small (upper red curve). When the dimer-noise interaction is strong, and both bands are narrow (blue lower curve in Fig. \ref{Fig13a}), the results become close to those of the two-level system. In this case, the populations of donor and acceptor bands become close.  

When the donor band is wider than  the acceptor band (red curve in Fig. \ref{Fig12a}), the acceptor is not efficiently populated. In this case, $\Gamma\approx 3$, $\nu_1\approx 239$, and $\nu_2\approx 90$. Then, in this case, both dynamical rates, $\mathfrak R_{1,2}$ decay fast, and the efficiency of acceptor population is small. 

 Finally, as the bandwidth of the acceptor increases, the population of the acceptor becomes more efficient (blue lower curve in Fig. \ref{Fig12a}). Similar results, with  different intensity of noise, are presented in Fig. \ref{Fig12c}. More efficient population of the acceptor, for $D\sigma=20$, occurs because this case is closer to the resonant one, $p=\sqrt{2}\varepsilon$.\\
\begin{figure}[tbh]
\scalebox{0.4}{\includegraphics{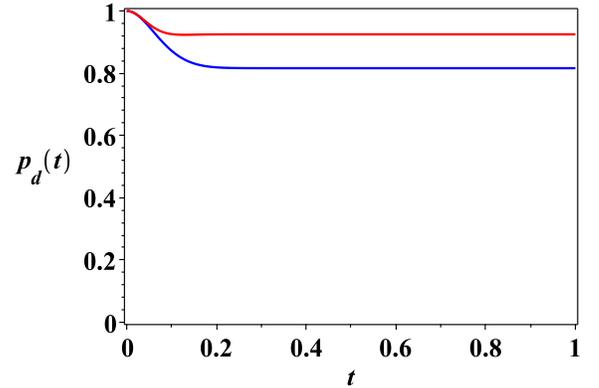}}
\caption{(Color online) Dependence of $p_d$ on time, ($\varepsilon=0$, $v=5$, 
$D\sigma =1$, $\gamma = 10$).  From the top to bottom: $ 
(\delta_d ,  
\delta_a )=[(100,50), (50,100)]$. 
\label{Fig12a}}
\end{figure}
\begin{figure}[tbh]
\scalebox{0.425}{\includegraphics{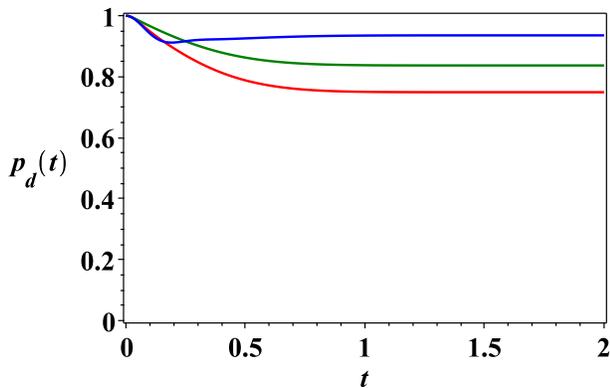}}
\caption{(Color online) Dependence of $p_d$ on time, ($\varepsilon=20$, $v=3$, 
$\delta_d =10$,  $\delta_a =25$, $\gamma = 75$).  From the top to bottom: $  D\sigma 
=5,50,20$. 
\label{Fig12c}}
\end{figure}

{\bf \em Uphill ET dynamics.}  Since the dynamical rates, ${\mathfrak R}_1(t)$ and ${\mathfrak R}_2(t)$, {\eqref{TB1},} do not depend on the sign of $\varepsilon$, the results for the ET dynamics are the same for the acceptor band located above or below the donor band (see Figs. \ref{Uph} and \ref{S1A}). 
In Fig. \ref{Fig14a}, we present the results of numerical 
simulations for the uphill ET ($\varepsilon<0$), for different values of the dimer-noise 
interaction constant, $D\sigma$. For chosen parameters, indicated by the green curve, the 
efficiency of the uphill acceptor population reaches $60\%$.
\begin{figure}[tbh]
\scalebox{0.425}{\includegraphics{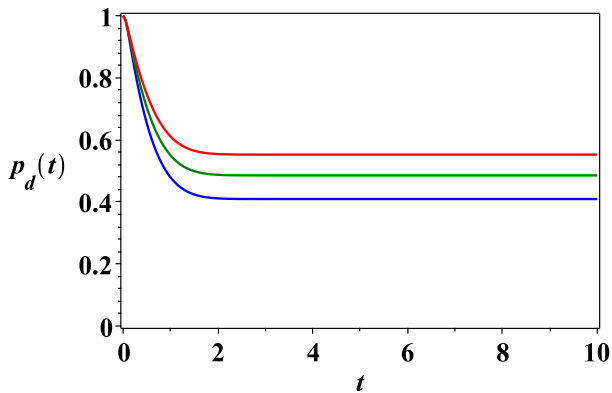}}
\caption{(Color online) Uphill ET for two finite bands. Dependence of $p_d$ on time, 
($\varepsilon=-5$, $v=2.5$, $\delta_d =5$,  $\delta_a =10$, $\gamma = 15$).  From the top 
to bottom: $D\sigma =20,15,10$.
\label{Fig14a}}
\end{figure}

\section{Conclusion}

   In this paper, we have studied the exciton and electron transfer (ET) dynamics in bio-complexes by using a donor-acceptor model in which both {the} donor and {the} acceptor are  represented by continuous energy bands of finite widths. Direct interactions between the donor and the acceptor {are described by matrix elements in the Hamiltonian}. Instead of the thermal bath, we model the protein-solvent environment by a stochastic process  which acts on both the donor and  the acceptor energy levels.

 The usefulness of our approach is that it allowed us (i) to derive a set of simplified rate-type differential equations, which describe the ET dynamics, and (ii) to present  analytically the generalized expressions for the dynamical {(time dependent)} ET rates. {They} are characterized by the redox potential, a sum of the contributions from the dimer-noise interaction constant and the bandwidths of donor and acceptor, and some time-independent decay rates. These generalized ET rates allowed us to derive analytically the ET rates for  Wigner-Weisskopf, F\"{o}rster-type, and Marcus-type limits. 
 
 We presented numerical simulations which illustrate and confirm our analytical results. We demonstrated that by manipulating the bandwidths of donor and acceptor, high efficiency of acceptor population can be achieved for both downhill and uphill, sharp and flat redox potentials.

{We have paid particular attention} to the formulation of the conditions of applicability of the simplified rate-type equations for the ET dynamics.

   Experimental tests of our results would be very useful for a simplified description of complex bio-systems. One possibility is to separate the contributions to the exciton (electron) transfer {coming} from (i) {the} thermal bath and noise, associated with the protein-solvent environment and (ii) {the} entropy factor -- the contribution to the ET rates related to the finite widths of electron donor and acceptor bands. Indeed, as we have demonstrated, the dimer-noise interaction constant and the bandwidths enter the asymptotic expression for the ET rate through dimensionless parameter {called $p$}.  Experiments measuring the ET rate and the efficiency (in which the bandwidths of both donor and acceptor and their overlap can be controlled) can be performed, for example, by using the photosynthetic bio-complexes based on chlorophyll molecules.  Another possibility is to use the artificial nano-systems considered in \cite{MA1,MA2}. In this case, both donor and acceptor are two dye molecules embedded in a} DNA-engineered environment \cite{NS}. In these experiments, the Wigner-Weisskopf, F\"{o}rster-type, and  Marcus-type limits can be studied for downhill and uphill, sharp and flat redox potentials.

\begin{acknowledgements}

  A.I.N. acknowledges the support from the CONACyT. Work by M.M. is supported by {an NSERC  Discovery Grant}. A.I.N. and M.M. are thankful to the Center for Nonlinear Studies at Los Alamos National Laboratory, for support of their visits. This work was supported in part by the U.S. Department of Energy. 

\end{acknowledgements}

\newpage

\begin{widetext}

\section* {\huge Supplemental Material}

In the Supplemental Material we present the technical details of our work. Starting from the 
Liouville equation, $i\dot{ 
\rho} 
= [\tilde{\mathcal H},\rho]$, we derive the dynamics of the average density matrix. We find the 
conditions of validity of the approximations that allow us to replace the exact dynamics (a set 
of coupled integro-differential equations) by rate-type differential equations. We use the 
interaction representation, viewing the 
off-diagonal elements as perturbations. The noise, $\xi(t)$, we consider is stationary, given by 
the  random telegraph process (RTP). It is centered, $\langle\xi(t)\rangle=0$, with correlation 
function,  $\chi(t-t')=\langle \xi(t)\xi(t')\rangle$.

\appendix

\section{A single donor level and a finite acceptor bandwidth}

We start with the simplified model consisting of a single electron energy level of the donor and 
an acceptor consisting of continuous band. Writing the Hamiltonian of the system as, 
$\tilde{\mathcal H}={\mathcal H}_0 + W$, where
\begin{align}\label{C1}
 {\mathcal H}_0\equiv {\mathcal H}_0(t) = & (E^{(d)}_0+\lambda_d(t))|d\rangle\langle d|  + \int 
 (E 
 +\lambda_a(t))|E\rangle\langle E|\varrho(E)dE  \nonumber \\
W=&   \int V|d\rangle\langle E|\varrho(E) dE + \rm h.c.,
 \end{align}
 $\varrho(E)$ being the density of electron states of the acceptor, we define the interaction 
 picture density matrix $\tilde \rho$ and operator $\tilde W$ by,
\begin{align}
\tilde \rho= e^{i\int_0^t {\mathcal H}_0(\tau) d \tau}\rho\, e^{-i\int_0^t{\mathcal H}_0(\tau) 
d\tau },\\
 \tilde W(t)=  e^{i\int_0^t {\mathcal H}_0(\tau) d \tau}W e^{-i\int_0^t {\mathcal H}_0(\tau) 
 d\tau }.
\end{align}
A computation yields,
\begin{align}
\tilde V(E,t) :=& \tilde W_{1E}(t) = V e^{i\varphi(t)} e^{i(E^{(d)}_0 -E)t},\label{m4}\\
 {\tilde \rho}_{1E} = & { \rho}_{1E}  e^{i\varphi(t)} e^{i(E^{(d)}_0 -E)t}, \\
 {\tilde \rho}_{EE'} = & { \rho}_{EE'} e^{i(E -E')t},
 \label{SV1}
\end{align}
where $\varphi(t) =D\int_0^t \xi (t')dt'$ and $D=g_d-g_a$. The subscript $1$ denotes the 
donor level, while the subscript $E$ refers to the continuous energy levels. Note, that 
$\tilde\rho_{11}(t)=\rho_{11}(t)$ and $\tilde\rho_{EE}(t)=\rho_{EE}(t)$ (diagonal density matrix 
elements). 

In the interaction representation,  we obtain the following equations of motion: 
\begin{align}  \label{A1aa}
\frac{d}{dt} \tilde \rho_{11} =& i\int dE \varrho(E)({\tilde \rho}_{1E}{\tilde V}^\ast(E,t)- {\tilde 
V}(E,t) 
{\tilde \rho}_{E1}), \\
\frac{d}{dt} \tilde \rho_{EE'} = &i{\tilde \rho}_{E1}{\tilde V}(E',t)-i  {\tilde \rho}_{1E'} {\tilde 
V}^\ast(E,t),  \\
\frac{d}{dt} \tilde \rho_{1E} =& i {\tilde V}(E,t)\rho_{11}-i \int dE' \varrho(E'){\tilde V}(E',t) {\tilde 
\rho}_{E'E}, \\
\frac{d}{dt}\tilde \rho_{E1} =& -i{\tilde V}^\ast(E,t){\tilde \rho}_{11}+ i \int dE' \varrho(E'){\tilde 
V}^\ast(E',t) {\tilde \rho}_{EE'}.
\label{A1bb}
\end{align}

We take the initial condition  ${\tilde \rho}_{1E}(0)={\tilde \rho}_{E1}(0)=0$ for all $E$. Then, 
using Eqs. (\ref{A1aa}) - (\ref{A1bb}), we obtain,
\begin{align}\label{C1a}
 {\tilde \rho}_{1E} =& i \int_0^t dt'\Big({\tilde V}(E,t'){\tilde \rho}_{11}(t')-\int dE' 
 \varrho(E'){\tilde V}(E',t') {\tilde \rho}_{E'E}(t')\Big), \\
 {\tilde \rho}_{E1} =&- i \int_0^t dt'\Big({\tilde V}^\ast(E,t'){\tilde \rho}_{11}(t')-\int dE' 
 \varrho(E'){\tilde V}^\ast(E',t') {\tilde \rho}_{EE'}(t')\Big).
 \label{C1b}
\end{align}
Now, inserting  (\ref{C1a}) and (\ref{C1b}) into Eqs. (\ref{A1aa}) - (\ref{A1bb}), 
 we obtain the following system of integro-differential equations,
\begin{align}\label{B3a}
\frac{d}{dt}{\tilde \rho}_{11}(t)=&- \int_0^t dt'\int dE \varrho(E)\Big 
({\tilde V}^\ast(E,t){\tilde V}(E,t') + {\tilde V}^\ast(E,t'){\tilde V}(E,t) \Big) 
\tilde \rho_{11}(t') \nonumber \\
&+\int_0^t dt'\iint dE dE' \varrho(E) \varrho(E') {\tilde V}^\ast(E,t) {\tilde V}(E',t') 
{\tilde \rho}_{E'E}(t') +  {\tilde V}(E,t){\tilde V}^\ast(E',t') {\tilde \rho}_{EE'}(t'),\\
\frac{d}{dt}{\tilde \rho}_{22}(t)\rangle =& \int_0^t dt'\int dE \varrho(E)\Big 
({\tilde V}^\ast(E,t){\tilde V}(E,t') + {\tilde V}^\ast(E,t'){\tilde V}(E,t) \Big) \tilde \rho_{11}(t')
\nonumber \\
&- \int_0^t dt'\iint dE dE' \varrho(E) \varrho(E'){\tilde V}^\ast(E,t) {\tilde V}(E',t') 
{\tilde \rho}_{E'E}(t') +  {\tilde V}(E,t){\tilde V}^\ast(E',t') {\tilde \rho}_{EE'}(t'),\\
\frac{d}{dt}{{\tilde\rho}}_{EE'}(t) =& \int_0^t dt'\Big ({\tilde 
V}^\ast(E,t'){\tilde V}(E',t) + {\tilde V}^\ast(E,t){\tilde V}(E',t') \Big) \tilde \rho_{11}(t')
\nonumber \\
&- \int_0^t dt'\int dE'' \varrho(E'') {\tilde V}^\ast(E,t) {\tilde V}(E'',t') {\tilde 
\rho}_{E''E'}(t') +  {\tilde V}(E',t){\tilde V}^\ast(E'',t') {\tilde \rho}_{EE''}(t') .
\label{B4a}
\end{align}
In this supplementary material, it is convenient to denote the average of the total acceptor 
probability by $\langle \tilde\rho_{22}(t) \rangle= \int \langle \tilde\rho_{EE }(t) 
\rangle\varrho(E) dE$. In the main text, we used the symbol $p_a(t)$ for this quantity. 
Averaging over the noise gives
\begin{align}\label{B3}
\frac{d}{dt}{\langle\tilde \rho}_{11}(t)\rangle =&- \int_0^t dt'\int dE \varrho(E)\big\langle\Big 
({\tilde V}^\ast(E,t){\tilde V}(E,t') + {\tilde V}^\ast(E,t'){\tilde V}(E,t) \Big) 
\tilde \rho_{11}(t')\big\rangle \nonumber \\
&+\int_0^t dt'\iint dE dE' \varrho(E) \varrho(E') \big\langle{\tilde V}^\ast(E,t) {\tilde V}(E',t') 
{\tilde \rho}_{E'E}(t') +  {\tilde V}(E,t){\tilde V}^\ast(E',t') {\tilde \rho}_{EE'}(t')\big\rangle,\\
\frac{d}{dt}{\langle\tilde \rho}_{22}(t)\rangle =& \int_0^t dt'\int dE \varrho(E)\big\langle\Big 
({\tilde V}^\ast(E,t){\tilde V}(E,t') + {\tilde V}^\ast(E,t'){\tilde V}(E,t) \Big) \tilde 
\rho_{11}(t')\big\rangle 
\nonumber \\
&- \int_0^t dt'\iint dE dE' \varrho(E) \varrho(E') \big\langle{\tilde V}^\ast(E,t) {\tilde V}(E',t') 
{\tilde \rho}_{E'E}(t') +  {\tilde V}(E,t){\tilde V}^\ast(E',t') {\tilde \rho}_{EE'}(t')\big\rangle,\\
\frac{d}{dt}{\langle{\tilde\rho}}_{EE'}(t)\rangle =& \int_0^t dt'\big\langle\Big ({\tilde 
V}^\ast(E,t'){\tilde V}(E',t) + {\tilde V}^\ast(E,t){\tilde V}(E',t') \Big) \tilde \rho_{11}(t')\big\rangle 
\nonumber \\
&- \int_0^t dt'\int dE'' \varrho(E'')\big\langle {\tilde V}^\ast(E,t) {\tilde V}(E'',t') {\tilde 
\rho}_{E''E'}(t') +  {\tilde V}(E',t){\tilde V}^\ast(E'',t') {\tilde \rho}_{EE''}(t')\big\rangle .
\label{B4}
\end{align}
The terms  in (\ref{B3}) - (\ref{B4}), containing products of the form ${\tilde V}^\ast(E,t){\tilde 
V}(E,t')$, give rise to correlators of the form, $\big\langle e^{-i\varphi(t)} 
e^{-i\varphi(t')} \tilde \rho_{11}(t')\big\rangle$ (see \eqref{m4}). To proceed further, we must 
split these correlators. Consider $\big\langle e^{i\varphi(t)} 
e^{-i\varphi(t')} \tilde \rho_{\alpha \beta}(t')\big\rangle $, where the indices $\alpha, \beta$ 
take values in  $\{ 1,2,E,E'\}$. For the random telegraph process (RTP) we have (for details see 
Corollary \ref{mcorlabel} in Appendix A), 
\begin{align}
\big\langle e^{i\varphi(t)} e^{-i\varphi(t')} \tilde \rho_{\alpha \beta}(t')\big\rangle =
\Phi(t -t')  \big\langle\tilde \rho_{\alpha \beta}(t')\big\rangle 
  - \frac{1}{iD\sigma}\frac{d}{dt'} \Phi(t- t')   \big\langle\tilde 
 \rho^\xi_{\alpha \beta}(t')\big\rangle,
\label{RTP1c}
\end{align}
where $\big\langle\tilde 
 \rho^\xi_{\alpha \beta}(t')\big\rangle = (1/\sigma)\big\langle \xi(t')\tilde  
 \rho_{\alpha \beta}(t')\big\rangle $, $\Phi(t - t') =\langle 
 e^{i\varphi(t)}e^{-i\varphi(t')}\rangle$,  and 
 we set $\sigma^2 =  \chi(0)$. 

Using the differential formula (\ref{SD1}), we obtain,
\begin{align}
\Big(\frac{d}{dt} +2\gamma \Big)\big\langle\tilde 
 \rho^\xi_{\alpha \beta}(t)\big\rangle =\Big\langle  {\xi}(t)\frac{d}{dt}\tilde 
 \rho_{\alpha \beta}(t) \Big\rangle.
\label{D1}
\end{align}
 The r.h.s. of this equation can be obtained for Eqs. (\ref{B3a}) -- (\ref{B4a}) by multiplying  
 both sides of these equations  with $\xi(t)$ and then averaging over the random process. 
 The result can be written as, 
 \begin{align}
 \Big\langle  {\xi}(t)\frac{d}{dt}\tilde 
  \rho_{\alpha \beta}(t) \Big\rangle =  |V|^2\Psi_{\alpha\beta}\Big( \big\langle\tilde 
   \rho_{\mu \nu}(t)\big\rangle ,\big\langle\tilde 
   \rho^\xi_{\mu \nu}(t)\big\rangle  \Big ),
 \end{align}
 where we denote by, $ |V|^2\Psi_{\alpha\beta}\Big( \big\langle\tilde 
    \rho_{\mu \nu}(t)\big\rangle ,\big\langle\tilde 
    \rho^\xi_{\mu \nu}(t)\big\rangle  \Big )$, the  result of the procedure described above. 
 Then,  Eq. (\ref{D1}) can be recast as,
 \begin{align}
\frac{d}{dt} \big\langle\tilde 
 \rho^\xi_{\alpha \beta}(t)\big\rangle
  =- 2\gamma \big\langle\tilde 
 \rho^\xi_{\alpha \beta}(t)\big\rangle + |V|^2\Psi_{\alpha\beta}\Big( \big\langle\tilde 
 \rho_{\mu \nu}(t)\big\rangle ,\big\langle\tilde 
 \rho^\xi_{\mu \nu}(t)\big\rangle  \Big ).
 \label{D2}
\end{align}

  As one can see, the solution for $\big\langle\tilde 
 \rho^\xi_{\alpha \beta}(t)\big\rangle $ can be written as,
 \begin{align}
\big\langle\tilde 
 \rho^\xi_{\alpha \beta}(t)\big\rangle = |V|^2 e^{-2\gamma t} F_{\alpha \beta}(t) ,
 \end{align}
where $F_{\alpha \beta}(t)$ obyes the following eqution:
\begin{align}
\frac{d}{dt} F_{\alpha \beta}(t)
  = e^{2\gamma t}\Psi_{\alpha\beta}\Big( \big\langle\tilde 
 \rho_{\mu \nu}(t)\big\rangle ,\big\langle\tilde 
 \rho^\xi_{\mu \nu}(t)\big\rangle  \Big ).
 \label{D3}
\end{align}

Thus, we conclude that Eq. (\ref{RTP1c}) can be recast as,
\begin{align}
\big\langle e^{i\varphi(t)} e^{-i\varphi(t')} \tilde \rho_{\alpha \beta}(t')\big\rangle =
\Phi(t-t') \big\rangle \big\langle\tilde \rho_{\alpha \beta}(t')\big\rangle  + {\mathcal O} (|V|^2).
\label{SC}
\end{align}
 In what follows, we neglect by higher terms, $ {\mathcal O}(|V|^2)$, and write
 \begin{align}\label{SP1}
 \big\langle e^{i\varphi(t)} e^{-i\varphi(t')} \tilde \rho_{\alpha \beta}(t')\big\rangle  \approx
 \Phi(t ,t')\big\langle\tilde \rho_{\alpha \beta}(t')\big\rangle. 
 \end{align}
After splitting of correlations and neglecting the terms ${\mathcal O} 
(|V|^4)$ in Eqs. (\ref{B3}) --  (\ref{B4}), we obtain for the average components of the 
density matrix the system of integro-differential equations:
\begin{align}\label{Eq3}
\frac{d}{dt}{\langle\tilde \rho}_{11}(t)\rangle =&- \int_0^t dt'\int dE \varrho(E)\Big 
(\big\langle{\tilde V}^\ast(E,t){\tilde V}(E,t')\big\rangle + \big\langle {\tilde V}^\ast(E,t'){\tilde 
V}(E,t)\big\rangle \Big) \big\langle\tilde \rho_{11}(t') \big\rangle   \nonumber \\
&+ \int_0^t dt'\iint dE dE' \varrho(E) \varrho(E') \Big (\big\langle{\tilde V}^\ast(E,t) {\tilde 
V}(E',t') \big\rangle \big\langle{\tilde \rho}_{E'E}(t')\big\rangle + \big\langle {\tilde V}(E,t){\tilde 
V}^\ast(E',t') \big\rangle  \big\langle {\tilde \rho}_{EE'}(t')\big\rangle\Big), \\
\frac{d}{dt}{\langle\tilde \rho}_{22}(t)\rangle =&\int_0^t dt'\int dE \varrho(E)\Big 
(\big\langle{\tilde 
V}^\ast(E,t){\tilde V}(E,t')\big\rangle + \big\langle {\tilde V}^\ast(E,t'){\tilde V}(E,t)\big\rangle 
\Big) \big\langle\tilde \rho_{11}(t') \big\rangle   \nonumber \\
&- \int_0^t dt'\iint dE dE' \varrho(E) \varrho(E') \Big (\big\langle{\tilde V}^\ast(E,t) {\tilde 
V}(E',t') \big\rangle \big\langle{\tilde \rho}_{E'E}(t')\big\rangle + \big\langle {\tilde V}(E,t){\tilde 
V}^\ast(E',t') \big\rangle  \big\langle {\tilde \rho}_{EE'}(t')\big\rangle\Big), \\
\frac{d}{dt}{\langle{\tilde\rho}}_{EE'}(t)\rangle =& \int_0^t dt'\Big (\big\langle{\tilde 
V}^\ast(E,t'){\tilde V}(E',t) \big\rangle + \big\langle{\tilde V}^\ast(E,t){\tilde V}(E',t') \big\rangle 
\Big)\big\langle \tilde \rho_{11}(t')\big\rangle \nonumber  \\
&- \int_0^t dt'\int dE'' \varrho(E'') \Big (\big\langle{\tilde V}^\ast(E,t) {\tilde V}(E'',t') \big\rangle 
\big\langle{\tilde \rho}_{E''E'}(t') \big\rangle + \big\langle {\tilde V}(E',t){\tilde V}^\ast(E'',t') 
\big\rangle\big\langle{\tilde \rho}_{EE''}(t')\big\rangle \Big ).
 \label{Eq4}
 \end{align}
 Then, using Eqs. (\ref{Eq3}) - (\ref{Eq4}) and the definition $\tilde V(E,t) 
 =Ve^{i(E^{(d)}_0 
-E)t} e^{i\varphi(t)}$, we obtain
\begin{eqnarray}\label{EqA5}
\frac{d}{dt}{\langle\tilde \rho}_{11}(t)\rangle &=&- 2|V|^2{\rm Re} \int_0^t 
dt'\Phi(t -t')\int dE 
\varrho(E)e^{i(E^{(d)}_0-E)(t-t')} \big\langle\tilde \rho_{11}(t') \big\rangle \nonumber   \\
&&+ 2|V|^2{\rm Re} \int_0^t dt'\Phi(t -t')\iint dE dE' \varrho(E) \varrho(E') 
e^{i(E^{(d)}_0-E)t}  
e^{-i(E^{(d)}_0-E')t'} \big\langle{\tilde \rho}_{EE'}(t')\big\rangle, \\
\frac{d}{dt}{\langle\tilde \rho}_{22}(t)\rangle &=& 2|V|^2{\rm Re} \int_0^t 
dt'\Phi(t -t')\int dE 
\varrho(E)e^{i(E^{(d)}_0-E)(t-t')} \big\langle\tilde \rho_{11}(t') \big\rangle \nonumber   \\
&&- 2|V|^2{\rm Re} \int_0^t dt'\Phi(t -t')\iint dE dE' \varrho(E) \varrho(E') 
e^{i(E^{(d)}_0-E)t}  
e^{-i(E^{(d)}_0-E')t'} \big\langle{\tilde \rho}_{EE'}(t')\big\rangle, \\
\frac{d}{dt}{\langle{\tilde \rho}}_{EE'}(t)\rangle &=&|V|^2\int_0^t dt'\Big[ 
\Phi(t -t') e^{-i(E^{(d)}_0-E)t'} 
e^{i(E^{(d)}_0-E')t}  +  \Phi(t' -t) e^{-i(E^{(d)}_0-E)t}  e^{i(E^{(d)}_0-E')t'}\Big] 
\big\langle\tilde \rho_{11}(t') 
\big\rangle   \nonumber \\
&&- |V|^2\int_0^t dt' \Big[ \Phi(t -t')\int  dE'' \varrho(E'') e^{i(E^{(d)}_0-E')t}  
e^{-i(E^{(d)}_0-E'')t'} 
\big\langle{\tilde \rho}_{E E''}(t')\big\rangle\nonumber\\
&&\qquad\qquad\quad   +\ \Phi(t' -t) \int dE'' \varrho(E'') 
e^{-i(E^{(d)}_0-E)t}  e^{i(E^{(d)}_0-E'')t'} 
\big\langle{\tilde \rho}_{E'' E'}(t')\big\rangle \Big],
\label{EqA6}
\end{eqnarray}
where $\Phi(t -t')=\Big \langle \exp\{iD \int\limits_0^{t-t'}\xi(\tau) d\tau\}\Big 
\rangle  $.

In what follows, we consider a Gaussian  density of states in the  acceptor 
band, centered at some point, $E^{(a)}_0$,
\begin{align}\label{G8}
\varrho(E) =   \varrho_0 e^{-\alpha \big (E-E^{(a)}_0 \big)^2}.
\end{align}
Then we compute
\begin{eqnarray}
\label{A16a}
&\langle \tilde \rho_{22}(t')\rangle = \int\langle\tilde\rho_{E E}(t')\rangle\varrho(E) dE \approx 
\langle\tilde\rho_{E^{(a)}_0E^{(a)}_0}(t')\rangle \varrho_0\sqrt{\frac{\pi }{\alpha}}, \\
 &\int  dE \varrho(E) e^{i(E^{(d)}_0-E)(t-t')} = \varrho_0\sqrt{\frac{\pi 
 }{\alpha}}e^{i\varepsilon(t-t')}  e^{-\frac{(t-t')^2}{4\alpha}},
\end{eqnarray}
and
\begin{eqnarray}
\lefteqn{\iint dE dE'\varrho(E)  \varrho(E')  e^{i(E^{(d)}_0-E)t}  e^{-i(E^{(d)}_0-E')t'} 
\big\langle{\tilde 
\rho}_{EE'}(t')\big\rangle}\nonumber\\
&& \approx \Big(\varrho_0\sqrt{\frac{\pi 
}{\alpha}}\langle\tilde\rho_{E^{(a)}_0E^{(a)}_0}(t')\rangle + 
\frac{i}{2\alpha}t'\langle \tilde X(t')\rangle - \frac{i}{2\alpha}t\langle{\tilde X}^\ast(t')\rangle  
\Big )\varrho_0\sqrt{\frac{\pi }{\alpha}}e^{i\varepsilon(t-t')} 
e^{-\frac{t^2+t'^2}{4\alpha}}\nonumber\\
&& \approx \Big(\langle\tilde\rho_{22}(t')\rangle + \frac{i}{2\alpha}t'\langle \tilde X(t')\rangle - 
\frac{i}{2\alpha}t\langle{\tilde X}^\ast(t')\rangle  \Big )\varrho_0\sqrt{\frac{\pi 
}{\alpha}}e^{i\varepsilon(t-t')} e^{-\frac{t^2+t'^2}{4\alpha}},
  \label{A16b}
\end{eqnarray}
where $\varepsilon = E^{(d)}_0-E^{(a)}_0$ is the difference between the donor energy and the 
center of the acceptor energy band and,
\begin{align}
\langle \tilde X(t')\rangle=\varrho_0\sqrt{\frac{\pi }{\alpha}}\frac{\partial}{\partial 
E'}{\langle\tilde \rho}_{EE'}(t')\rangle\Big |_{E=E'=E^{(a)}_0}.
\end{align}
Using these results and  Eq. (\ref{EqA5}),  we obtain the following integro-differential equations: 
\begin{eqnarray}\label{B4ar}
\frac{d}{dt}{\langle\tilde \rho}_{11}(t)\rangle &=&-2{\rm Re} \int_0^t dt'{\tilde 
K}_1(t,t')\big\langle\tilde \rho_{11}(t')\big\rangle  +2{\rm Re} \int_0^t dt' {\tilde 
K}_2(t,t')\big\langle\tilde \rho_{22}(t')\big\rangle \nonumber\\
&& -\frac{1}{\alpha}{\rm Im} \int_0^t dt'\tilde K_2(t,t') \big( t'\langle \tilde X(t')\rangle 
-t\langle\tilde X^*(t')\rangle\big), \\
\frac{d}{dt}{\langle\tilde \rho}_{22}(t)\rangle &=&2{\rm Re} \int_0^t dt'{\tilde 
K}_1(t,t')\big\langle\tilde \rho_{11}(t')\big\rangle  -2{\rm Re} \int_0^t dt' {\tilde 
K}_2(t,t')\big\langle\tilde \rho_{22}(t')\big\rangle \nonumber\\
&& +\frac{1}{\alpha}{\rm Im} \int_0^t dt'\tilde K_2(t,t') \big( t'\langle \tilde X(t')\rangle 
-t\langle\tilde X^*(t')\rangle\big),
\label{B4br}
\end{eqnarray}
where
\begin{align}\label{AK1a}
  {\tilde K}_1(t,t') = & v^2\Phi(t -t') \exp\Big 
  ({i\varepsilon(t-t')-\frac{(t-t')^2}{4\alpha}}\Big) , \\
 {\tilde K}_2(t,t') =  &v^2\Phi(t -t') \exp\Big ({i\varepsilon(t-t')-\frac{t^2 
 +t'^2}{4\alpha}}\Big).
   \label{AK2a}
\end{align}
Here we denote  ${v}^2 =|V|^2\varrho_0 \sqrt{\pi/\alpha}$. Next, we expand, 
\begin{align}\label{B22a}
\langle\tilde \rho_{11}(t')\rangle \approx&\langle\rho_{11}(t)\rangle - 
(t-t')\frac{d}{dt}{\langle \rho}_{11}(t)\rangle.
\end{align}
Using this expansion and neglecting the higher order terms, we 
find that 
(\ref{B4br})  can be approximated by the ordinary differential equations,
\begin{align} \label{SBN6ar}
\frac{d}{dt}{\langle \rho}_{11}\rangle =&- {\mathfrak R}_1(t)\big\langle 
\rho_{11}\big\rangle - D\sigma \frac{\partial }{\partial  \varepsilon}{\mathfrak 
R}_1(t)\big\langle \rho^\xi_{11}\big\rangle  + {\mathfrak R}_2(t)\big\langle 
\rho_{22}\big\rangle  +  D\sigma \frac{\partial }{\partial  \varepsilon}{\mathfrak 
R}_2(t)\big\langle \rho^\xi_{22}\big\rangle  , \\
\frac{d}{dt}{\langle \rho}_{22}\rangle = & - {\mathfrak R}_2(t)\big\langle 
\rho_{22}\big\rangle -  D\sigma \frac{\partial }{\partial  \varepsilon}{\mathfrak 
R}_2(t)\big\langle \rho^\xi_{22}\big\rangle + {\mathfrak R}_1(t)\big\langle\tilde 
\rho_{11}\big\rangle  + D\sigma \frac{\partial }{\partial  \varepsilon}{\mathfrak 
R}_1(t)\big\langle \rho^\xi_{11}\big\rangle   , \\
\frac{d}{dt}{\langle \rho}^\xi_{11}\rangle =&- 2\gamma {\langle \rho}^\xi_{11}\rangle 
- {\mathfrak R}_1(t)\big\langle 
\rho^\xi_{11}\big\rangle - D\sigma \frac{\partial }{\partial  \varepsilon}{\mathfrak 
R}_1(t)\big\langle \rho_{11}\big\rangle  + {\mathfrak R}_2(t)\big\langle 
\rho^\xi_{22}\big\rangle  +  D\sigma \frac{\partial }{\partial  \varepsilon}{\mathfrak 
R}_2(t)\big\langle \rho_{22}\big\rangle  , \\
\frac{d}{dt}{\langle \rho}^\xi_{22}\rangle = &  - 2\gamma {\langle 
\rho}^\xi_{22}\rangle - {\mathfrak R}_2(t)\big\langle 
\rho^\xi_{22}\big\rangle -  D\sigma \frac{\partial }{\partial  \varepsilon}{\mathfrak 
R}_2(t)\big\langle \rho_{22}\big\rangle +  {\mathfrak R}_1(t)\big\langle\tilde 
\rho^\xi_{11}\big\rangle  + D\sigma \frac{\partial }{\partial  \varepsilon}{\mathfrak 
R}_1(t)\big\langle \rho_{11}\big\rangle , \\
\label{SBN7r}
\end{align}
where $\mathfrak {R}_{1,2}(t) =2{\rm Re} \int_0^t {\tilde K}_{1,2}(t,t') dt'$, ${\langle 
\rho}_{11}\rangle + {\langle \rho}_{22}\rangle =1$ and ${\langle 
\rho}^\xi_{11}(0)\rangle = {\langle \rho}^\xi_{22}(0)\rangle =0$. 

Our numerical simulations show that if the conditions of validity, (\ref{Dsigma}) given below, 
are satisfied, then one can simplify this system as follows:
\begin{align} \label{BN6ar}
\frac{d}{dt}{\langle \rho}_{11}\rangle =&- {\mathfrak R}_1(t)\big\langle
\rho_{11}\big\rangle 
+ {\mathfrak R}_2(t)\big\langle \rho_{22}\big\rangle , \\
\frac{d}{dt}{\langle \rho}_{22}\rangle = & {\mathfrak R}_1(t)\big\langle\tilde 
\rho_{11}\big\rangle - 
{\mathfrak R}_2(t)\big\langle \rho_{22}\big\rangle.
\label{BN7r}
\end{align}
The solution of these equations with the initial condition $\langle{\rho_{11}}(0)\rangle=1$ is 
\begin{align}
{\langle \rho}_{11}(t)\rangle = & e^{-f(t)}\big ( 1 + \int^t_0{\mathfrak R}_2(s) e^{f(s)} 
ds\big ),\\
{\langle \rho}_{22}(t)\rangle =&1 -\langle \rho_{11}(t)\rangle,
\label{DC6b}
\end{align}
where $f(t) = \int^t_0\big({\mathfrak R}_1(\tau)+ {\mathfrak R}_2(\tau)\big)d\tau$.

To proceed further, one needs to know the explicit expression for the 
characteristic functional $\Phi(t -t')$. Using  (\ref{SP1}), we find that 
$\Phi(t)$ obeys 
the following integro-differential equation \cite{KV2,KV3}:
\begin{align}
\frac{d}{dt} \Phi(t) = -D^2\int_0^t \chi(t-t') \Phi(t')dt'  .
\end{align}
One can show that in the time interval, $0<t<\infty$, the Gaussian approximation is valid 
yielding,
\begin{eqnarray}\label{G1}
\frac{d}{dt} \Phi(t) \approx -\Phi(t) D^2\int_0^t \chi(t-t') dt' ,
\end{eqnarray}
 if $D^2\int_0^\infty  t\chi(t)dt \ll 1$. This condition can be recast as follows: 
 $D\sqrt{\chi(0)}\ll 
 1/\tau_{c}$, where  $\tau_c$ is the correlation time \cite{NB1},
\begin{align}
\tau_c = \frac{1}{\chi(0)}\int_0^\infty  \chi(t)dt.
\end{align}

 The solution of Eq. (\ref{G1}) can be written as,
\begin{align}
\Phi(t) =\exp \Big (- D^2 \int_0^t dt'  \int_0^{t'}  dt'' \chi (t'-t'') \Big ).
\end{align}
To obtain the analytic expressions for  the dynamical rates, $\mathfrak {R}_{1,2}(t) =2{\rm Re} 
\int_0^t {\tilde K}_{1,2}(t,t')dt'$, we write
\begin{align}
\Phi(t)=  \exp\Big (-\frac{D^2\chi(0)}{2 }t^2- \frac{D^2\dot\chi(0)}{3! }t^3 + \dots\Big ).
\label{C5}
\end{align}
Performing the integration, one can neglect the higher orders, if  $|\dot\chi(0)t_0 |\ll 3$, where 
$t_0 ={i2\varepsilon}/{p^2}$. Then, the leading contribution in (\ref{C5}) is,
\begin{align}\label{C4}
\Phi(t')= \exp\Big(-\frac{D^2\chi(0) }{2}t^2\Big).
\end{align}
Thus, the Gaussian approximation is valid if the following conditions hold: 
\begin{align}
|D\sqrt{\chi(0)}| \ll \frac{d}{dt}\ln \chi (t)\Big |_{t=0} \ll
\frac{3p^2}{2\varepsilon},
\label{Dsigma}
\end{align}
where $p=\sqrt{1/\alpha +2D^2\chi(0)}$. The parameter, $p$, includes contributions of the 
acceptor bandwidth and the intensity of noise. Next, employing  Eqs. (\ref{AK1a}), (\ref{AK2a}) 
and (\ref{C4}), we find
\begin{align}
{\mathfrak R}_1(t)=& 
\frac{\sqrt{\pi}\,{v}^2}{p}\exp\bigg(-\frac{\varepsilon^2}{p^2}\bigg)\bigg({\rm 
erf}\Big(\frac{pt}{2} + i\frac{\varepsilon}{p} \Big )+{\rm erf}\Big(\frac{pt}{2} - 
i\frac{\varepsilon}{p} \Big )\bigg),
\label{BR1}\\
{\mathfrak R}_2(t)=& \frac{\sqrt{\pi}\,{v}^2}{p}\Bigg\{\exp\Big( 
-\frac{t^2}{2\alpha}+\frac{(t/2\alpha- i\varepsilon)^2}{p^2}\Big)\bigg({\rm erf}\Big(\frac{pt}{2} 
- \frac{(t/2\alpha-
 i\varepsilon)}{p} \Big ) + {\rm erf}\Big(\frac{(t/2\alpha- i\varepsilon)}{p} \Big )\bigg)  +\rm  
 c.c.\Bigg\},
\label{BR1a}
\end{align}
where,  ${\rm erf}(z)= (2/\sqrt{\pi} )\int_0^z e^{-t^2} dt$, is the error function \cite{abr}.

\subsection{Conditions for validity of the approximation}

In this Section, we estimate the validity of the approximation leading to the Eqs. (\ref{BN6ar}) 
and  
(\ref{BN7r}):
\begin{align} \label{SMr}
\frac{d}{dt}{\langle \rho}_{11}\rangle =&- {\mathfrak R}_1(t)\big\langle
\rho_{11}\big\rangle 
+ {\mathfrak R}_2(t)\big\langle \rho_{22}\big\rangle , \\
\frac{d}{dt}{\langle \rho}_{22}\rangle = & {\mathfrak R}_1(t)\big\langle\tilde 
\rho_{11}\big\rangle - 
{\mathfrak R}_2(t)\big\langle \rho_{22}\big\rangle.
\label{SM1a}
\end{align}
We start with Eqs. (\ref{B4ar}) and  (\ref{B4br}) written as,
\begin{align} \label{B21a}
\frac{d}{dt}{\langle\tilde \rho}_{11}(t)\rangle =&- \int_0^t { K}_1(t,t') 
\big\langle\tilde \rho_{11}(t')\big\rangle dt' +  \int_0^t { K}_2(t,t') 
\big\langle\tilde \rho_{22}(t')\big\rangle dt'  + \zeta (t), \\
\frac{d}{dt}{\langle\tilde \rho}_{22}(t)\rangle = &\int_0^t { K}_1(t,t') 
\big\langle\tilde \rho_{11}(t')\big\rangle dt'  - \int_0^t { K}_2(t,t') 
\big\langle\tilde \rho_{22}(t')\big\rangle dt' -  \zeta (t), 
\label{B21b}
\end{align}
where $ \zeta (t) = \tilde \zeta (t) +{\tilde \zeta}^\ast (t)$, and
\begin{align}
\tilde\zeta (t) = i\int_0^t {\tilde K}_2(t,t')\Big (\frac{t'}{2\alpha}\langle{\tilde 
X}^\ast(t')\rangle- 
\frac{t}{2\alpha}\langle \tilde X(t')\rangle  \Big )  dt' .
\end{align}
The solution of the Eqs. (\ref{B21a}) and (\ref{B21b}) can be written as, 
${\langle\tilde \rho}_{11}(t)\rangle = {\langle \rho}_{11}(t)\rangle +\Delta \rho(t)$ and 
${\langle\tilde \rho}_{22}(t)\rangle = {\langle \rho}_{22}(t)\rangle - \Delta \rho(t)$, where 
\begin{align}
\Delta \rho (t) = \int\limits_0^t \zeta(t') dt', 
\end{align}
 and ${\langle \rho}_{11}(t)\rangle$  and  ${\langle \rho}_{22}(t)\rangle$ obey the system of the 
 non-perturbative 
 integro-differential equations:
\begin{align} \label{B10a}
\frac{d}{dt}{\langle  \rho}_{11}(t)\rangle =&- \int_0^t { 
K_1}(t,t')\big\langle  \rho_{11}(t')\big\rangle dt' +\int_0^t { 
K_2}(t,t')\big\langle  \rho_{22}(t')\big\rangle dt' ,\\
\frac{d}{dt}{\langle  \rho}_{22}(t)\rangle =& \int_0^t { 
K_1}(t,t')\big\langle  \rho_{11}(t')\big\rangle dt'  -\int_0^t 
{K_2}(t,t')\big\langle  \rho_{22}(t')\big\rangle dt'.
\label{B10b}
\end{align}
From here it follows that if 
\begin{align}
 \frac{| \Delta \rho(t)|}{{\langle \rho}_{22}(t)\rangle}  \ll 1,\quad 0< 
 t < 
 \infty,
\end{align}
then  Eqs. (\ref{B21a}) and (\ref{B21b})  can be replaced by Eqs.  (\ref{B10a}) and (\ref{B10b}).

Employing Eqs. (\ref{EqA6}) and (\ref{SMr}) and (\ref{SM1a}), we find that $\langle \tilde 
X(t)\rangle$ obeys the equation:
\begin{align} \label{B20a}
\frac{d}{dt}\langle \tilde X(t)\rangle =&- i\int_0^t\big(t{\tilde K}_1(t,t')+ t'{\tilde 
K}^\ast_1(t,t')\big)\big\langle \rho_{11}(t')\big\rangle  dt' +it\int_0^t {\tilde 
K}_2(t,t')\big\langle \rho_{22}(t')\big\rangle dt' .
\end{align}

Next, expanding $\langle \rho_{11}(t')\rangle $ and $\langle \rho_{22}(t')\rangle $ as,
\begin{align}\label{B22c}
\langle   \rho_{11}(t')\rangle \approx&\langle  \rho_{11}(t)\rangle 
-\frac{d}{dt}{\langle  \rho}_{11}(t)\rangle (t-t') ,\\
\langle  \rho_{22}(t')\rangle \approx&\langle  \rho_{22}(t)\rangle 
-\frac{d}{dt}{\langle\tilde \rho}_{22}(t)\rangle (t-t'),
\label{B22d}
\end{align}
 we find that equations (\ref{B10a}) and (\ref{B10b}) can be recast as, 
\begin{align} \label{B24a}
\frac{d}{dt}\langle { \rho}_{11}\rangle =&-{\mathfrak R}_1(t)(1+Y_1(t))\big\langle
\rho_{11}\big\rangle + {\mathfrak R}_2(t)(1+Y_2(t))\big \langle \rho_{22}\big\rangle ,\\
\frac{d}{dt}{ \langle \rho}_{22}\rangle = &{\mathfrak R}_1(t)(1+Y_1(t))\big\langle\tilde 
\rho_{11}\big\rangle - {\mathfrak R}_2(t)(1+Y_2(t))\big \langle \rho_{22}\big\rangle ,
\label{B24b}
\end{align}
where $Y_{1,2}(t) =\int_0^t(t-t'){ K}_{1,2}(t,t')dt'$. The latter can be rewritten as, $Y_{1,2}(t) 
=i(\partial {\tilde{ \mathfrak R}}^\ast_{1,2}(t)/\partial \varepsilon -\partial {\tilde {\mathfrak 
R}}_{1,2}(t)/\partial \varepsilon)$. If 
\begin{align}
\Big |\int_{0}^{t} (t-t') K_{1,2}(t,t') d t' \Big | \ll 1,~~t\in (0, \infty),
 \end{align}
 one can neglect the contribution of the  terms,  $Y_{1,2}(t)$, and the system of 
  integro-differential equations  (\ref{B24a}) and  (\ref{B24b}) can be approximated 
 by  the following pair of ordinary differential equations,
\begin{align} \label{B25a}
\frac{d}{dt}{ \langle \rho}_{11}\rangle =&- {\mathfrak R}_1(t)\big\langle\tilde 
\rho_{11}\big\rangle 
+ {\mathfrak R}_2(t)\big \langle \rho_{22}\big\rangle , \\
\frac{d}{dt}{ \langle \rho}_{22}\rangle = & {\mathfrak R}_1(t)\big\langle\tilde 
\rho_{11}\big\rangle - {\mathfrak R}_2(t)\big \langle \rho_{22}\big\rangle.
\label{B25b}
\end{align}
In the same order of expansion as above, we have
\begin{align} \label{A30a}
\frac{d}{dt}\langle \tilde X(t)\rangle=& -it {\mathfrak 
R}_1(t) \big \langle \rho_{11}(t)\big\rangle - \frac{\partial\tilde {\mathfrak R}^\ast_1(t) 
}{\partial 
\varepsilon} \big \langle \rho_{11}(t)\big\rangle + it\tilde {\mathfrak 
R}_2(t) \big \langle \rho_{22}(t)\big\rangle  , \\
\frac{d}{dt}\Delta \rho(t)  =&-\frac{1}{2\alpha} \frac{\partial\tilde {\mathfrak R}_2(t) 
}{\partial 
\varepsilon} \langle {\tilde X}^\ast(t)\rangle
-\frac{i t}{2\alpha} {\tilde {\mathfrak R}_2(t) }\big (\langle{\tilde X}(t)\rangle- \langle {\tilde 
X}^\ast(t)\rangle  \big )  + \rm c.c.,
\label{A30b}
\end{align}
 where
 \begin{align}
 \tilde {\mathfrak R}_1(t)=& 
 \frac{\sqrt{\pi}\,{v}^2}{p}\exp\bigg(-\frac{\varepsilon^2}{p^2}\bigg)\bigg({\rm 
 erf}\Big(\frac{pt}{2} + i\frac{\varepsilon}{p} \Big )-{\rm erf}\Big(  i\frac{\varepsilon}{p} \Big 
 )\bigg),
 \label{BR1r}\\
 \tilde {\mathfrak R}_2(t)=& \frac{\sqrt{\pi}\,{v}^2}{p}\exp\Big( 
 -\frac{t^2}{2\alpha}+\frac{(t/2\alpha- i\varepsilon)^2}{p^2}\Big)\bigg({\rm erf}\Big(\frac{pt}{2} 
 - \frac{(t/2\alpha-
  i\varepsilon)}{p} \Big ) + {\rm erf}\Big(\frac{(t/2\alpha- i\varepsilon)}{p} \Big )\bigg). 
 \label{BR1ar}
 \end{align}
Combining all results, one can see that in the time interval, $0<t<\infty$, the original system of 
integro-differential equations (\ref{B4ar}) and (\ref{B4br}) can be approximated by the system of 
ordinary differential equations,
 (\ref{B25a}) and (\ref{B25b}), provided
\begin{align}
\Big|\Big(\frac{\partial \tilde{\mathfrak R}_{1,2}(t) }{\partial \varepsilon}-\frac{\partial\tilde 
{\mathfrak R}^\ast_{1,2}(t) }{\partial \varepsilon}\Big)\Big| \ll 1, \quad {\rm and 
}\quad  \frac{| \Delta \rho(t)|}{{\langle \rho}_{22}(t)\rangle}  \ll 1.
\label{B26g}
\end{align}

To proceed further, it is convenient to define a scaled time, $\tau = p t$, and  the 
dimensionless, 
complex dynamical rates, $\tilde {\mathfrak r}_{1,2}(\tau)=(1/p)\tilde {\mathfrak R}_{1,2}(\tau) 
$:
\begin{align}\label{AR1}
\tilde{\mathfrak r}_1 (\tau) =& \sqrt{\pi}a^2 \exp (-\eta^2)\Big ( {\rm erf}\Big(\frac{\tau}{2} + 
i\eta\Big ) -{\rm erf}\big( i\eta\big )  \Big ), \\
\tilde{\mathfrak r}_2(\tau)=& \sqrt{\pi} a^2\exp\Big( 
-\frac{\mu^2\tau^2}{2}+{\Big(\frac{\mu\tau}{2}- i\eta \Big )^2}\Big)\Bigg ({\rm 
erf}\Big(\frac{\tau}{2} - \frac{\mu\tau}{2} + i\eta\Big ) + {\rm erf}\Big( \frac{\mu\tau}{2}- 
i\eta\Big )\Big),
\label{AR2}
\end{align}
where $a=v/p$, $\eta = \varepsilon/p$ and $\mu= 1/\sqrt{\alpha} p$.

Then, one can rewrite the conditions of Eq. (\ref{B26g}) as,
\begin{align}
a^2 \ll W_{1,2}(\tau; \eta,\mu) \quad {\rm and }\quad Z(\tau; a,\eta,\mu) \ll 1, \quad 0 
< 
\tau < \infty,
\label{A1g}
\end{align}
where
\begin{align}
W_{1,2}(\tau;\eta,\mu) =  &\frac{a^2}{\displaystyle\bigg |\frac{\partial \tilde{\mathfrak 
r}_{1,2}(\tau) 
}{\partial \eta}-\frac{\partial\tilde {\mathfrak r}^\ast_{1,2}(\tau) }{\partial \eta}\bigg |}, \\
 Z(\tau; a,\eta,\mu)= &\frac{| \Delta \rho(\tau)|}{{\langle \rho}_{22}(\tau)\rangle}= 
 \frac{\mu^2}{{ 
 \langle \rho}_{22}(\tau)\rangle }\bigg |{\rm 
 Re}\int\limits_{0}^{\tau}d \tau'\bigg ( 
 \displaystyle\frac{\partial\tilde 
 {\mathfrak r}_2(\tau') }{\partial \eta} \langle {\tilde X}^\ast(\tau')\rangle
 +i \tau' {\tilde {\mathfrak r}_2(\tau') }\big (\langle{\tilde X}(\tau')\rangle-\langle {\tilde 
 X}^\ast(\tau')\rangle  \big ) \bigg )\bigg |.
\end{align}
Now Eqs. (\ref{B25a}) -- (\ref{A30a}) can be rewritten as,
\begin{align} \label{B25ag}
\frac{d}{d \tau}{ \langle \rho}_{11}(\tau)\rangle =&- {\mathfrak 
r}_1(\tau)\big \langle \rho_{11}(\tau)\big\rangle 
+ {\mathfrak r}_2(\tau)\big \langle \rho_{22}(\tau)\big\rangle , \\
\frac{d}{d\tau}{ \langle \rho}_{22}(\tau)\rangle = & {\mathfrak 
r}_1(\tau)\big \langle \rho_{11}(\tau)\big\rangle - 
{\mathfrak r}_2(\tau)\big \langle \rho_{22}(\tau)\big\rangle , \\
\frac{d}{d\tau}\langle \tilde X(\tau)\rangle=& -i\tau {\mathfrak 
r}_1(\tau) \big \langle \rho_{11}(\tau)\big\rangle - \frac{\partial\tilde {\mathfrak 
r}^\ast_1(\tau) }{\partial \eta} \big \langle \rho_{11}(\tau)\big\rangle + i\tau\tilde {\mathfrak 
r}_2(\tau) \big \langle \rho_{22}(\tau)\big\rangle, \\
\frac{d}{d\tau}\Delta \rho(\tau)  =&-\frac{1}{2\alpha} \frac{\partial\tilde {\mathfrak r}_2(\tau) 
}{\partial \eta} \langle {\tilde X}^\ast(\tau)\rangle
-\frac{i \tau}{2\alpha} {\tilde {\mathfrak 2}_2(\tau) }\big (\langle{\tilde X}(\tau)\rangle- \langle 
{\tilde X}^\ast(\tau)\rangle  \big )  + \rm c.c.,
\label{B25hh}
\end{align}
where we set ${\mathfrak r}_{1,2} =2{\rm Re}\, \tilde {\mathfrak r}_{1,2} $.

Let us denote the real and imaginary parts of $\langle \tilde X(\tau)\rangle$  as, $U (\tau) = 
{\rm Re} \langle \tilde X(\tau)\rangle$ and $V (\tau) = {\rm Im} 
\langle \tilde X(\tau)\rangle$. Then, employing Eq. 
(\ref{B25hh}), one can show that the functions $U(\tau)  $ and $V(\tau)  $ obey the following 
differential equations:
\begin{align}
\frac{d}{d\tau} U(\tau)=& - { \rm Re}\frac{\partial\tilde {\mathfrak r}_1(\tau) 
}{\partial \eta} \big \langle \rho_{11}(\tau)\big\rangle -  \tau{ \rm Im}\,\tilde 
{\mathfrak r}_2(\tau)  \big \langle \rho_{22}(\tau)\big\rangle, \\
\frac{d}{d\tau} V(\tau)=& \Big ( { \rm Im}\frac{\partial\tilde {\mathfrak r}_1(\tau) 
}{\partial \eta}    -\tau {\mathfrak r}_1(\tau)  
\Big)\big \langle \rho_{11}(\tau)\big\rangle +  \tau{ \rm Re}\,\tilde 
{\mathfrak r}_2(\tau)  \big \langle \rho_{22}(\tau)\big\rangle.  
\label{Z1}
\end{align}
 In addition we obtain
 \begin{align}
|\Delta \rho(\tau; a,\eta,\mu)|= & \displaystyle{\mu^2\bigg |\int\limits_{0}^{\tau}d \tau'\bigg 
( 
 \displaystyle {\rm Re}\frac{\partial\tilde 
 {\mathfrak r}_2(\tau') }{\partial \eta}  U(\tau') +  \displaystyle {\rm Im}\frac{\partial\tilde 
  {\mathfrak r}_2(\tau') }{\partial \eta}  V(\tau') 
 -\tau' { {\mathfrak r}_2(\tau') }V(\tau')\bigg )\bigg |}.
 \end{align}
 A straightforward computation shows that 
$|W_{1,2}(\tau; \mu, \eta)| \geq W(\eta) $, where (See Fig. \ref{SFig1})
\begin{align}
W(\eta) = \frac{1}{{|\rm Re(1-i\sqrt{\pi} \eta e^{-\eta^2} {\rm erfc }(i  \eta)) |}} \geq 1.
\end{align} 
Thus, the first condition from Eq. (\ref{A1g}) can be replaced by the stronger inequality
\begin{align}
\frac{v}{p}\ll 1.
\end{align} 
 \begin{figure}[tbh]
 \scalebox{0.4}{\includegraphics{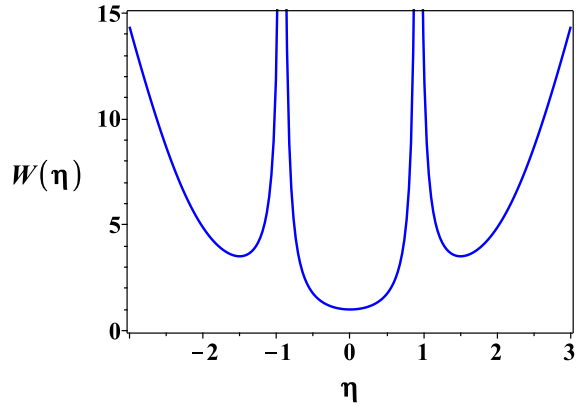}}
 \caption{(Color online) Dependence of $W$ on $\eta$.
 \label{SFig1}}
 \end{figure}
Gathering all results, we conclude that in the time interval, $0<t<\infty$, the original 
system 
of  integro-differential equations (\ref{B4ar} and (\ref{B4br}) can be approximated by the system 
of the rate-type ordinary differential equations:
\begin{align} \label{B25c}
\frac{d}{dt}{ \langle \rho}_{11}\rangle =&- {\mathfrak R}_1(t)\big\langle\tilde 
\rho_{11}\big\rangle 
+ {\mathfrak R}_2(t)\big \langle \rho_{22}\big\rangle , \\
\frac{d}{dt}{ \langle \rho}_{22}\rangle = & {\mathfrak R}_1(t)\big\langle\tilde 
\rho_{11}\big\rangle - {\mathfrak R}_2(t)\big \langle \rho_{22}\big\rangle ,
\label{B25d}
\end{align}
  if the following conditions hold:
\begin{align}
&|D\sqrt{\chi(0)}| \ll \frac{d}{dt}\ln \chi (t)\Big |_{t=0} \ll
\frac{3p^2}{2\varepsilon}, \\
&\frac{v}{p}\ll 1, \quad {\rm and 
}\quad  \frac{| \Delta \rho(t)|}{{\langle \rho}_{22}(t)\rangle}  \ll 1.
\label{B26b}
\end{align}

\section{Finite electron bands}

The Hamiltonian describing finite electron bands for both donor and acceptor,  in the 
continuous limit, has the form $\mathcal H (t) ={\mathcal H}_0(t) + W$, where
\begin{align}\label{DAH1}
{\mathcal H}_0 (t)= &  \int (E_d +\lambda_d(t))|E_d\rangle\langle E_d|\varrho(E_d)dE_d
+ \int (E_a +\lambda_a(t))|E_a\rangle\langle E_a|\varrho(E_a)dE_a , \nonumber \\
W=&   \int\int  dE_d dE_a \varrho(E_d)\varrho(E_a) \Big(V(E_d,E_a)|E_d\rangle\langle E_a| + 
\rm h.c.\Big),
 \end{align}
 $\varrho(E_d)$ and  $\varrho(E_a)$ are the densities of electron states of the donor and  
 acceptor, respectively.  We assume that the amplitude of transition is a smoothly varying 
 function of energy, so that one can approximate, $V(E_d,E_a)\approx V=\rm const$.
In the interaction representation, $\tilde \rho= e^{i\int_0^t  {\mathcal H}_0(\tau) d \tau}\rho 
e^{-i\int_0^t {\mathcal 
		H}_0(\tau) 
	d\tau }$,  the equations of motion are given by 
\begin{align}  \label{DA1a}
\frac{d}{dt}\tilde \rho_{E_d E'_d} =& i\int dE_a \varrho(E_a)\big({\tilde \rho}_{E_dE_a}{\tilde 
V}^\ast(E_a,E'_d,t)- {\tilde V}(E_d,E_a,t) {\tilde \rho}_{E_aE'_d}\big), \\
\frac{d}{dt} \tilde \rho_{E_aE_a'} = &i\int dE''_d \varrho(E''_d)\big({\tilde \rho}_{E_a 
E''_d}{\tilde V}^\ast(E''_d,E_a',t)- {\tilde V}(E_a,E''_d,t){\tilde \rho}_{E''_dE_a'} \big),  \\
\frac{d}{dt} \tilde \rho_{E_dE_a} =& i\int dE''_d \varrho(E''_d){\tilde \rho}_{E_d E''_d} {\tilde 
V}(E''_d,E_a,t)-i \int dE''_a \varrho(E''_a){\tilde V}(E_d, E''_a,t) {\tilde 
\rho}_{E''_aE_a}, \\
\frac{d}{dt} \tilde \rho_{E_aE_d} =&i \int dE''_a \varrho(E''_a){\tilde \rho}_{E_aE''_a}{\tilde 
V}^\ast(E''_a,E_d,t)  
-i\int dE''_d \varrho(E''_d) {\tilde V}^\ast(E_a,E''_d,t){\tilde \rho}_{E''_d E_d} ,
\label{DA1b}
\end{align}
where  $\tilde V(E_d,E_a, t) =Ve^{i(E_d -E_a)t} e^{i\varphi_{da}(t)}$ and $\varphi_{da}(t) 
=(g_d-g_a)\int_0^t \xi (t')dt'$.

We assume that initially,  ${\tilde \rho}_{E_dE_a}(0)={\tilde \rho}_{E_d E_a}(0)=0$, and employ 
the same procedure as in the previous section. After splitting of correlations, we obtain
\begin{align}\label{DEq3}
&\frac{d}{dt}\big{\langle\tilde \rho}_{E_d,E'_d}(t)\big \rangle = &{} \nonumber \\
&- \int_0^t dt'\iint dE_a \varrho(E_a)dE''_d \varrho(E''_d)\Big (  
\big\langle{\tilde\rho}_{E_dE''_d}(t') \big\rangle  \big\langle {\tilde V}(E''_d,E_a,t'){\tilde 
V}^\ast(E_a,E'_d,t)\big\rangle + \big\langle {\tilde V}(E_d,E_a,t){\tilde 
V}^\ast(E_a,E''_d,t')\big\rangle  \big\langle{\tilde\rho}_{E''_dE'_d}(t') \big\rangle  \Big) 
\nonumber \\
&+\int_0^t dt'\iint dE_a \varrho(E_a)dE''_a \varrho(E''_a)\Big (  
\big\langle{\tilde\rho}_{E''_aE_a}(t') \big\rangle  \big\langle {\tilde V}(E_d,E''_a,t'){\tilde 
V}^\ast(E_a,E'_d,t)\big\rangle + \big\langle {\tilde V}(E_d,E_a,t){\tilde 
V}^\ast(E''_a,E'_d,t')\big\rangle  \big\langle{\tilde\rho}_{E_aE''_a}(t') \big\rangle  \Big) 
, 
\\
&\frac{d}{dt}{\langle{\tilde\rho}}_{E_aE_a'}(t)\rangle =&{} \nonumber \\
& \int_0^t dt'\iint dE'_d \varrho(E'_d)dE''_d \varrho(E''_d)\Big (  
\big\langle{\tilde\rho}_{E'_dE''_d}(t') \big\rangle  \big\langle {\tilde 
V}^\ast(E''_d,E'_a,t){\tilde V}(E_a,E'_d,t')\big\rangle + \big\langle {\tilde 
V}(E'_d,E'_a,t){\tilde V}^\ast(E_a,E''_d,t')\big\rangle  
\big\langle{\tilde\rho}_{E''_dE'_d}(t') 
\big\rangle  \Big) \nonumber \\
&-\int_0^t dt'\iint dE'_d \varrho(E'_d)dE''_a \varrho(E''_a)\Big (  
\big\langle{\tilde\rho}_{E'_aE''_a}(t') \big\rangle  \big\langle {\tilde 
V}^\ast(E''_d,E'_a,t){\tilde V}(E_a,E''_d,t')\big\rangle + \big\langle {\tilde 
V}(E''_d,E''_a,t'){\tilde V}^\ast(E_a,E''_d,t)\big\rangle  
\big\langle{\tilde\rho}_{E''_aE'_a}(t') 
\big\rangle  \Big).
 \label{DEq4}
 \end{align}
These equations can be recast as,
\begin{align}\label{DEq5a}
&\frac{d}{dt}\big{\langle\tilde \rho}_{E_d,E'_d}(t)\big \rangle = &{} \nonumber \\
&- |V|^2\int_0^t dt' \Phi(t,t')\iint dE_a \varrho(E_a)dE''_d \varrho(E''_d)\Big (  
\big\langle{\tilde\rho}_{E_dE''_d}(t') 
\big\rangle   e^{-i(E_a -E'_d)t} e^{i(E''_d-E_a)t'}  +   e^{i(E_d -E_a)t}  e^{-i(E_a-E''_d)t'}
\big\langle{\tilde\rho}_{E''_dE'_d}(t') 
\big\rangle  \Big) \nonumber \\
&+|V|^2\int_0^t dt \Phi(t,t')'\iint dE_a \varrho(E_a)dE''_a \varrho(E''_a)\Big (  
\big\langle{\tilde\rho}_{E''_aE_a}(t') 
\big\rangle     e^{-i(E_a -E'_d)t} e^{i(E_d- E''_a)t'}+  e^{i(E_d -E_a)t}  e^{-i(E''_a -E'_d)t'}
 \big\langle{\tilde\rho}_{E_aE''_a}(t') 
\big\rangle  \Big) , \\
&\frac{d}{dt}{\langle{\tilde\rho}}_{E_aE_a'}(t)\rangle =&{} \nonumber \\
& |V|^2\int_0^t dt' \Phi(t,t')\iint dE'_d \varrho(E'_d)dE''_d \varrho(E''_d)\Big (  
\big\langle{\tilde\rho}_{E'_dE''_d}(t') 
\big\rangle   e^{-i(E''_d-E'_a)t}  e^{i(E_a-E'_d)t'} +  e^{i(E'_d - E'_a)t}  e^{-i(E_a- E''_d)t'}
 \big\langle{\tilde\rho}_{E''_dE'_d}(t') 
\big\rangle  \Big) \nonumber \\
&- |V|^2\int_0^t dt'\Phi(t,t')\iint dE'_d \varrho(E'_d)dE''_a \varrho(E''_a)\Big (  
\big\langle{\tilde\rho}_{E'_aE''_a}(t') 
\big\rangle  e^{-i(E''_d -E'_a)t}  e^{i(E_a -E''_d)t'} +  e^{-i(E_a -E''_d)t}  e^{i(E''_d-E''_a)t'}
  \big\langle{\tilde\rho}_{E''_aE'_a}(t') 
\big\rangle  \Big),
 \label{DEq5b}
 \end{align}
where  $\Phi(t,t')=\langle e^{i\varphi_{da}(t)}e^{-i\varphi_{da}(t')}\rangle$.

Further, we consider a Gaussian  density of states in the donor and acceptor bands centered at 
some 
energies, $E^{(a)}_0$, for the acceptor, and, $E^{(d)}_0$, for the donor, 
\begin{align}\label{AB8}
\varrho(E_d) =   \varrho_d e^{-\alpha_1(E_d-E^{(d)}_0)^2} , \\
\varrho(E_a) =   \varrho_a e^{-\alpha_2 (E_a-E^{(a)}_0)^2} .
\end{align}

We denote by ${\langle\tilde \rho}_{11}(t)\rangle= \int \langle\tilde \rho_{E_dE_d } 
\rangle (t)\varrho(E_d) dE_d$ and ${\langle\tilde \rho}_{22}(t)\rangle= \int 
\langle\tilde \rho_{E_aE_a} \rangle (t)\varrho(E_a) dE_a$  the probabilities to populate 
the donor and 
the acceptor, respectively. 
A computation yields
\begin{align}\label{D7a}
&{\langle\tilde \rho}_{11}(t)\rangle= \int \langle\tilde \rho_{E_dE_d } 
\rangle (t)\varrho(E_d) dE_d
\approx 
\langle\tilde\rho_{E^{(d)}_0E^{(d)}_0}(t')\rangle \varrho_0\sqrt{\frac{\pi }{\alpha}}, \\
&{\langle\tilde \rho}_{22}(t)\rangle= \int 
\langle\tilde \rho_{E_aE_a} \rangle (t)\varrho(E_a) dE_a
\approx 
\langle\tilde\rho_{E^{(a)}_0E^{(a)}_0}(t')\rangle \varrho_0\sqrt{\frac{\pi }{\alpha}}, \\
&\int dE_a \varrho(E_a)\iint dE_d dE'_d \varrho(E_d) \varrho(E'_d)e^{i(E_a-E_d)t}  e^{i(E'_d-E_a)t'} 
\big\langle{\tilde\rho}_{E_dE'_d}(t') \big\rangle  {} \nonumber \\
&\approx {\langle\tilde \rho}_{11}(t)\rangle \exp\bigg (i\varepsilon(t-t') 
{-\frac{(t-t')^2}{4\alpha_2}}- \frac{t^2+t'^2}{4\alpha_1}\bigg )\\
&\int dE_d \varrho(E_d)\iint dE_a dE'_a \varrho(E_a) \varrho(E'_a)e^{i(E_d-E_a)t}  
e^{i(E'_a-E_d)t'}  
\big\langle{\tilde \rho}_{E_aE'_a}(t')\big\rangle    {} \nonumber \\
&\approx  {\langle\tilde \rho}_{22}(t)\rangle \exp\bigg (i\varepsilon(t-t') 
{-\frac{(t-t')^2}{4\alpha_1}}- \frac{t^2+t'^2}{4\alpha_2}\bigg ),
\label{D7b}
\end{align}
where $\varepsilon = E^{(d)}_0 - E^{(a)}_0$. Next, employing Eqs. (\ref{D7a}) -- 
(\ref{D7b}), we obtain from Eqs. (\ref{DEq5a}) -- (\ref{DEq5b}) the following system of 
integro-differential equations for the diagonal components of the density matrix,
\begin{align} \label{DB4a}
\frac{d}{dt}{\langle\tilde \rho}_{11}(t)\rangle =&- \int_0^t { K_1}(t,t')\big\langle\tilde 
\rho_{11}(t')\big\rangle dt'+\int_0^t { K_2}(t,t')\big(\big\langle\tilde \rho_{22}(t')\big\rangle 
dt',\\
\frac{d}{dt}{\langle\tilde \rho}_{22}(t)\rangle =& \int_0^t { K_1}(t,t')\big\langle\tilde 
\rho_{11}(t')\big\rangle dt' -\int_0^t {K_2}(t,t')\big\langle\tilde \rho_{22}(t')\big\rangle dt',
\label{DB4b}
\end{align}
 where
 \begin{align}\label{DAK1}
   K_1(t,t') = & 2v^2\Phi(t,t') \cos (\varepsilon(t-t'))  \exp\Big (-\frac{(t-t')^2}{4\alpha_2} 
   -\frac{t^2  +t'^2}{4\alpha_1}\Big) , \\
    K_2(t,t') = & 2v^2\Phi(t,t') \cos (\varepsilon(t-t'))  \exp\Big (-\frac{(t-t')^2}{4\alpha_1} 
    -\frac{t^2  +t'^2}{4\alpha_2}\Big).
    \label{DAK2}
 \end{align}

For $|\int_{0}^{t} (t-t') K_{1,2}(t,t') d t'  | \ll 1$,  the system of integro-differential equations 
(\ref{DB4a}) and (\ref{DB4b})   can be approximated by  the system of ordinary differential 
equations,
\begin{align} \label{DN7a}
\frac{d}{dt}{ \langle \rho}_{11}\rangle =&- {\mathfrak R}_1(t)\big\langle
\rho_{11}\big\rangle + {\mathfrak R}_2(t)\big \langle \rho_{22}\big\rangle , \\
\frac{d}{dt}{ \langle \rho}_{22}\rangle = & {\mathfrak R}_1(t)\big\langle 
\rho_{11}\big\rangle - {\mathfrak R}_2(t)\big \langle \rho_{22}\big\rangle,
\label{DN7b}
\end{align}
where $\mathfrak {R}_{1,2}(t) =\int_0^t K_{1,2}(t,t') d t' $.
A computation yields
\begin{align}
{\mathfrak R}_1(t)=& \frac{\sqrt{\pi}\,{v}^2}{p}\Bigg\{\exp\Big( 
-\frac{t^2}{2\alpha_1}+\frac{(t/2\alpha_1- i\varepsilon)^2}{p^2}\Big)\bigg({\rm 
erf}\Big(\frac{pt}{2} - \frac{(t/2\alpha_1-
 i\varepsilon)}{p} \Big ) + {\rm erf}\Big(\frac{(t/2\alpha_1- i\varepsilon)}{p} \Big )\bigg)  +\rm  
 c.c.\Bigg\}, 
\label{DBR1}\\
{\mathfrak R}_2(t)=& \frac{\sqrt{\pi}\,{v}^2}{p}\Bigg\{\exp\Big( 
-\frac{t^2}{2\alpha_2}+\frac{(t/2\alpha_2- i\varepsilon)^2}{p^2}\Big)\bigg({\rm 
erf}\Big(\frac{pt}{2} - \frac{(t/2\alpha_2-
 i\varepsilon)}{p} \Big ) + {\rm erf}\Big(\frac{(t/2\alpha_2- i\varepsilon)}{p} \Big )\bigg)  +\rm  
 c.c.\Bigg\},
\label{DBR1a}
\end{align}
where $p=\sqrt{1/\alpha_1 +1/\alpha_2 +2(D\sigma)^2}$.

Following the same procedure as in the Section I, one can show that the condition for validity of 
approximating the integro-differential equations by the ordinary differential equations is 
modified as follows:
\begin{align}
&|D\sqrt{\chi(0)}| \ll \frac{d}{dt}\ln \chi (t)\Big |_{t=0} \ll
\frac{3p^2}{2\varepsilon}, \\
&\Big|\Big(\frac{\partial \tilde{\mathfrak R}_{1,2}(t) }{\partial \varepsilon}-\frac{\partial\tilde 
{\mathfrak R}^\ast_{1,2}(t) }{\partial \varepsilon}\Big)\Big| \ll 1, \quad {\rm and 
}\quad  Z(t) =\frac{| \Delta \rho(t)|}{{\langle \rho}_{22}(t)\rangle}  \ll 1,
\label{B28b}
\end{align}
where
\begin{align}
\tilde {\mathfrak R}_1(t)=& \frac{\sqrt{\pi}\,{v}^2}{p}\exp\Big( 
-\frac{t^2}{2\alpha_1}+\frac{(t/2\alpha_1- i\varepsilon)^2}{p^2}\Big)\bigg({\rm 
erf}\Big(\frac{pt}{2} - \frac{(t/2\alpha_1-
 i\varepsilon)}{p} \Big ) + {\rm erf}\Big(\frac{(t/2\alpha_1- i\varepsilon)}{p} \Big )\bigg) , 
\label{SR1}\\
\tilde {\mathfrak R}_2(t)=& \frac{\sqrt{\pi}\,{v}^2}{p}\exp\Big( 
-\frac{t^2}{2\alpha_2}+\frac{(t/2\alpha_2- i\varepsilon)^2}{p^2}\Big)\bigg({\rm 
erf}\Big(\frac{pt}{2} - \frac{(t/2\alpha_2-
 i\varepsilon)}{p} \Big ) + {\rm erf}\Big(\frac{(t/2\alpha_2- i\varepsilon)}{p} \Big )\bigg),
\label{SR2} \\
| \Delta \rho(t)|= &\bigg |{\rm  Re}\int\limits_{0}^{t}d t'\bigg ( 
 \displaystyle\frac{\partial\tilde 
 {\mathfrak R}_1(t') }{\partial \varepsilon} \langle {\tilde X}_1^\ast(t')\rangle
 +i t' {\tilde {\mathfrak R}_1(t') }\big (\langle{\tilde X}_1(t')\rangle-\langle {\tilde 
 X}_1^\ast(t')\rangle  \big ) \bigg ) \nonumber \\
&+{\rm Re}\int\limits_{0}^{t}d t'\bigg ( 
  \displaystyle\frac{\partial\tilde 
  {\mathfrak R}_2(t') }{\partial \varepsilon} \langle {\tilde X}_2^\ast(t')\rangle
  +i t' {\tilde {\mathfrak R}_2(t') }\big (\langle{\tilde X}_2(t')\rangle-\langle {\tilde 
  X}_2^\ast(\tau')\rangle  \big ) \bigg )\bigg | .
\label{SR3}
\end{align}

Here we denote,
\begin{align}
&\langle \tilde X_1(t)\rangle=\varrho_0\sqrt{\frac{\pi }{\alpha_1}}\frac{\partial}{\partial 
E'_d}{\langle\tilde \rho}_{E_dE'_d}(t)\rangle\bigg |_{E_d=E'_d = E^{(a)}_0} \quad {\rm and } 
\quad \langle \tilde X_2(t)\rangle=\varrho_0\sqrt{\frac{\pi }{\alpha_2}}\frac{\partial}{\partial 
E'_a}{\langle\tilde \rho}_{E_aE'_a}(t)\rangle\bigg |_{E_a=E'_a = E^{(a)}_0}.
\end{align}

To proceed further, it is convenient to define a scaled time, $\tau = p t$, and  the 
dimensionless complex dynamical rates, $\tilde {\mathfrak r}_{1,2}(\tau)=(1/p)\tilde {\mathfrak 
R}_{1,2}(\tau) 
$:
\begin{align}\label{SR1a}
\tilde{\mathfrak r}_1(\tau)=& \sqrt{\pi} a^2\exp\Big( 
-\frac{\mu_1^2\tau^2}{2}+{\Big(\frac{\mu_1\tau}{2}- i\eta \Big )^2}\Big)\Bigg ({\rm 
erf}\Big(\frac{\tau}{2} - \frac{\mu_1\tau}{2} + i\eta\Big ) + {\rm erf}\Big( \frac{\mu_1\tau}{2}- 
i\eta\Big )\Big) , \\
\tilde{\mathfrak r}_2(\tau)=& \sqrt{\pi} a^2\exp\Big( 
-\frac{\mu_2^2\tau^2}{2}+{\Big(\frac{\mu_2\tau}{2}- i\eta \Big )^2}\Big)\Bigg ({\rm 
erf}\Big(\frac{\tau}{2} - \frac{\mu_2\tau}{2} + i\eta\Big ) + {\rm erf}\Big( \frac{\mu_2\tau}{2}- 
i\eta\Big )\Big),
\label{SR2a}
\end{align}
where, $a=v/p$, $\eta = \varepsilon/p$ and $\mu_{1,2}= 1/\sqrt{\alpha_{1,2}} p$. We obtain
\begin{align}
| \Delta \rho(\tau)|= &\bigg |\mu_1^2{\rm  Re}\int\limits_{0}^{\tau}d \tau'\bigg ( 
 \displaystyle\frac{\partial\tilde 
 {\mathfrak r}_1(\tau') }{\partial \eta} \langle {\tilde X}_1^\ast(\tau')\rangle
 +i \tau' {\tilde {\mathfrak r}_1(\tau') }\big (\langle{\tilde X}_1(\tau')\rangle-\langle {\tilde 
 X}_1^\ast(\tau')\rangle  \big ) \bigg ) \nonumber \\
&+\mu_2^2{\rm 
  Re}\int\limits_{0}^{\tau}d \tau'\bigg ( 
  \displaystyle\frac{\partial\tilde 
  {\mathfrak r}_2(\tau') }{\partial \eta} \langle {\tilde X}_2^\ast(\tau')\rangle
  +i \tau' {\tilde {\mathfrak r}_2(\tau') }\big (\langle{\tilde X}_2(\tau')\rangle-\langle {\tilde 
  X}_2^\ast(\tau')\rangle  \big ) \bigg )\bigg | .
\end{align}
The functions, ${ \langle \rho}_{11}(\tau)\rangle$,  ${ \langle \rho}_{22}(\tau)\rangle$ and 
$\langle{\tilde X}_{1,2} (\tau)\rangle$, obey the the following differential equations
\begin{align} \label{B25g}
\frac{d}{d \tau}{ \langle \rho}_{11}(\tau)\rangle =&- {\mathfrak 
r}_1(\tau)\big \langle \rho_{11}(\tau)\big\rangle 
+ {\mathfrak r}_2(\tau)\big \langle \rho_{22}(\tau)\big\rangle , \\
\frac{d}{d\tau}{ \langle \rho}_{22}(\tau)\rangle = & {\mathfrak 
r}_1(\tau)\big \langle \rho_{11}(\tau)\big\rangle - 
{\mathfrak r}_2(\tau)\big \langle \rho_{22}(\tau)\big\rangle , \\
\frac{d}{dt}\langle \tilde X_1(\tau)\rangle=& -i\tau {\mathfrak 
r}_1(\tau) \big \langle \rho_{11}(\tau)\big\rangle - \frac{\partial\tilde {\mathfrak 
r}^\ast_1(\tau) 
}{\partial \eta} \big \langle \rho_{11}(\tau)\big\rangle + i\tau\tilde {\mathfrak 
r}_2(\tau) \big \langle \rho_{22}(\tau)\big\rangle,  \\
\frac{d}{dt}\langle \tilde X_2(\tau)\rangle=& -i\tau {\mathfrak 
r}_2(\tau) \big \langle \rho_{22}(\tau)\big\rangle - \frac{\partial\tilde {\mathfrak 
r}^\ast_2(\tau) }{\partial \eta} \big \langle \rho_{22}(\tau)\big\rangle + i\tau\tilde {\mathfrak 
r}_1(\tau) \big \langle \rho_{11}(\tau)\big\rangle.
\label{B25h}
\end{align}
\begin{figure}[tbh]
\begin{center}
\scalebox{0.425}{\includegraphics{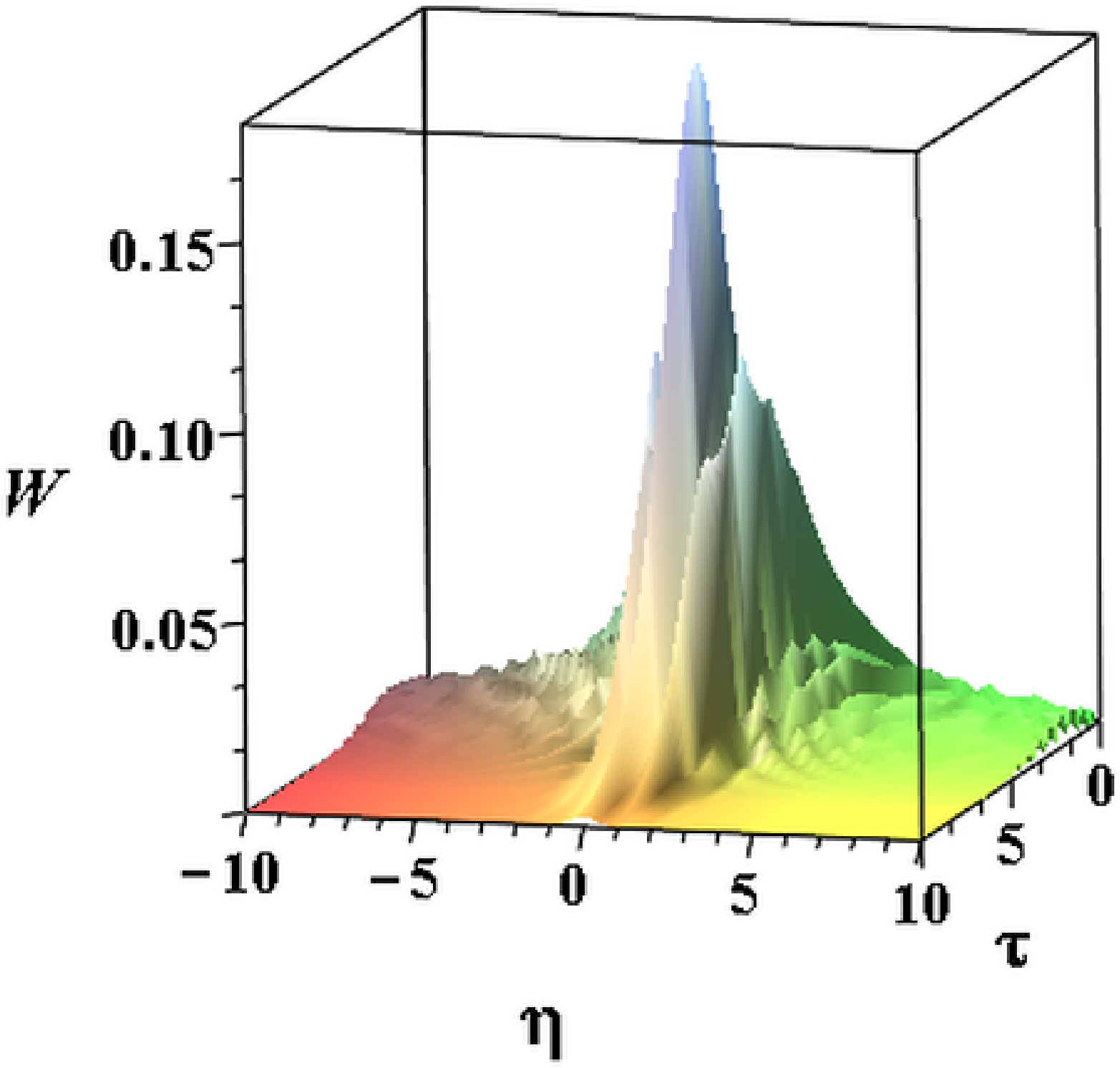}}
\scalebox{0.375}{\includegraphics{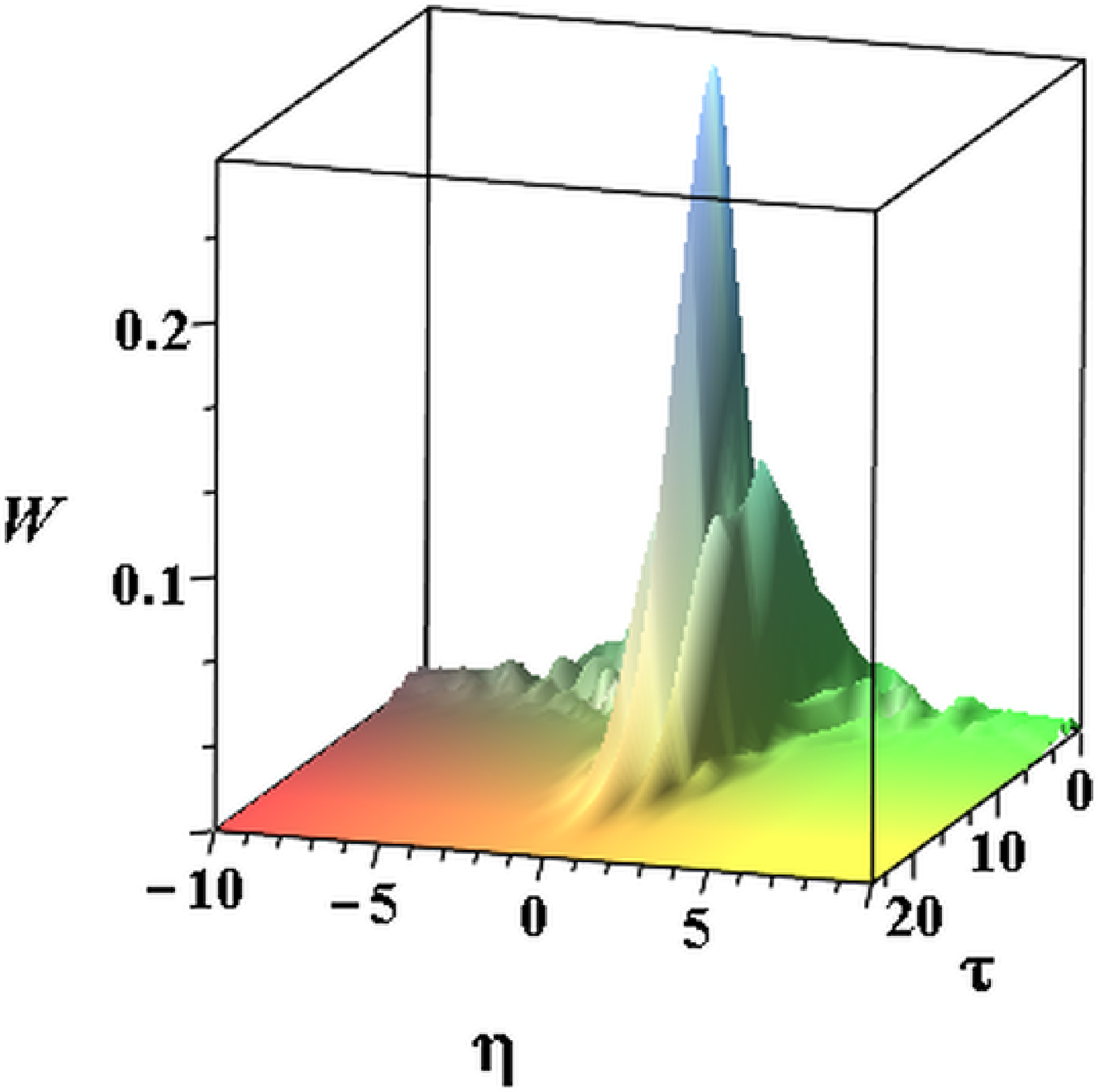}}
\end{center}
\caption{(Color online) Plot of the function $W$. Dependence of $W$ on the scaled time, $\tau$, 
and parameter, $\eta = \varepsilon/p$. Left: $\mu =0.5 $, $v/p =0.2$. Right: $\mu =0.25 $, 
$v/p =0.2$.   
\label{SMW}}
\end{figure}

\begin{figure}[tbh]
\begin{center}
\scalebox{0.325}{\includegraphics{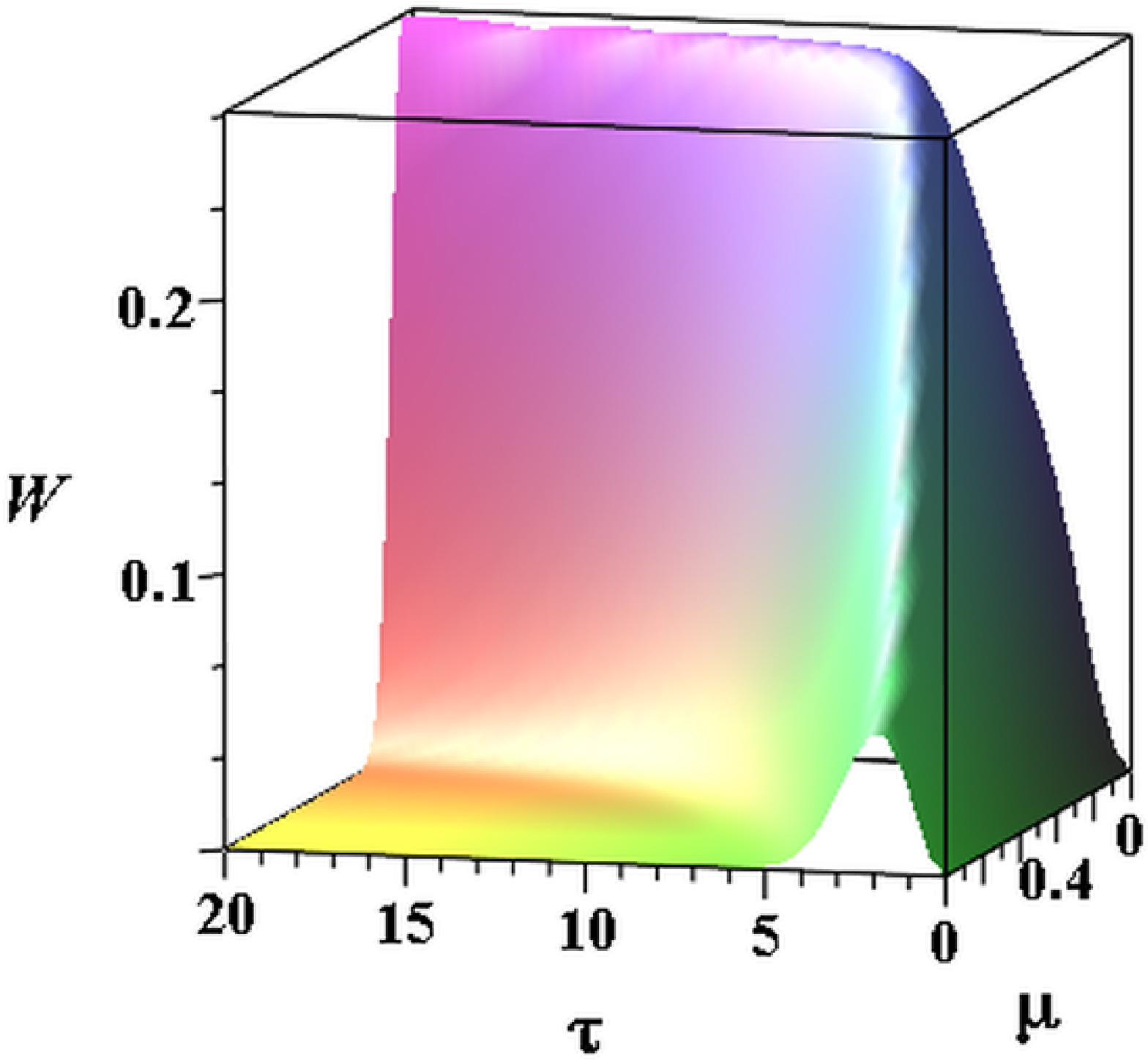}}
\scalebox{0.35}{\includegraphics{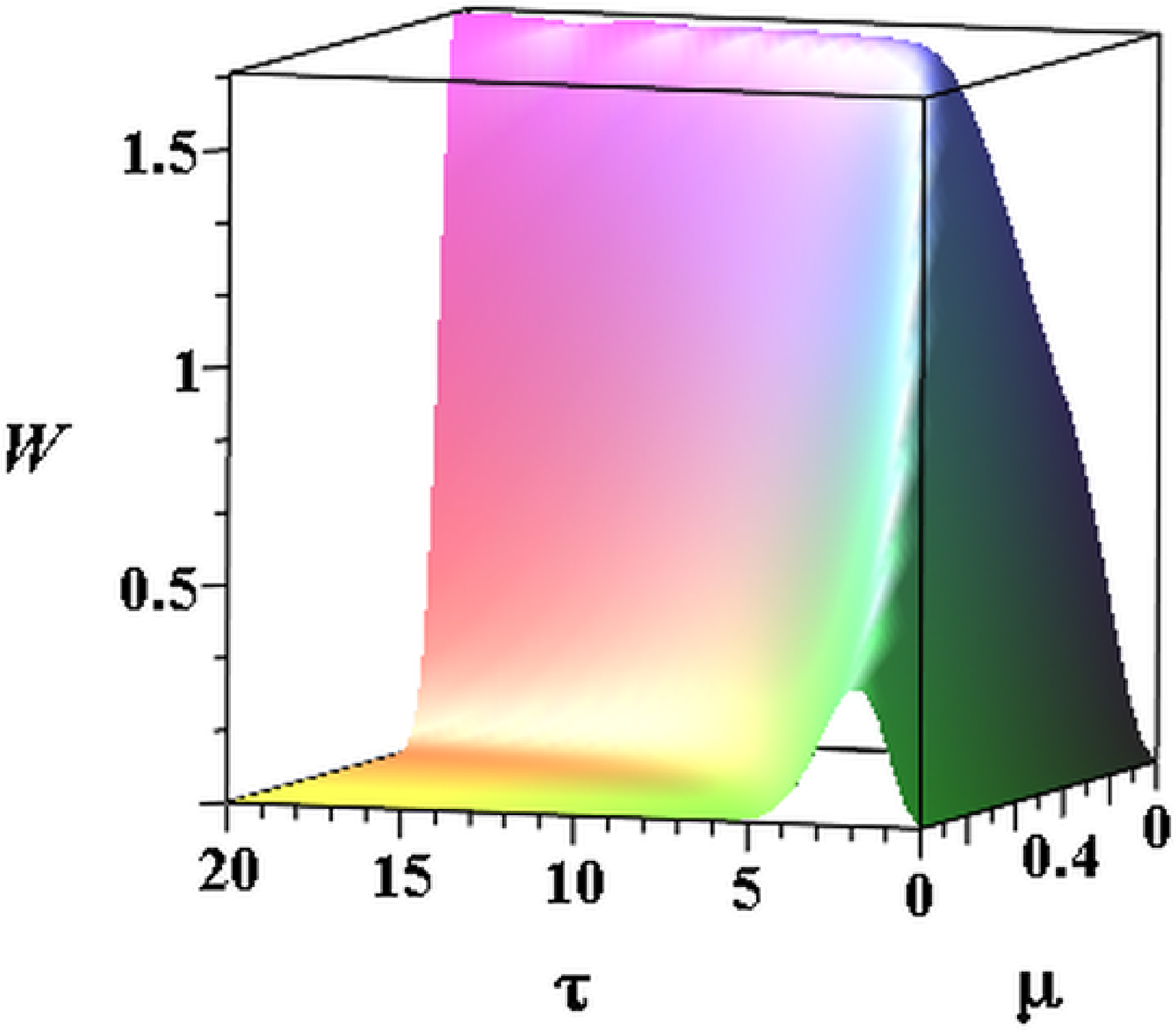}}
\end{center}
\caption{(Color online) Plot of the function $W$. Dependence $W$ on the scaled time $\tau$ 
and parameter $\mu$ ($\eta =0$). Left: $v/p =0.2$. Right: $v/p =0.5$.
\label{SMW1}}
\end{figure}

Putting all results together, we conclude that in the time interval, $0<t<\infty$, the original 
system 
of  integro-differential equations (\ref{D7a}) and (\ref{D7b}) can be approximated by the system 
of 
ordinary differential equations
 (\ref{DN7a}) and (\ref{DN7b}), if the following conditions hold:
\begin{align}
&|D\sqrt{\chi(0)}| \ll \frac{d}{dt}\ln \chi (t)\Big |_{t=0} \ll
\frac{3p^2}{2\varepsilon}, \\
& W(\tau, a,\mu,\eta)=\Big|\Big(\frac{\partial \tilde{\mathfrak r}_{1,2}(\tau) }{\partial 
\eta}-\frac{\partial\tilde 
{\mathfrak r}^\ast_{1,2}(\tau) }{\partial \eta}\Big)\Big| \ll 1, \label{SW}\\
& Z(\tau, a,\mu,\eta) = \frac{| \Delta \rho(\tau)|}{{\langle \rho}_{22}(\tau)\rangle}  \ll 1.
\label{B26q}
\end{align}
In Figs. \ref{SMW} and \ref{SMW1}, the function, $W$, is presented. As one can see, for the 
chosen values of parameters, the condition (\ref{SW}) is satisfied. As it follows from analysis of 
Fig. \ref{SMW1}, and taking into account that in our perturbative theory the ``small'' parameter 
is, $v/p$, the condition (\ref{SW}) should be replaced by a more strong inequality: $v/p \ll 1 $. 
Thus, the conditions  of validity of approximation leading to the  rate-type equations 
(\ref{DN7a}) and (\ref{DN7b}) can be written in its final form as,
\begin{align}\label{SQA1a}
 &|D\sqrt{\chi(0)}| \ll \frac{d}{dt}\ln \chi (t)\Big |_{t=0} \ll
\frac{3p^2}{2\varepsilon}, \\
&\frac{v}{p}\ll 1, \quad {\rm and 
}\quad  \frac{| \Delta \rho(t)|}{{\langle \rho}_{22}(t)\rangle}  \ll 1.
 \label{SQA1}
	\end{align}\\

\section{Correlation splitting}

In this section, we consider the correlation 
splitting for the random telegraph process (RTP) 
 $\xi(t)$ having the property $\langle  \xi(t) \rangle =0$ and the correlation function
\begin{eqnarray}
\label{SA1b}
\chi(t-t') = \langle \xi(t)\xi(t')\rangle=  \sigma^2 e^{-2\gamma(t-t')}, \quad t \geq t'.
\end{eqnarray}
Thus $\chi(0) =\sigma^2$.  We also have $\xi(t)^2=\sigma^2$. Let $M_n (t_1,t_2,\dots, t_n) = 
\langle  \xi(t_1)\dots \xi(t_n)
\rangle$ for ordered times $t_1\geq t_2\geq \dots \geq t_n $. Then we have (see 
\cite{KV2,KV3}) 
\begin{align}
&M_n (t_1,t_2,\dots, t_n) =  \langle  \xi(t_1) \xi(t_2)
\rangle M_{n-2} (t_3,\dots, t_n).
\label{SA1h}
\end{align}
The RTP is conveniently described by its characteristic function \cite{KV3}, defined by
\begin{eqnarray}
\Phi[t;\nu (\tau)] = \Big\langle \exp\Big \{i\int\limits_{0}^{t}
 d\tau \xi(\tau) \nu (\tau)\Big \}\Big\rangle .
\label{SA3}
\end{eqnarray}
Here, $\nu(\tau)$ is any function s.t. the integral in the exponent of \eqref{SA3} exists.  
Applying Eq. (\ref{SA1h})  and using the Taylor expansion  of Eq. (\ref{SA3}), we obtain 
an exact integral equation for the characteristic
function $\Phi[t;\nu(\tau)]$,
\begin{eqnarray}
\Phi[t;\nu (\tau)]  = 1 - \int\limits_{0}^{t} dt_1
\int\limits_{0}^{t_1} dt_2 \chi(t_1-t_2) \nu(t_1) \nu(t_2)
\Phi[t_2;\nu (\tau)].
\end{eqnarray}
One can transform this integral equation into the integro-differential equation,
\begin{align}
\label{A4} \frac{d}{dt}\Phi[t;\nu (\tau)] = -
\nu(t)\int\limits_{0}^{t} dt_1  \chi(t-t_1) \nu(t_1)
\Phi[t_1;\nu (\tau)].
\end{align}
Let $R[t;\xi(\tau)]$ be an arbitrary functional. Then, for correlators, $\langle \xi(t) R[t;\xi(\tau)] 
\rangle $ and $\langle  {\xi}(t_1){\xi}(t_2)R[t;\xi(\tau)] \rangle $,  one can show that the 
following  correlation splitting formulas hold \cite{KV3}:
\begin{eqnarray} \label{SA4r}
\langle \xi(t) R[t;\xi(\tau)] \rangle = \int_0^t dt_1 \chi(t- t_1) \bigg\langle \frac{\delta}{\delta 
\xi(t_1)} {\tilde R} [t, t_1;\xi(\tau)] \bigg\rangle, \\
\langle  {\xi}(t_1){\xi}(t_2)R[t;\xi(\tau)] \rangle =
\chi(t_1-t_2)\langle R[t;\xi(\tau)] \rangle, \;  t_1\geq
t_2\geq \tau,
\label{A4r}
\end{eqnarray}
where ${\tilde R} [t, t_1;\xi(\tau)]  = { R} [t;\xi(\tau)] \theta(t_1-\tau +0)$, and $\theta(z)$ is 
the Heaviside step function. The differentiation of  (\ref{SA4r}) yields the differential formula 
\cite{KV2,KV3},
\begin{align}
\Big(\frac{d}{dt} +2\gamma \Big)\langle  {\xi}(t)R[t;\xi(\tau) ]
\rangle =\Big\langle  {\xi}(t)\frac{d}{dt}R[t;\xi(\tau)]
\Big\rangle.
\label{SD1}
\end{align}

{\thm Let  $R[t;\xi(\tau) ]$ be an arbitrary functional. Then, if $\xi(t)$ is the RTP the following 
correlation splitting holds:
\begin{align}
\Big\langle \exp\Big \{i\int\limits_{0}^{t}
 d\tau \xi(\tau) \nu (\tau)\Big \}R[t';\xi(\tau) ] \Big\rangle  = \Phi[t;\nu (\tau)]  \big\langle 
 R[t';\xi(\tau) ] \big\rangle  + \Psi[t,t';\nu (\tau)] \big\langle R^\xi[t';\xi(\tau) ] \big\rangle, 
 \quad t \geq t',
\label{A20}
\end{align}
where $\big\langle R^\xi[t';\xi(\tau) ] \big\rangle  = \big\langle \xi(t')R[t';\xi(\tau) ] 
\big\rangle$ and
\begin{align}
	\Psi[t,t';\nu (\tau)] = \frac{\chi(t-t')}{i\nu(t) \chi^2(0)}\frac{d}{dt}\Phi[t;\nu (\tau)] .
\end{align}}

\proof Expanding the functional, $\exp\Big \{i\int\limits_{0}^{t}
 d\tau \xi(\tau) \nu (\tau)\Big \}$, in  the Taylor series, we obtain
\begin{align}
\Big\langle \exp\Big \{i\int\limits_{0}^{t}
 d\tau \xi(\tau) \nu (\tau)\Big \}R[t';\xi(\tau) ] \Big\rangle   =\Big\langle \Big ( 
 \sum\limits_{n=0}^{\infty} \frac{i^{2n}}{2n!} Z_{2n}(t) +  \sum\limits_{n=0}^{\infty} 
 \frac{i^{2n+1}}{(2n+1)!} Z_{2n+1}(t) \Big) R[t';\xi(\tau) ] \Big\rangle,
 \label{SZ1}
\end{align}
where $Z_n (t) = \prod^n_{k=1} \int^t_0 \nu(t_k)\xi(t_k) dt_k $,  $(t_1\geq t_2\geq 
\dots \geq t_n)$. Using the property of the RTP, $\xi^2(t') = \chi(0)$, and  the 
recurrence relationship (\ref{SA1h}), one can recast (\ref{SZ1}) as,
\begin{align}
\Big\langle \exp\Big \{i\int\limits_{0}^{t}
& d\tau \xi(\tau) \nu (\tau)\Big \}R[t';\xi(\tau) ] \Big\rangle   =\bigg(  
 \sum\limits_{n=0}^{\infty} \frac{i^{2n}}{2n!} \Big\langle Z_{2n} (t) \Big\rangle \bigg ) 
 \Big\langle R[t';\xi(\tau) ] \Big\rangle \nonumber \\
& +  \frac{1}{\chi(0)}\bigg(  \sum\limits_{n=0}^{\infty} \frac{i^{2n+1}}{(2n+1)!} \Big\langle 
Z_{2n+1}(t) \xi (t')\Big\rangle \bigg ) \Big\langle R^\xi [t';\xi(\tau) ] \Big\rangle.
 \label{SZ2}
\end{align}
Next, expanding the  characteristic function $\Phi[t;\nu (\tau)]$ in  its Taylor series and 
taking into account that $\langle Z_{2n+1} (t) \rangle =0$, we obtain
\begin{align}
	\Phi[t;\nu (\tau)] = \sum\limits_{n=0}^{\infty} \frac{i^n}{n!} \Big\langle Z_{n} (t) 
	\Big\rangle = \sum\limits_{n=0}^{\infty} \frac{i^{2n}}{2n!} \Big\langle Z_{2n} (t) \Big\rangle .
	\label{SZ3}
\end{align}
A similar consideration yields
\begin{align}
\Big\langle \exp\Big \{i\int\limits_{0}^{t}
 d\tau \xi(\tau) \nu (\tau)\Big \}\xi(t')\Big\rangle   = \sum\limits_{n=0}^{\infty} \frac{i^n}{n!} 
 \Big\langle Z_{n} (t) \xi(t')\Big\rangle =  \sum\limits_{n=0}^{\infty} \frac{i^{2n+1}}{(2n+1)!} 
 \Big\langle Z_{2n+1}(t) \xi (t')\Big\rangle. 
 \label{SZ3g}
\end{align}
Using Eqs. (\ref{SZ3}) and (\ref{SZ3g}), one can rewrite  (\ref{SZ2}) as,
\begin{align}
\Big\langle \exp\Big \{i\int\limits_{0}^{t}
 d\tau \xi(\tau) \nu (\tau)\Big \}R[t';\xi(\tau) ] \Big\rangle   = \Phi[t;\nu (\tau)] 
 \Big\langle R[t';\xi(\tau) ] \Big\rangle  + \frac{1}{\chi(0)}  \Big\langle \exp\Big 
 \{i\int\limits_{0}^{t} d\tau \xi(\tau) \nu (\tau)\Big \}\xi(t')\Big\rangle \Big\langle R^\xi 
 [t';\xi(\tau)]. \Big\rangle
 \label{SZ2a}
\end{align}
Once again using the relationship $\xi^2(t) =  \chi(0)$ and Eq. (\ref{SA1h}), we obtain
\begin{align}
 \sum\limits_{n=0}^{\infty} \frac{i^n}{n!} 
 \Big\langle Z_{n} (t) \xi(t')\Big\rangle  = \frac{1}{\chi(0)} \sum\limits_{n=0}^{\infty} 
 \frac{i^n}{n!} 
 \Big\langle \xi(t) Z_{n} (t) \xi(t)\xi(t')\Big\rangle =  \frac{1}{\chi(0)} \sum\limits_{n=0}^{\infty} 
 \frac{i^{2n+1}}{(2n+1)!} 
 \Big\langle \xi(t )Z_{2n+1}(t)\Big\rangle  \langle \xi(t)\xi (t')\rangle   .
 \label{SZ4}
\end{align}
The last term can be rewritten as,
\begin{align}
 \frac{1}{\chi(0)} \sum\limits_{n=0}^{\infty} \frac{i^{2n+1}}{(2n+1)!} 
 \Big\langle \xi(t )Z_{2n+1}(t)\Big\rangle  \langle \xi(t)\xi (t')\rangle  = \frac{\chi(t- 
 t')}{\nu(t)\chi(0)} \sum\limits_{n=0}^{\infty} \frac{i^n}{n!} \frac{d}{dt}
 \Big\langle Z_{n}(t)\Big\rangle =  \frac{\chi(t - t')}{i\nu(t)\chi(0)}  \frac{d}{dt} \Phi[t;\nu (\tau)] .
 \label{SZ5}
\end{align}

From here it follows,
\begin{align}
 \Big\langle \exp\Big \{i\int\limits_{0}^{t}d\tau \xi(\tau) \nu (\tau)\Big \}\xi(t')\Big\rangle  = 
 \frac{\chi(t - t')}{i\nu(t)\chi(0)}  \Big\langle \exp\Big \{i\int\limits_{0}^{t}d\tau \xi(\tau) \nu 
 (\tau)\Big \}\xi(t')\Big\rangle 
 =  \frac{\chi(t - t')}{i\nu(t)\chi(0)}  \frac{d}{dt} \Phi[t;\nu (\tau)] .
 \label{S5a}
\end{align}
Inserting this result into the r.h.s. of Eq. (\ref{SZ2}), we obtain
\begin{align}
\Big\langle \exp\Big \{i\int\limits_{0}^{t}
 d\tau \xi(\tau) \nu (\tau)\Big \}R[t';\xi(\tau) ] \Big\rangle  = \Phi[t;\nu (\tau)]  \big\langle  , 
 R[t';\xi (\tau) ] \big\rangle  + \Psi [t,t';\nu (\tau)] \big\langle R^\xi[t';\xi(\tau) ] \big\rangle,
 \quad t \geq t',
\end{align}
where \begin{align}
	\Psi[t,t';\nu (\tau)] = \frac{\chi(t-t')}{i\nu(t) \chi^2(0)}\frac{d}{dt}\Phi[t;\nu (\tau)] .
\end{align} 

\qed

{\thm Let  $R[t;\xi(\tau) ]$ be an arbitrary functional. Then, if $\xi(t)$ is the RTP, the following 
correlation splitting holds:
\begin{align}
\Big\langle \exp\Big \{i\int\limits_{0}^{t-t'}
 d\tau \xi(\tau) \nu (\tau)\Big \}R[t';\xi(\tau) ] \Big\rangle  = \Phi[t-t';\nu (\tau)]  \big\langle 
 R[t';\xi(\tau) ] \big\rangle  -\frac{1}{i\nu(t) \chi(0)}\frac{d}{dt'}\Phi[t-t';\nu (\tau)] \big\langle 
 R^\xi[t';\xi(\tau) ] \big\rangle , 
 \quad t \geq t' ,
\label{SA1g}
\end{align}
where $\big\langle R^\xi[t';\xi(\tau) ] \big\rangle  = \big\langle \xi(t')R[t';\xi(\tau) ] \big\rangle 
$ and $\Phi[t-t';\nu (\tau)]= \Big\langle \exp\Big \{i\int\limits_{0}^{t-t'}
 d\tau \xi(\tau) \nu (\tau)\Big \}\Big\rangle$.}

\proof  Since for the stationary random process, $\xi(t) = \xi(t-t')$, one can recast the 
functional $\Phi[t-t';\nu (\tau)]$ as follows:
$\Phi[t-t';\nu (\tau)]=\langle e^{i\varphi(t)} e^{-i\varphi(t')}\rangle$, where $\varphi(t) 
=\int_0^t d\tau \xi (\tau)\nu(\tau)$. Next, expanding $e^{i\varphi(t)}$ and 
$e^{-i\varphi(t')}\rangle$ in  the Taylor series, one can write
\begin{align}
\Big\langle \exp\Big \{i\int\limits_{0}^{t-t'}
 d\tau \xi(\tau) \nu (\tau)\Big \}R[t';\xi(\tau) ] \Big\rangle   =\Big\langle \Big ( 
 \sum\limits_{n=0}^{\infty} \frac{i^{n}}{n!} Z_{n}(t) \Big ) \Big(  \sum\limits_{m=0}^{\infty} 
 \frac{(-i)^{m}}{m!} Z_{m}(t') \Big) R[t';\xi(\tau) ] \Big\rangle,
 \label{SZ1a}
\end{align}
where $Z_n (t) = \prod^n_{k=1} \int^t_0 \nu(t_k)\xi(t_k) dt_k $ and  $Z_m (t) = 
\prod^m_{k=1} \int^t_0 \nu(\tau_k)\xi(\tau_k) d\tau_k $, $(t_1\geq t_2\geq \dots \geq t_n 
\geq \tau_1\geq \tau_2\geq \dots \geq \tau_m)$. 

Using the results of Theorem 1.1, after some algebra we obtain,
\begin{align}
&\Big\langle \exp\Big \{i\int\limits_{0}^{t-t'}
 d\tau \xi(\tau) \nu (\tau)\Big \}R[t';\xi(\tau) ] \Big\rangle   = \Phi[t-t';\nu (\tau)]  \big\langle 
 R[t';\xi(\tau) ] \big\rangle  \nonumber \\
 &+ \Big\langle \Big ( 
 \sum\limits_{m,n=0}^{\infty} \frac{i^{2n}}{2n!} \frac{(-i)^{2m+1}}{(2m+1)!} Z_{2n}(t) Z_{2m 
 +1}(t')  + \sum\limits_{m,n=0}^{\infty} \frac{i^{2n+1}}{(2n+1)!} \frac{(-i)^{2m}}{2m!} Z_{2n 
 +1}(t) Z_{2m}(t') \Big) R[t';\xi(\tau)] \Big\rangle.
 \label{SZ1b}
\end{align}
Further, employing the relationship $\xi^2(t') =  \chi(0)$, one can recast the r.h.s. of 
Eq.  (\ref{SZ1b}) as,
\begin{align}
 &\Big\langle \Big ( \sum\limits_{m,n=0}^{\infty} \frac{i^{2n}}{2n!} \frac{(-i)^{2m+1}}{(2m+1)!} 
 Z_{2n}(t) Z_{2m +1}(t')  + \sum\limits_{m,n=0}^{\infty} \frac{i^{2n+1}}{(2n+1)!} 
 \frac{(-i)^{2m}}{2m!} Z_{2n +1}(t) Z_{2m}(t') \Big) R[t';\xi(\tau) ] \Big\rangle \nonumber \\
  &=\frac{1}{\chi(0)}\Big (\Big\langle  \sum\limits_{m,n=0}^{\infty} \frac{i^{2n}}{2n!} 
  \frac{(-i)^{2m+1}}{(2m+1)!} Z_{2n}(t) Z_{2m +1}(t') \xi(t') + \sum\limits_{m,n=0}^{\infty} 
  \frac{i^{2n+1}}{(2n+1)!} \frac{(-i)^{2m}}{2m!} Z_{2n +1}(t) Z_{2m}(t') \xi(t') \Big)\Big\rangle 
  \Big\langle  R^\xi[t';\xi(\tau) ] \Big\rangle. 
 \label{SZ2b}
\end{align}

From Eq. (\ref{SZ1a}) it follows,
\begin{align}
\frac{d}{dt'} \Phi[t-t';\nu (\tau)]    =\Big\langle \Big ( 
 \sum\limits_{n=0}^{\infty} \frac{i^{n}}{n!} Z_{n}(t) \Big ) \Big ( \frac{d}{dt'} 
 \sum\limits_{m=0}^{\infty} 
 \frac{(-i)^{m}}{m!} Z_{m}(t') \Big)\Big\rangle.
 \label{SZ3a}
\end{align}
A short computation yields,
\begin{align}
&\frac{d}{dt'} \Phi[t-t';\nu (\tau)]    =-i \nu(t')\Big\langle \Big ( 
 \sum\limits_{n=0}^{\infty} \frac{i^{n}}{n!} Z_{n}(t) \Big ) \Big ( \sum\limits_{m=0}^{\infty} 
 \frac{(-i)^{m}}{m!} Z_{m}(t') \xi(t')\Big)  \Big\rangle \nonumber \\
&=-i \nu(t')\Big (\Big\langle  \sum\limits_{m,n=0}^{\infty} \frac{i^{2n}}{2n!} 
\frac{(-i)^{2m+1}}{(2m+1)!} Z_{2n}(t) Z_{2m +1}(t') \xi(t') + \sum\limits_{m,n=0}^{\infty} 
\frac{i^{2n+1}}{(2n+1)!} \frac{(-i)^{2m}}{2m!} Z_{2n +1}(t) Z_{2m}(t') \xi(t') \Big)\Big\rangle .
 \label{SZ4a}
\end{align}
Using this result in Eqs. (\ref{SA1g}) and  (\ref{SZ1a}), we obtain
\begin{align}
\Big\langle \exp\Big \{i\int\limits_{0}^{t-t'}
 d\tau \xi(\tau) \nu (\tau)\Big \}R[t';\xi(\tau) ] \Big\rangle  = \Phi[t-t';\nu (\tau)]  \big\langle 
 R[t';\xi(\tau) ] \big\rangle  -\frac{1}{i\nu(t) \chi(0)}\frac{d}{dt'}\Phi[t-t';\nu (\tau)] \big\langle 
 R^\xi[t';\xi(\tau) ] \big\rangle , 
 \quad t \geq t' .
\label{SA1a}
\end{align}
\qed

{\cor
	\label{mcorlabel} Let  $\varphi(t) = D\int_0^t \xi (\tau) d\tau $. Then, for an arbitrary 
	functional, $R[t;\xi(\tau) ]$, the following correlation splitting holds:
\begin{align}
\big\langle e^{i\varphi(t)} e^{-i\varphi(t')} R[t;\xi(\tau) ]\big\rangle =
\Phi(t-t') \big\langle R[t;\xi(\tau) ]\big\rangle 
 - \frac{1}{iD\sigma^2}\frac{d}{dt'} \Phi(t- t')  \big\langle R^\xi[t;\xi(\tau) ]\big\rangle,
\label{RTP}
\end{align}
where $\Phi(t-t') =\langle e^{i\varphi(t)} e^{-i\varphi(t')} \big\rangle $ and $\sigma^2 = 
\chi(0)$.}

\section{Population distribution inside acceptor band}

 The population distribution, $\langle{\rho}_{EE}(t)\rangle$, inside acceptor band can 
 be found as solution of the following differential equation,  obtained from Eqs. 
 (\ref{EqA5}) - (\ref{EqA6})  in the same approximation as Eqs. (\ref{BN6ar}) - 
 (\ref{BN7r}). The population density inside of the band is given by $\rho(E)  =\langle{  
 \rho}_{EE}\rangle \varrho(E)$. For  $\rho(E)  $ we obtain the following differential 
 equation:
\begin{align}
\frac{d}{dt}\rho(E,t)  =& \gamma_1(t) \big\langle{\rho}_{11}(t) \big\rangle   -
\gamma_2(t) {\langle{ \bar\rho}}_{22}(t)\rangle ,
\label{EqA9}
\end{align}
where 
\begin{align} \label{G2}
\gamma_1(t)  =&{v^2}e^{-\alpha (E- E_0)^2}\int_0^t dt' e^{-{\bar p}^2(t-t')^2/4 }\Big 
(e^{i(E_1-E)(t-t')} +  
e^{-i(E_1-E)(t-t')}  \Big). \\
\gamma_2(t)  =&{v^2}e^{-\alpha (E- E_0)^2}\int_0^t dt' e^{-{\bar p}^2(t-t')^2/4 } 
e^{-t'^2/4\alpha }\Big 
( e^{i\varepsilon(t-t')} e^{i(E_0-E)t}+ e^{-i\varepsilon(t-t')} e^{-i(E_0-E)t}  \Big),
\label{G2a}
\end{align}
and we set $\bar p = \sqrt{2} D\sigma$. Performing the integration, we obtain
\begin{align}
\gamma_1(t)  =& {\frac{\sqrt{\pi} v^2}{\bar p}} \exp\bigg(-\alpha \Delta_0^2 
-\frac{\Delta_1^2}{{\bar p}^2}\bigg)\bigg({\rm erf}\Big(\frac{\bar p t}{2} + 
i\frac{\Delta_1}{\bar p} \Big )-{\rm erf}\Big( i\frac{\Delta_1}{\bar p} \Big )+{\rm 
c.c.}\bigg), \nonumber \\
{\tilde \gamma}_2(t) 
=&{\frac{\sqrt{\pi} v^2}{p}} \exp\Big(-\alpha \Delta_0^2 -i\Delta_0 
t-\frac{t^2}{4\alpha}\Big)\bigg(
\Big( \exp\frac{(t/2\alpha -i\varepsilon )^2}{p^2} \Big)  {\rm 
erf}\Big(\frac{pt}{2}-\frac{t/2\alpha -i\varepsilon }{p} \Big) 
-\exp\Big(-\frac{\varepsilon ^2}{p^2} \Big)  {\rm erf}\Big(\frac{i\varepsilon }{p} 
\Big)\bigg) + {\rm c.c.} ,
\end{align}
where ${\Delta}_0 = E_0 - E$ and ${\Delta}_1 = E_1 - E$. 
\begin{align}
{\tilde \gamma}_2(t) 
&={\frac{A\sqrt{\pi}}{p}} \exp\Big(-i\Delta_0 t-\frac{t^2}{4\alpha}\Big)\bigg(
\Big( \exp\frac{(t/2\alpha -i\varepsilon )^2}{p^2} \Big)  {\rm 
erf}\Big(\frac{pt}{2}-\frac{t/2\alpha -i\varepsilon }{p} \Big) 
-\exp\Big(-\frac{\varepsilon ^2}{p^2} \Big)  {\rm erf}\Big(\frac{i\varepsilon }{p} 
\Big)\bigg) ,
\end{align}

The solution of Eq. (\ref{EqA9}) can be written as,
\begin{align}
{\langle{\bar \rho}}_{EE}(t)\rangle =\int_0^t(\gamma_1(\tau) + \gamma_2(\tau)  ) 
\langle{\rho}_{11}(\tau) \big\rangle d\tau - \int_0^t \gamma_2(\tau)  d\tau .
\end{align}

Using the asymptotic formula, $ \big\langle{\rho}_{11}(t) \big\rangle \sim 
e^{-\Gamma_1 t}$,  where
\begin{align}\label{G}
\Gamma_1= \frac{2\sqrt{\pi}\,{v}^2}{{p}}e^{-{\varepsilon^2}/{p^2}},
\end{align}
and expanding the limits of integration in the last integral as, $\tau 
\rightarrow\infty$, we obtain
\begin{align}\label{R2}
\rho(E)=\rho(E, t)\big |_{t \rightarrow \infty}  =\int_0^\infty(\gamma_1(t) + 
\gamma_2(t)  )e^{-\Gamma_1 t} dt  - \int_0^\infty \gamma_2(t)  dt.
\end{align}

The computation yields
\begin{align}\label{G12b}
\rho(E)= \frac{\sqrt{\pi/2} v^2}{D \sigma}  \Big(\frac{\delta}{2\sqrt{\pi} \Gamma_1} 
\Psi_1(E) + \Psi_2(E) - \Psi_0(E)\Big)\displaystyle e^{-4\pi (E_0 -E)^2/\delta^2},
\end{align}
where 
\begin{align}\label{G12a}
\Psi_0(E) = &2 {\rm Re}\bigg( w\Big(\frac{E-E_0 -\varepsilon }{{\sqrt{2} D\sigma 
}}\Big) w\Big(\frac{2\sqrt{\pi} (E-E _0)}{{\delta }}\Big) \bigg ) ,\\
\Psi_1(E) =& 2 {\rm Re} \bigg (w\Big(\frac{E-E_0 -\varepsilon +i\Gamma_1}{{\sqrt{2} 
D\sigma}}\Big) \bigg ) , \\
\Psi_2(E) = &2 {\rm Re}\bigg(w\Big(\frac{E-E_0 -\varepsilon+i\Gamma_1 }{{\sqrt{2} 
D\sigma}}\Big) w\Big(\frac{2\sqrt{\pi} (E-E _0 +i\Gamma_1)}{{\delta }}\Big)\bigg ) .
\end{align}
Here, $w(z) = e^{-z^2} {\rm erfc}(-iz)$ is a complex complementary error function 
\cite{abr}. 

To proceed further, we use the approximated formula \cite{abr}:
\begin{align}
w(z) \approx \left \{
\begin{array}{l}
1, \quad |z| \ll 1\\
\displaystyle\frac{i}{\sqrt{\pi}z}, \quad |z| \gg 1
\end{array}
\right .
\end{align}

In the limit, $\bar p \ll \Gamma_1$ (weak noise) and $\delta \ll\varepsilon$  (narrow 
zone), we obtain the leading term as,
\begin{align}
{\langle{ \bar \rho}}_{EE}\rangle  \sim \frac{{2v}^2}{\Gamma_1^2 +(E -E_1)^2 }.
\end{align}
For $\sqrt{\Gamma_1^2 +(\varepsilon +E_0-E)^2} \ll \bar p$ (strong noise), we find 
\begin{align}
{\langle{ \bar \rho}}_{EE}\rangle  \sim \frac{{2v}^2}{\bar p\,\Gamma_1}\Big 
(1+\frac{2{\Gamma_1}^2}{\Gamma_1^2 +(E -E_0)^2}\Big).
\end{align}

\subsection*{Details of computation}

Here we present the details of calculations for ${\langle{\rho}}_{EE}\rangle$. We start 
with some auxiliary integrals \cite{PR1,PR2}
\begin{align}\label{P1}
\int e^{-(ax^2+bx+c)} dx =\frac{1}{2}\sqrt{\frac{\pi}{a}}e^{(b^2-4ac)/(4a)}{\rm 
erf}(\sqrt{a}\, x+\frac{b}{2\sqrt{a}}), \\
\int_0^\infty e^{-px}{\rm erf}(cx +b) dx=\frac{1}{p} {\rm erf}(b) + \frac{1}{p} 
\exp\Big( \frac{p^2 +4pbc}{4c^2}\Big){\rm erfc}\Big(b+\frac{p}{2c}\Big), \\
\int e^{-px}{\rm erf}(cx +b) dx=-\frac{1}{p} e^{-px}{\rm erf}(cx+b) + \frac{1}{p} 
\exp\Big( \frac{p^2 +4pbc}{4c^2}\Big){\rm erf}\Big(cx +b+\frac{p}{2c}\Big). 
\label{P2}
\end{align}

Performing integration in (\ref{G2}) with help of (\ref{P1}), we obtain
\begin{align}
\gamma_1(t)  =&|V|^2\varrho_0\int_0^t  e^{-{\bar p}^2\tau^2/4 }\Big 
(e^{i(E_1-E)\tau} +  
e^{-i(E_1-E)\tau}  \Big)d\tau \nonumber \\
= & A\exp\bigg(-\frac{\Delta_1^2}{{\bar p}^2}\bigg)\bigg({\rm erf}\Big(\frac{\bar p 
t}{2} + i\frac{\Delta_1}{\bar p} \Big )-{\rm erf}\Big( i\frac{\Delta_1}{\bar p} \Big 
)+{\rm c.c.}\bigg),
\end{align}
where $\Delta_1 = E_1 - E$ and $A={2{v}^2}/({{\bar p}\delta}) $. 

Next step is to find $I_1(t) =\int_0^t\gamma_1(\tau)e^{-\Gamma_1 \tau d\tau} $. 
Using (\ref{P2}), we obtain
\begin{align}
I_1(t) =&-\frac{A}{\Gamma_1}e^{-\Gamma_1 t}\Big({\rm erfc}\Big(\frac{\bar p t}{2} 
+\frac{i\Delta_1}{\bar p} \Big )-{\rm erfc}\Big(\frac{i\Delta_1}{\bar p} \Big ) \Big) 
\nonumber \\
&+\frac{A}{\Gamma_1}\bigg(\exp\bigg(\frac{(\Gamma_1 +i\Delta_1)^2}{{\bar 
p}^2}\bigg)\bigg({\rm erfc}\Big(\frac{\bar p t}{2} +\frac{\Gamma_1 +i\Delta_1}{\bar 
p} \Big )-{\rm erfc}\Big(\frac{\Gamma_1 +i\Delta_1}{\bar p} \Big )\bigg)+{\rm 
c.c.}\bigg).
\end{align}

To calculate $\gamma_2(t)$, it is convenient to define a new function, 
\begin{align}
{\tilde \gamma}_2(t) = &\frac{A}{\varrho_0}\sqrt{\frac 
{\alpha}{\pi}}\int^{\infty}_{-\infty} dE' 
\varrho(E')  \int_0^t dt' e^{-{\bar p}^2(t-t')^2/4 } e^{-i(E_1-E)t}  e^{i(E_1-E')t'},
\label{G3}
\end{align}
so that ${ \gamma}_2={\tilde \gamma}_2+ {\tilde \gamma}_2^\ast$. Performing 
integration over band, we obtain
\begin{align}
{\tilde \gamma}_2(t) =& A \int_0^t dt'e^{-t'^2/4\alpha } e^{-{\bar p}^2(t-t')^2/4 } 
e^{i(E-E_1)t}  e^{i\varepsilon t'}\nonumber \\
& =A  e^{-t^2/4\alpha-i\Delta_0 t } \int_0^t dt'e^{-{ p}^2 \tau^2/4 } e^{(t/2\alpha 
-i\varepsilon )\tau} \nonumber \\
&={\frac{A\sqrt{\pi}}{p}} \exp\Big(-i\Delta_0 t-\frac{t^2}{4\alpha}\Big)\bigg(
\Big( \exp\frac{(t/2\alpha -i\varepsilon )^2}{p^2} \Big)  {\rm 
erf}\Big(\frac{pt}{2}-\frac{t/2\alpha -i\varepsilon }{p} \Big) 
-\exp\Big(-\frac{\varepsilon ^2}{p^2} \Big)  {\rm erf}\Big(\frac{i\varepsilon }{p} 
\Big)\bigg) ,
\end{align}
where ${\Delta}_0 = E_0 - E$.

Using  (\ref{P1}), we obtain,
\begin{align}
\int_0^t{\tilde \gamma}_2(\tau) d\tau = &{\frac{A\sqrt{\pi}}{p}}\int_0^t
 d\tau \exp\Big(-i\Delta_0 \tau -\frac{\tau ^2}{4\alpha}\Big) \Big( \exp\frac{(\tau 
 /2\alpha -i\varepsilon )^2}{p^2} \Big)  {\rm erf}\Big(\frac{p\tau }{2}-\frac{\tau 
 /2\alpha -i\varepsilon }{p} \Big) \nonumber \\
&+\frac{A\sqrt{\pi}}{p}\exp\Big(-\frac{\varepsilon ^2}{p^2} \Big)  {\rm 
erf}\Big(\frac{i\varepsilon }{p} \Big)\Big( {\rm erf}\Big(\frac{t}{2\alpha} 
+i\sqrt{\alpha}\Delta_0\Big)- {\rm erf}(i\sqrt{\alpha}\Delta_0)\Big).
\end{align}

Introducing a new variable $\tau = t- t'$, we can rewrite (\ref{G3}) as,
\begin{align}
{\tilde \gamma}_2(t) = &\frac{2{v}^2}{\sqrt{\pi}{\bar 
p}\varrho_0\delta}\int^{\infty}_{-\infty} 
dE' \varrho(E')   e^{-i(E'-E)t} \int_0^t  e^{-{\bar p}^2\tau^2/4 } e^{-i(E_1-E')\tau} 
d\tau,
\label{G4}
\end{align}
Then, integrating over $\tau$, we obtain
\begin{align}
{\tilde \gamma}_2(E,t) = &\frac{2{v}^2}{\sqrt{\pi}{\bar 
p}\varrho_0\delta}\int^{\infty}_{-\infty} 
dE' \varrho(E')  e^{i\Delta't}\exp\bigg(-\frac{{\Delta'}_1^2}{{\bar p}^2}\bigg)\bigg({\rm 
erf}\Big(\frac{\bar p t}{2} +i\frac{{\Delta'}_1}{\bar p} \Big )-{\rm erf}\Big( 
i\frac{{\Delta'}_1}{\bar 
p} \Big )\bigg),
\label{G5}
\end{align}
where $\Delta' = E - E'$ and ${\Delta'}_1 = E_1 - E'$.

Next step is to calculate the contributions: 
$I_{1,2}(\Gamma_1):=\int_0^\infty\gamma_{1,2}(E,t) e^{-\Gamma_1 t} dt $ and 
$I_0:=\int_0^\infty\gamma_{2}(E,t) dt $. As can be seen, $I_0 =I_{2}(0)$. We start 
with $I_1(\Gamma_1)$. Using (\ref{P2}), we obtain
\begin{align}
I_1 (\Gamma_1)=\frac{2{v}^2}{{\bar p}\Gamma_1\delta}\exp\bigg(\frac{(\Gamma_1 
+i\Delta_1)^2}{{\bar p}^2}\bigg)\bigg({\rm erfc}\Big(\frac{\Gamma_1 +i\Delta_1}{\bar 
p} \Big )+{\rm c.c.}\bigg).
\end{align}
This can be recast as follows: $I_1 (\Gamma_1)={\tilde I}_1 (\Gamma_1) + {\tilde 
I^\ast}_1 (\Gamma_1)$, 
\begin{align}
{\tilde I}_1(\Gamma_1) =\frac{2{v}^2}{{\bar p}\Gamma_1\delta} w\Big(\frac{ 
\Delta_1+ i\Gamma_1}{\bar p} \Big ),
\end{align}
where $w(z) = e^{-z^2} {\rm erfc}(-iz)$ \citep{abr}.

To calculate $I_2(\Gamma_1)$, we rewrite it as, $I_2(\Gamma_1)= {\tilde 
I}_2(\Gamma_1)+ {\tilde I}_2^\ast(\Gamma_1)$,
\begin{align}
 {\tilde I}_2(\Gamma_1)=\int_0^\infty {\tilde\gamma}_{2}(E,t) e^{-\Gamma_1 t} dt = 
 \int^{\infty}_{-\infty} dE' \varrho(E')  f(\Gamma_1),
\label{G6}
\end{align}
where 
\begin{align}
 f(\Gamma_1)=\frac{2{v}^2}{\sqrt{\pi}{\bar p}\varrho_0\delta}\,e 
 ^{-{\Delta'}_1^2/{\bar 
 p}^2}\int^{\infty}_0 dt\,e^{-(\Gamma_1 -i\Delta')t}\bigg({\rm erf}\Big(\frac{\bar p 
 t}{2} 
 +i\frac{{\Delta'}_1}{\bar p} \Big )-{\rm erf}\Big( i\frac{{\Delta'}_1}{\bar p} \Big )\bigg).
\label{G7}
\end{align}
Performing the integration, we obtain
\begin{align}
f(\Gamma_1)= \frac{2{v}^2}{\sqrt{\pi}{\bar p}\varrho_0\delta}\frac{1}{\Gamma_1 
-i\Delta'} 
\exp\bigg(\frac{(\Gamma_1 +i\Delta_1)^2}{{\bar p}^2}\bigg) {\rm erfc}\Big( 
\frac{\Gamma_1 
+i{\Delta}_1}{\bar p} \Big )=\frac{\Gamma_1 {\tilde 
I}_1(\Gamma_1)}{\sqrt{\pi}\varrho_0(\Gamma_1 -i\Delta')}.
\label{G9}
\end{align}
Inserting $f(\Gamma_1)$ into Eq. (\ref{G6}), we obtain
\begin{align}
 {\tilde I}_2(\Gamma_1)={\tilde I}_1(\Gamma_1)\frac{\Gamma_1}{\sqrt{\pi} 
 \varrho_0}  
 \int^{\infty}_{-\infty}  \frac{\varrho(E')   dE'}{\Gamma_1 -i\Delta'}.
\label{G10}
\end{align}

For the Gaussian  density of electron states in the  acceptor band, $\varrho(E)  =   
\varrho_0 
e^{-\alpha (E-E_0)^2}$, we have
\begin{align}
\int^{\infty}_{-\infty}  \frac{\varrho(E')   dE'}{\Gamma_1 -i\Delta'} 
=\int^{\infty}_{-\infty}  
\frac{ \varrho_0 e^{-\alpha (E'-E_0)^2}dE'}{\Gamma_1 -i(E-E')}.
\end{align}
This integral can be calculated using the following relation \cite{abr}:\begin{align}
w(z) = \frac{i}{\pi} \int^{\infty}_{-\infty} \frac{e^{-t^2} dt}{z-t} \quad (\Im z >0).
\end{align}
A computation yields
\begin{align}
\int^{\infty}_{-\infty}  \frac{ \varrho_0 e^{-\alpha (E'-E_0)^2}dE'}{\Gamma_1 -i(E-E')} 
= \pi 
\varrho_0 w\Big(\frac{ 2 \sqrt{\pi} ( i\Gamma_1- \Delta_0) }{\delta} \Big ).
\end{align}
Using this result, we obtain
\begin{align}
 {\tilde I}_2(\Gamma_1)=\sqrt{\pi} {\Gamma_1 } {\tilde I}_1(\Gamma_1)w\Big(\frac{ 
 2\sqrt{\pi} ( i\Gamma_1- \Delta_0) }{\delta} \Big ) = \frac{2 \sqrt{\pi} {v}^2}{{\bar 
 p}\delta} w\Big(\frac{ \Delta_1+ i\Gamma_1}{\bar p} \Big )w\Big(\frac{ 2\sqrt{\pi} ( 
 i\Gamma_1- \Delta_0)}{\delta} \Big ) 
\label{G11a}.
\end{align}
Since  $I_0 =I_{2}(0)$, we obtain
\begin{align}
 {\tilde I}_0= \frac{2\sqrt{\pi} {v}^2}{{\bar p}\delta} w\Big(\frac{ \Delta_1+ 
 i\Gamma_1}{\bar p} \Big )w\Big(\frac{ 2 \sqrt{\pi}( i\Gamma_1- \Delta_0)}{\delta} 
 \Big ) 
\label{G11}.
\end{align}

Performing the integration with $\Phi(t-t')$ taken from Eq. (\ref{C4}), after some 
calculations, we obtain
\begin{align}\label{G12}
{\langle{ \rho}}_{EE}\rangle  \sim b \Big(\frac{\delta}{2\sqrt{\pi} \Gamma_1} \Psi_1(E) 
+ \Psi_2(E) - \Psi_0(E)\Big),
\end{align}
where $b=\sqrt{\pi} {v}^2\varrho_0/{\bar p} $,  $\bar p = \sqrt{2} D\sigma$ and
\begin{align}\label{G12c}
\Psi_0(E) = &2 {\rm Re}\bigg( w\Big(\frac{E-E_1 }{{\bar p }}\Big) 
w\Big(\frac{2\sqrt{\pi} (E-E _0)}{{\delta }}\Big) \bigg ) ,\\
\Psi_1(E) =& 2 {\rm Re} \bigg (w\Big(\frac{E-E_1 +i\Gamma_1}{{\bar p }}\Big) \bigg ) 
, \\
\Psi_2(E) = &2 {\rm Re}\bigg(w\Big(\frac{E-E_1+i\Gamma_1 }{{\bar p }}\Big) 
w\Big(\frac{2\sqrt{\pi} (E-E _0 +i\Gamma_1)}{{\delta }}\Big)\bigg ) .
\end{align}

\section{Estimation of the approximations in the main text}

While the function, $W(\tau)$, is defined explicitly in terms of the 
special functions, the analytical expression for $Z(\tau)$ is unknown. Thus, we restrict 
ourselves to numerical simulations in order to estimate the validity of the approximation for the 
results 
obtained in the main text of the paper.  In Figs. \ref{SM9} and \ref{SM11}, we present the plots 
of the function, $Z(\tau)$, for the parameters used in Figs. 9 -- 16, 19 and 21 -- 26 in the 
paper. In the title to 
each figure below, the number of the corresponding figure in the main text is shown.  Only 
main parameters for identification of the corresponding figures in the main text are presented 
below in figure captions for functions, $Z(\tau)$.
These results allow us to obtain estimates for the accuracy of our approximation.
\begin{figure}[tbh]
\begin{center}
\scalebox{0.375}{\includegraphics{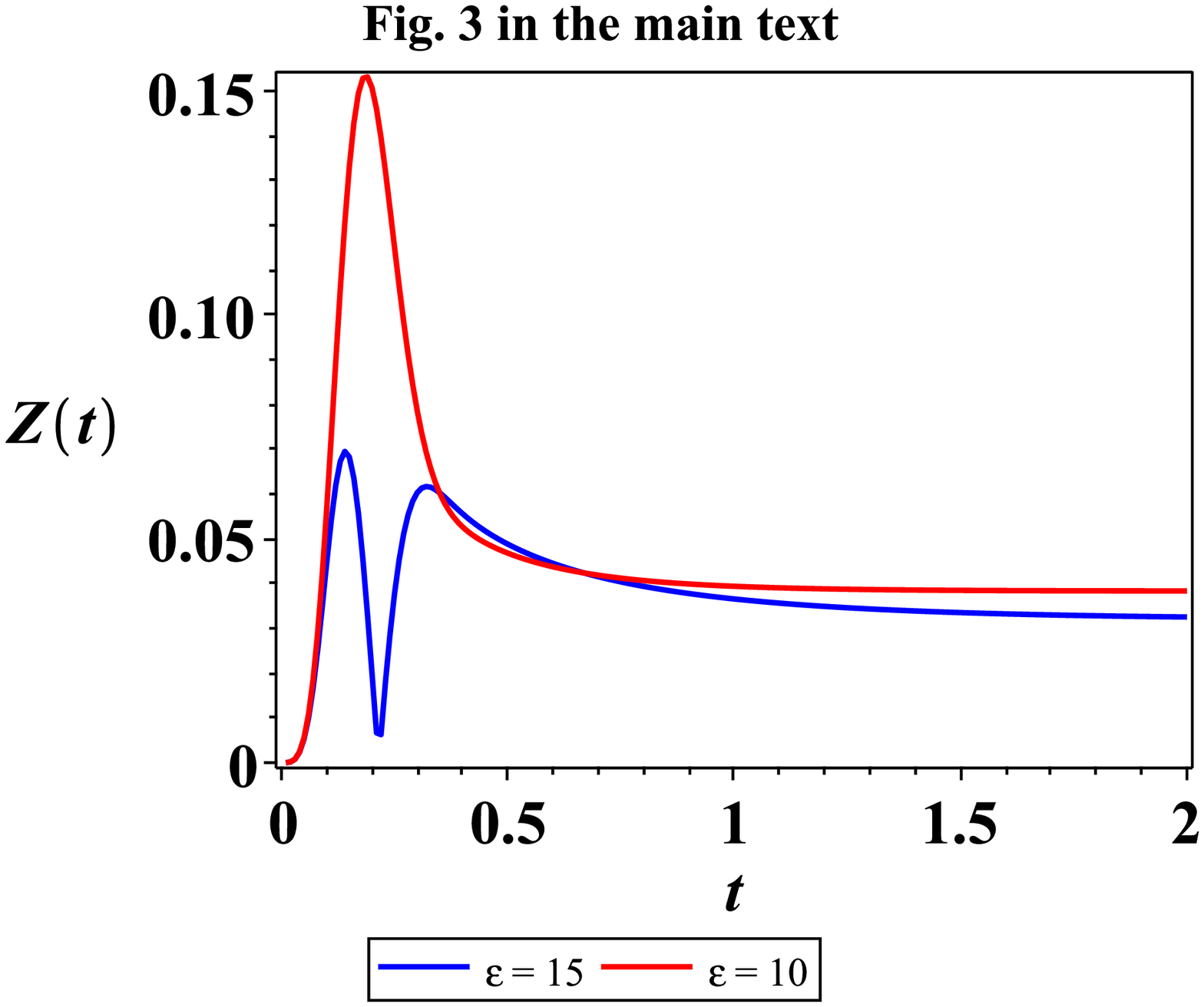}}
\scalebox{0.415}{\includegraphics{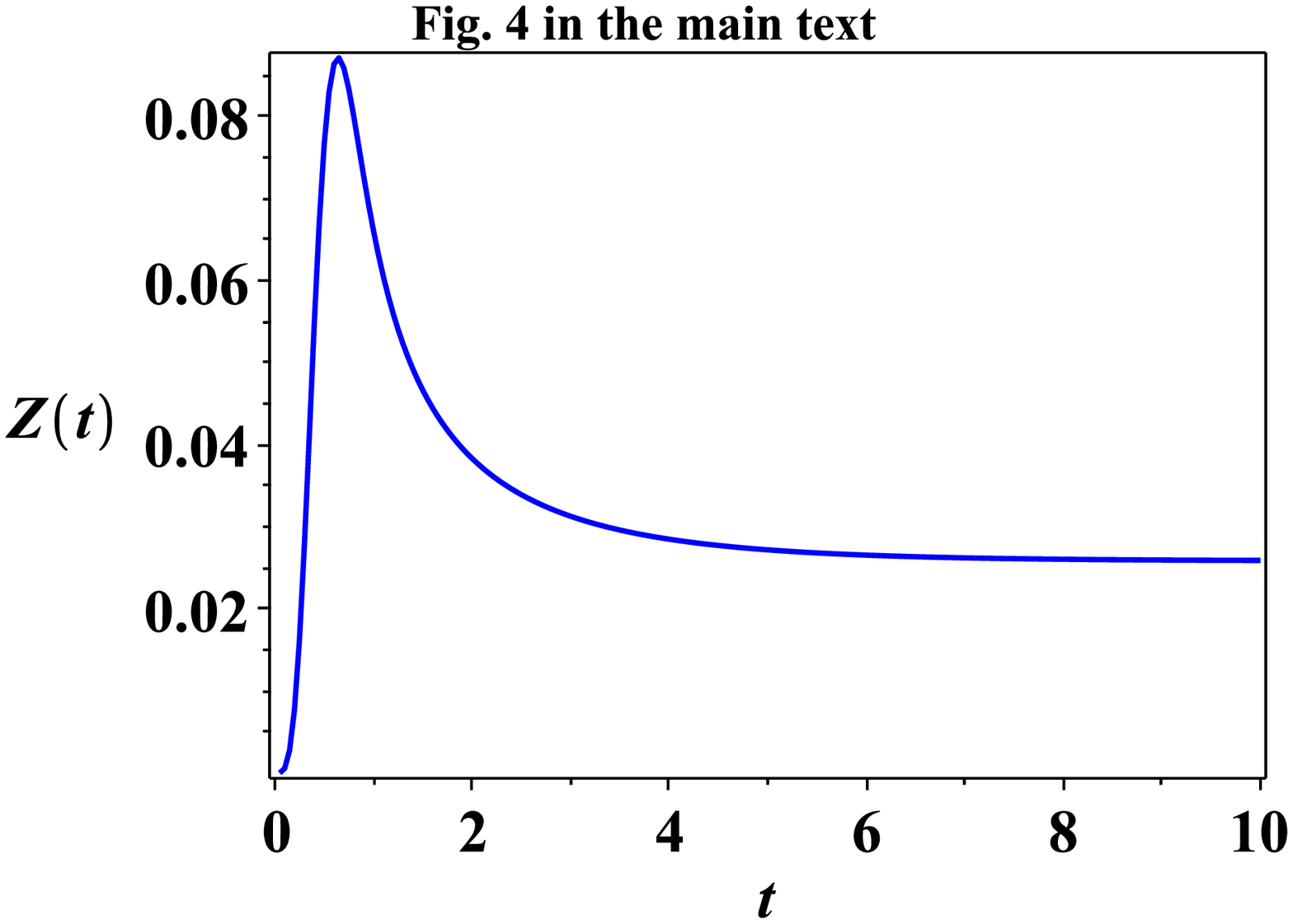}}
\scalebox{0.425}{\includegraphics{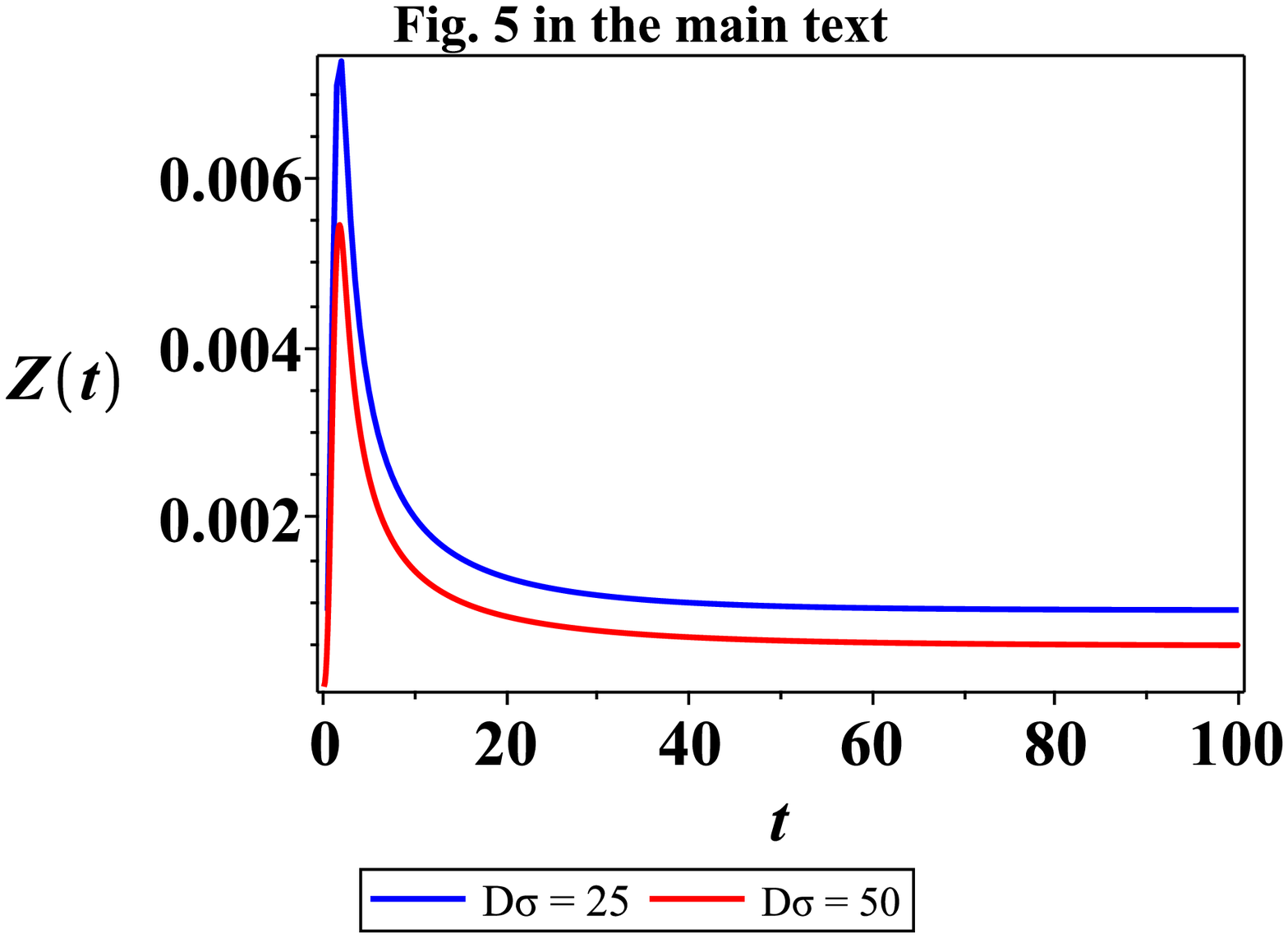}}
\scalebox{0.41}{\includegraphics{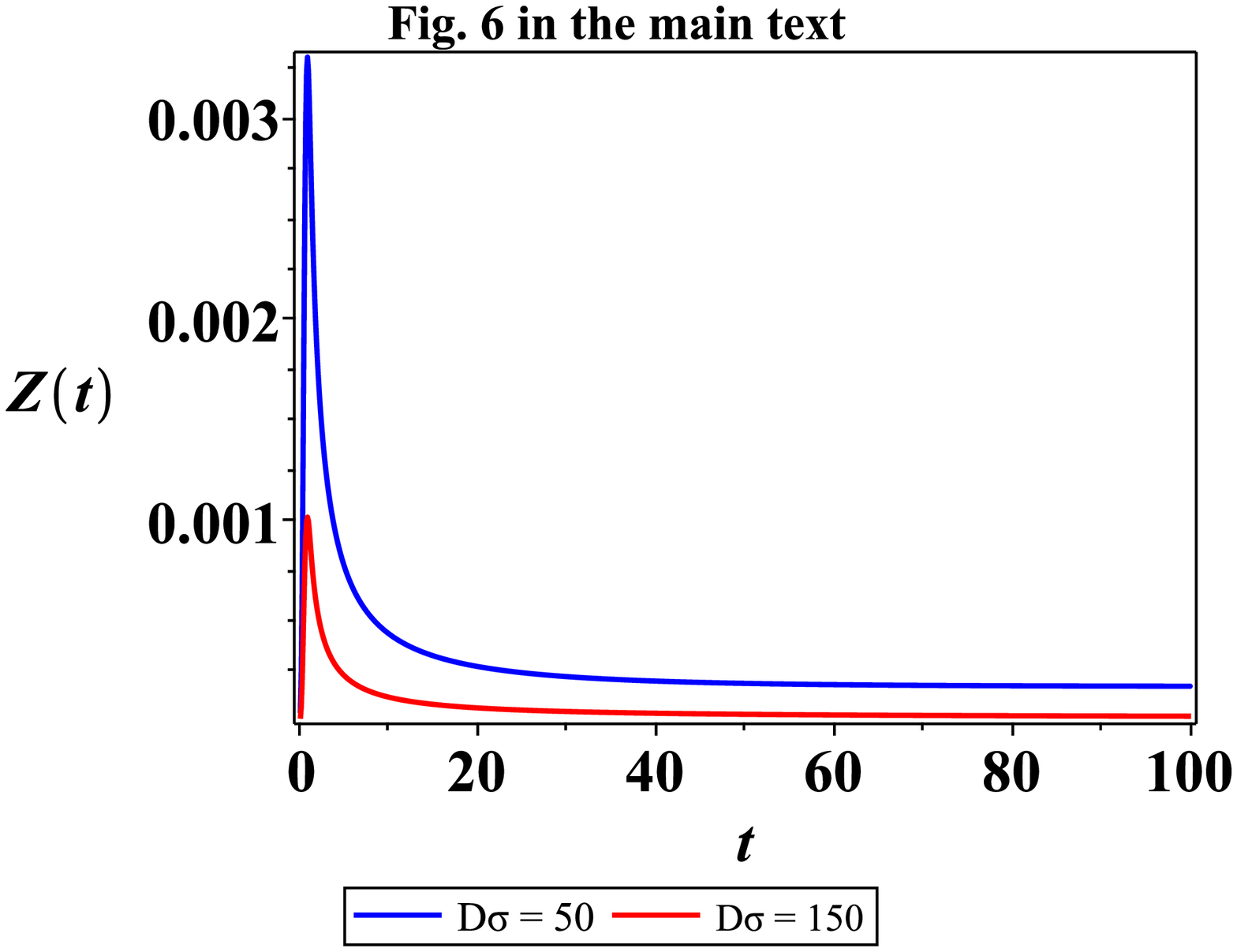}}
\scalebox{0.4}{\includegraphics{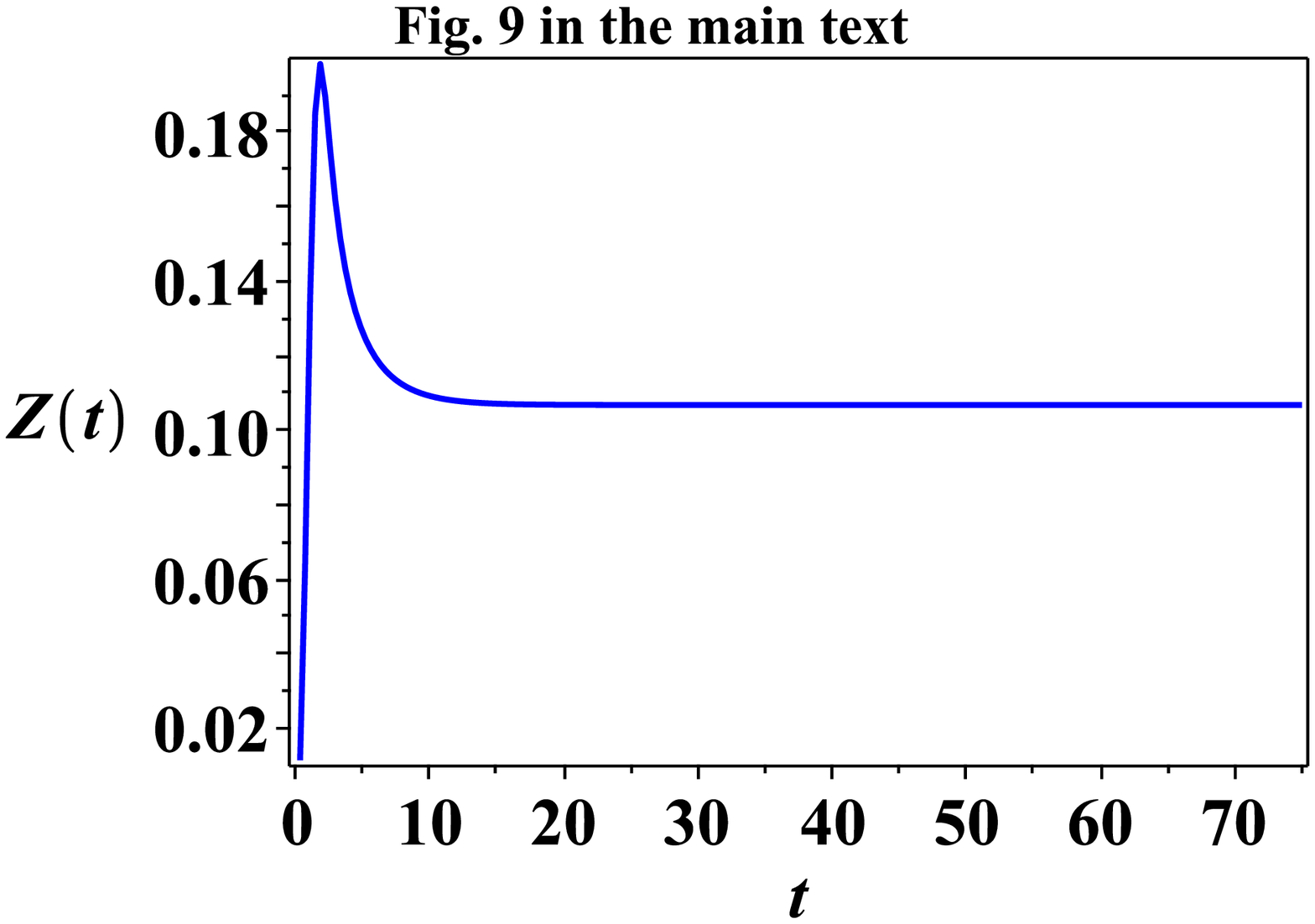}}
\scalebox{0.385}{\includegraphics{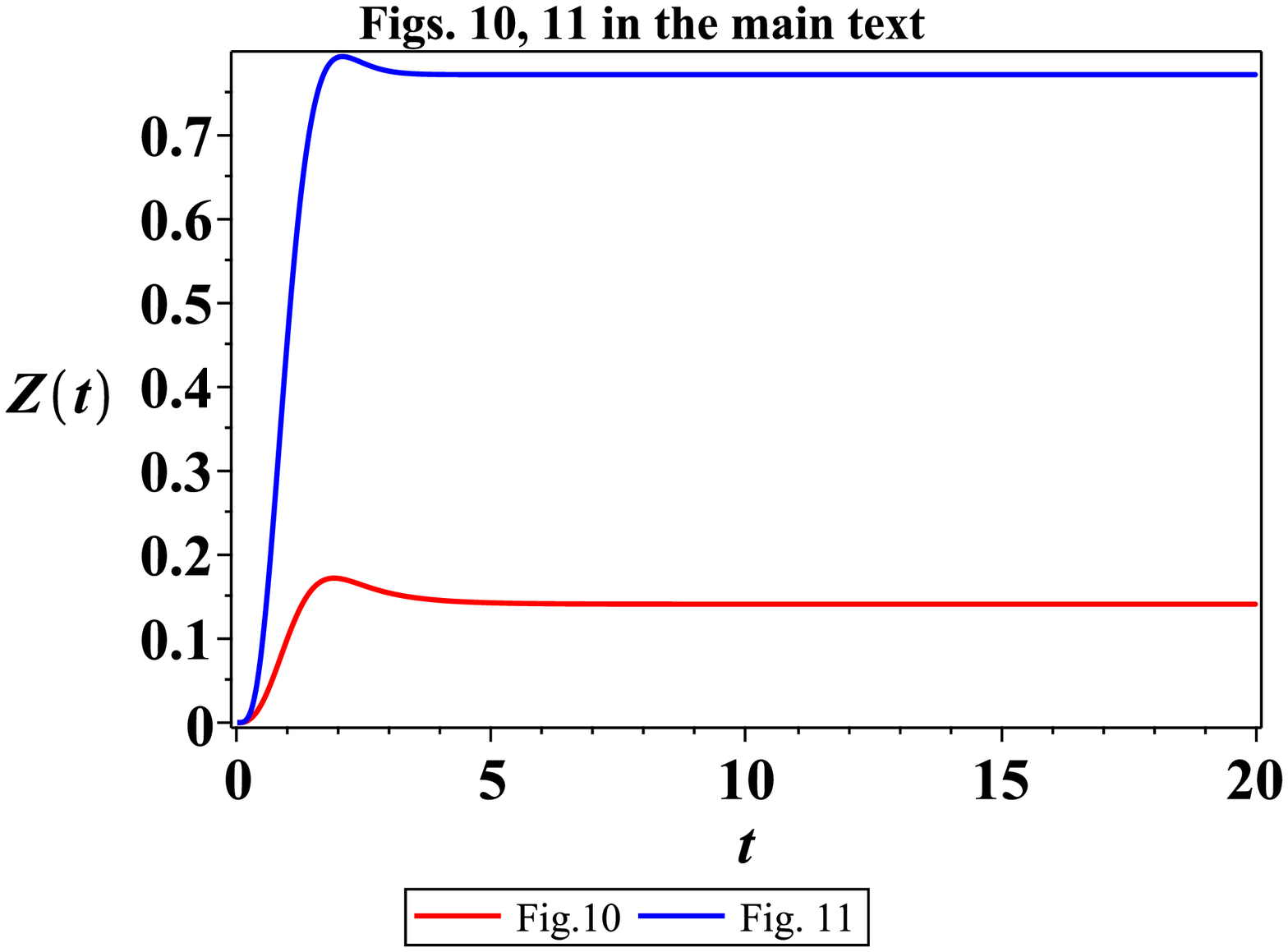}}
\end{center}
\caption{(Color online) Dependence of $Z$ on time $t$. Estimates are made 
for the results presented in  Figs. 3 -- 6 and Figs.  9 --11 in the main text.
\label{SM8}}
\end{figure}

\begin{figure}[tbh]
\begin{center}
\scalebox{0.4}{\includegraphics{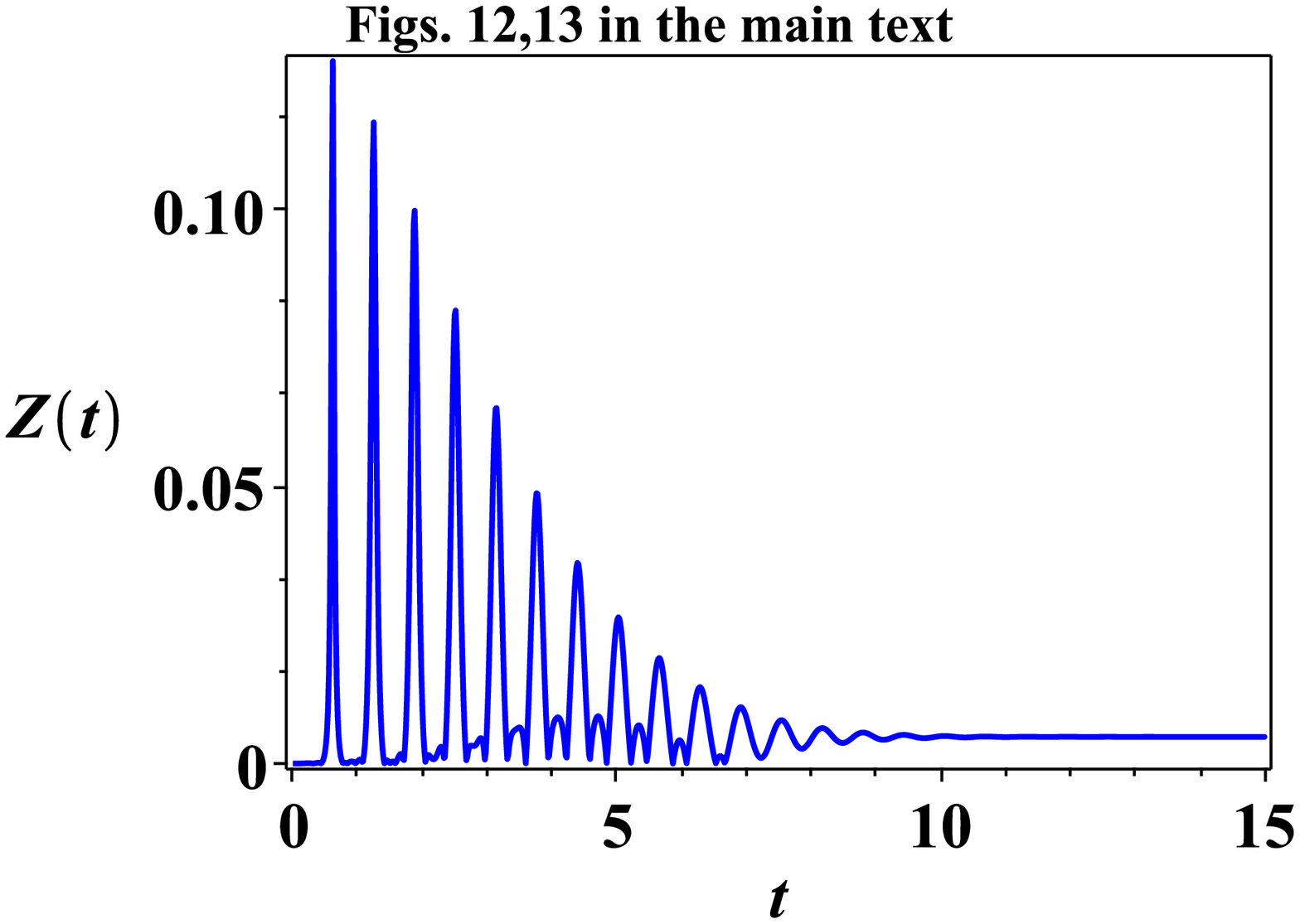}}
\scalebox{0.41}{\includegraphics{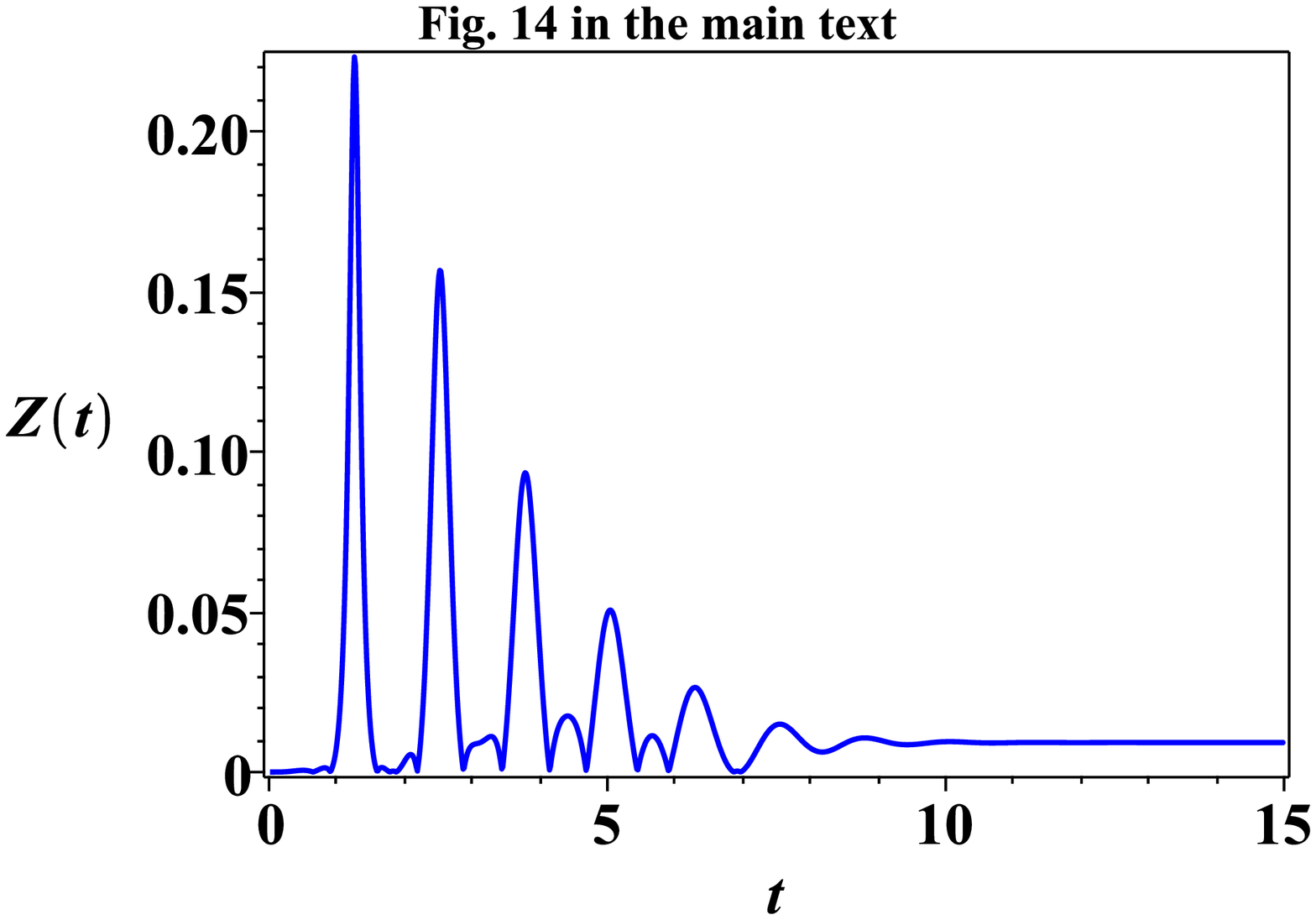}}
\scalebox{0.375}{\includegraphics{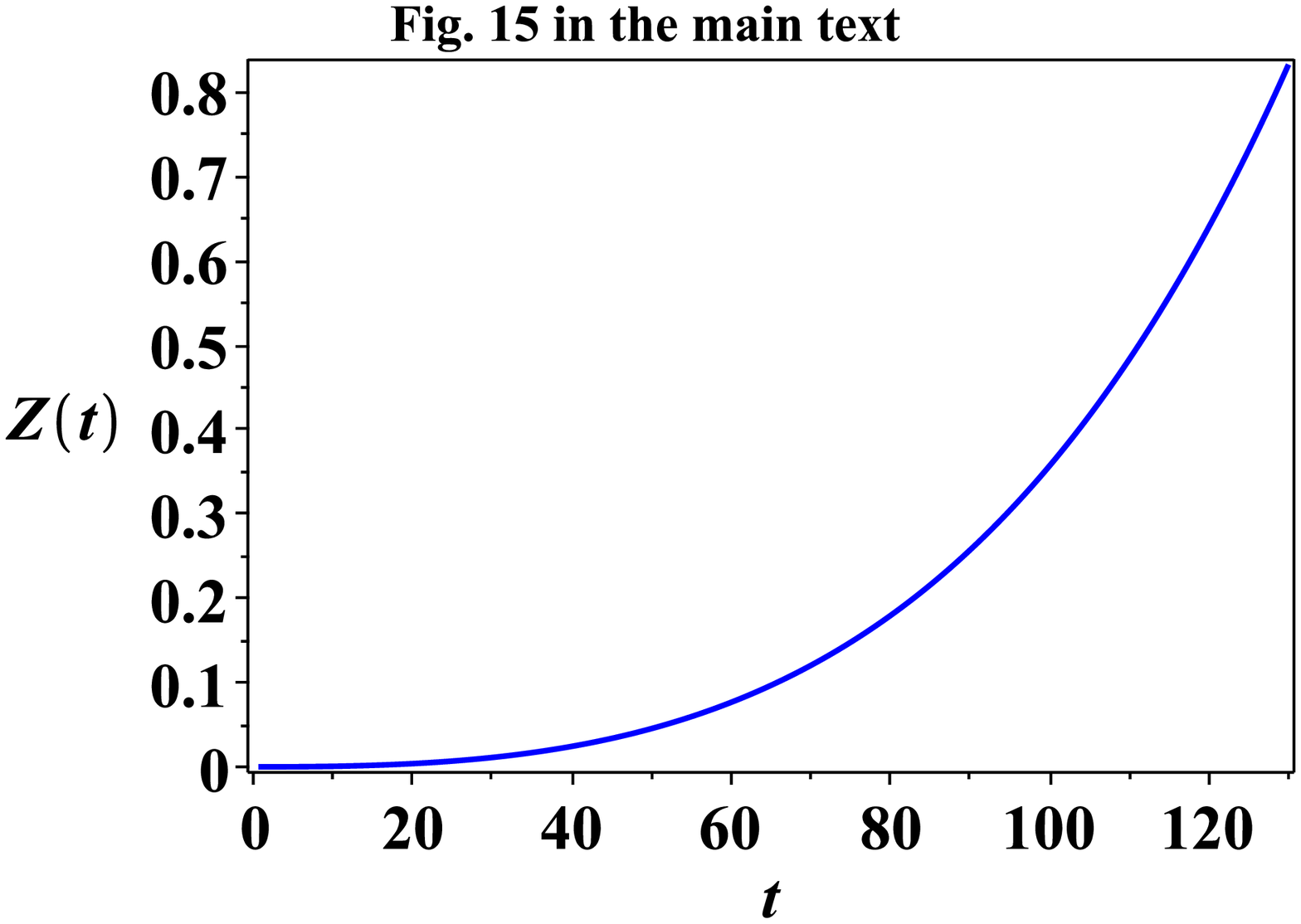}}
\scalebox{0.4}{\includegraphics{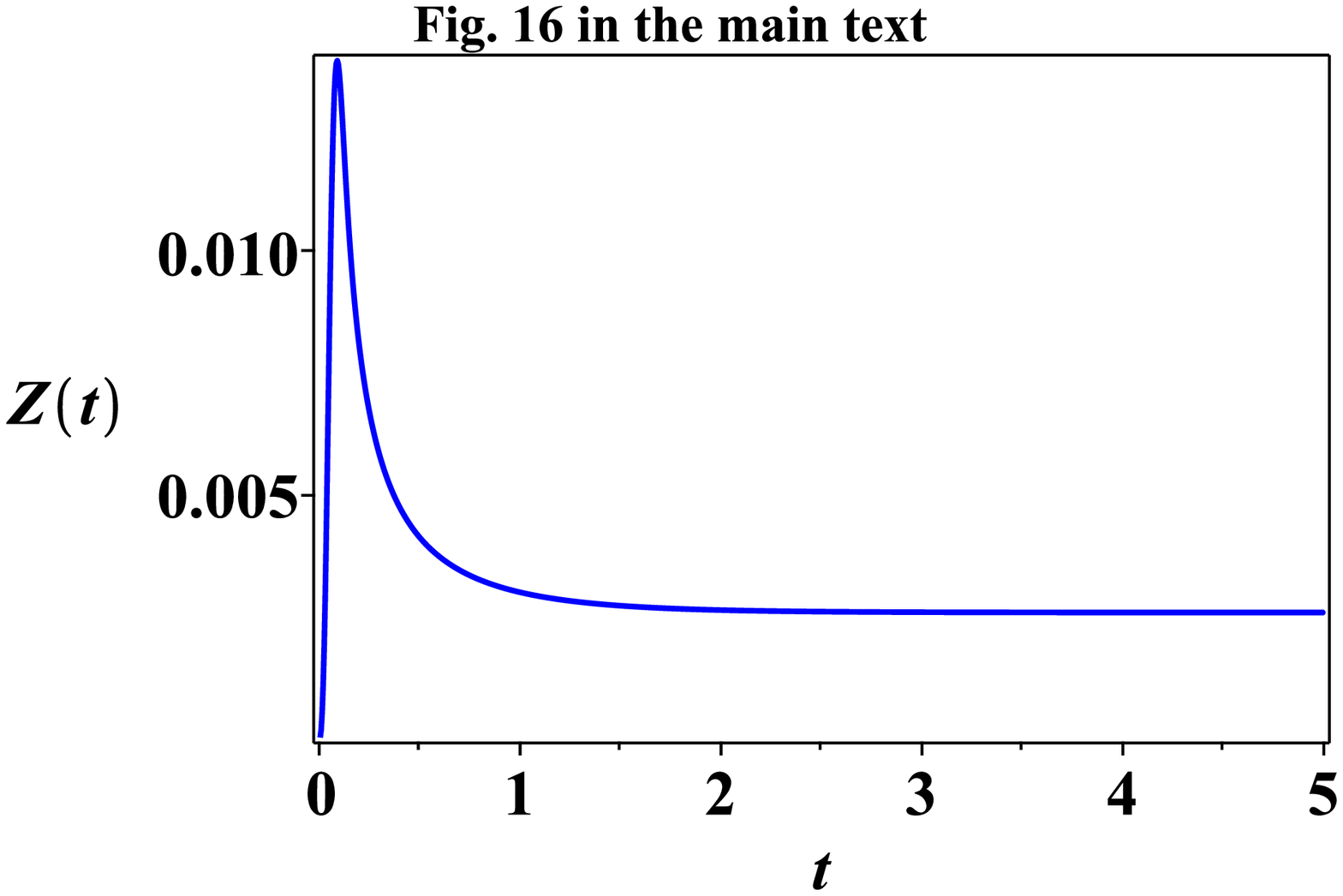}}
\scalebox{0.4}{\includegraphics{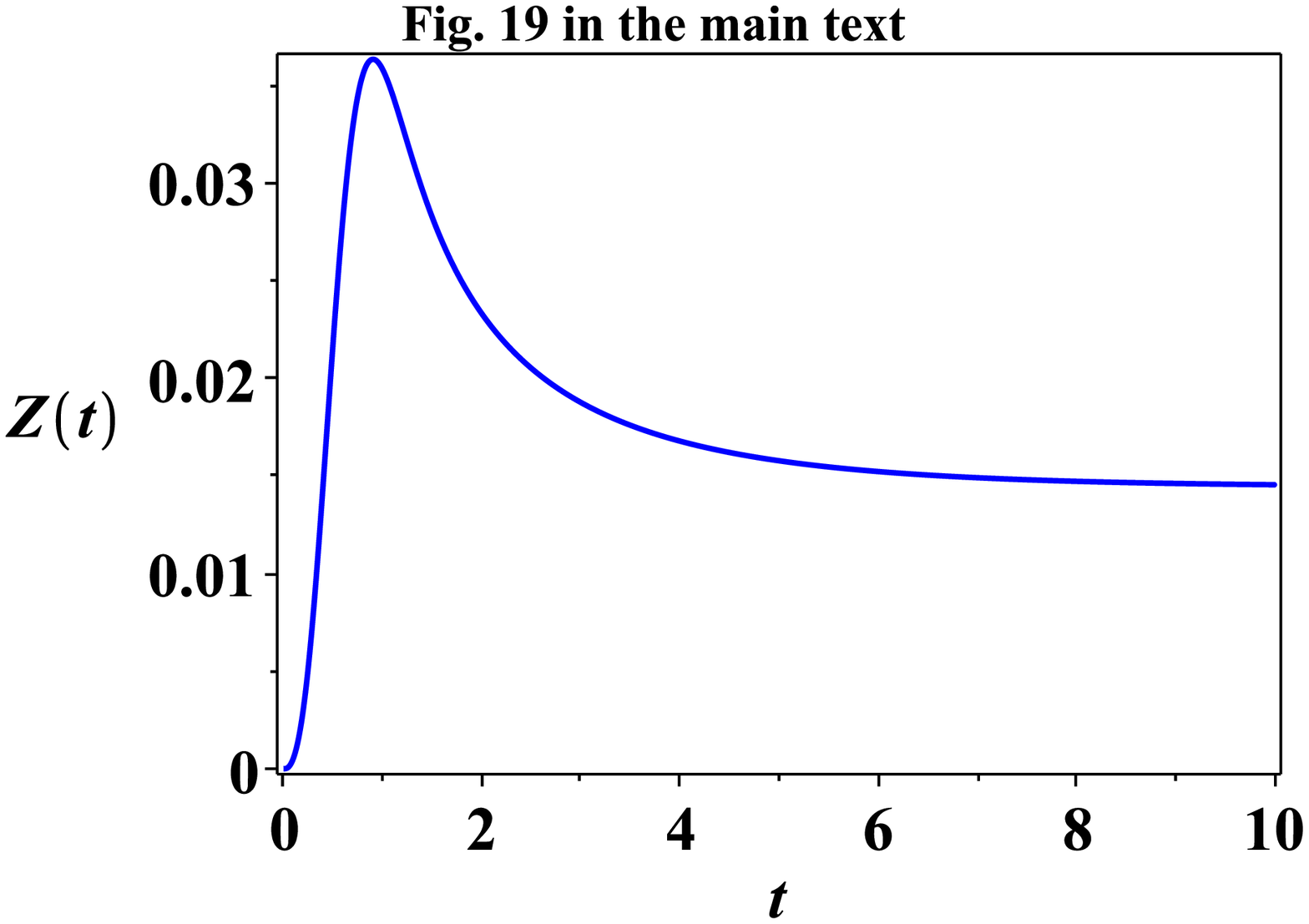}}
\scalebox{0.425}{\includegraphics{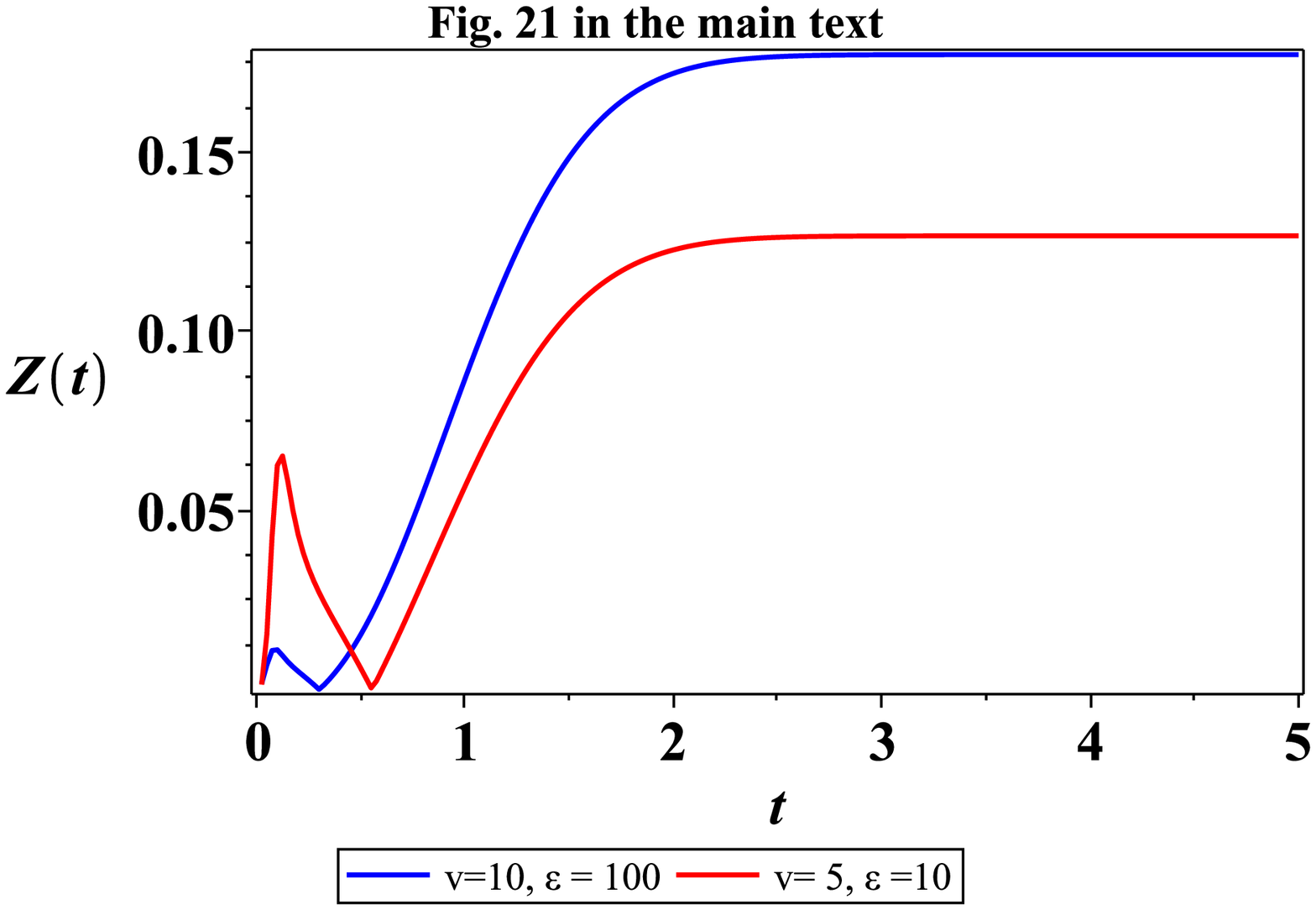}}
\end{center}
\caption{(Color online)  Dependence of $Z$ on time $t$. Estimates are made 
for the results presented in  Figs. 12 -- 16 and 19, 21 in the main text.
\label{SM9}}
\end{figure}

\begin{figure}
\begin{center}
\scalebox{0.4}{\includegraphics{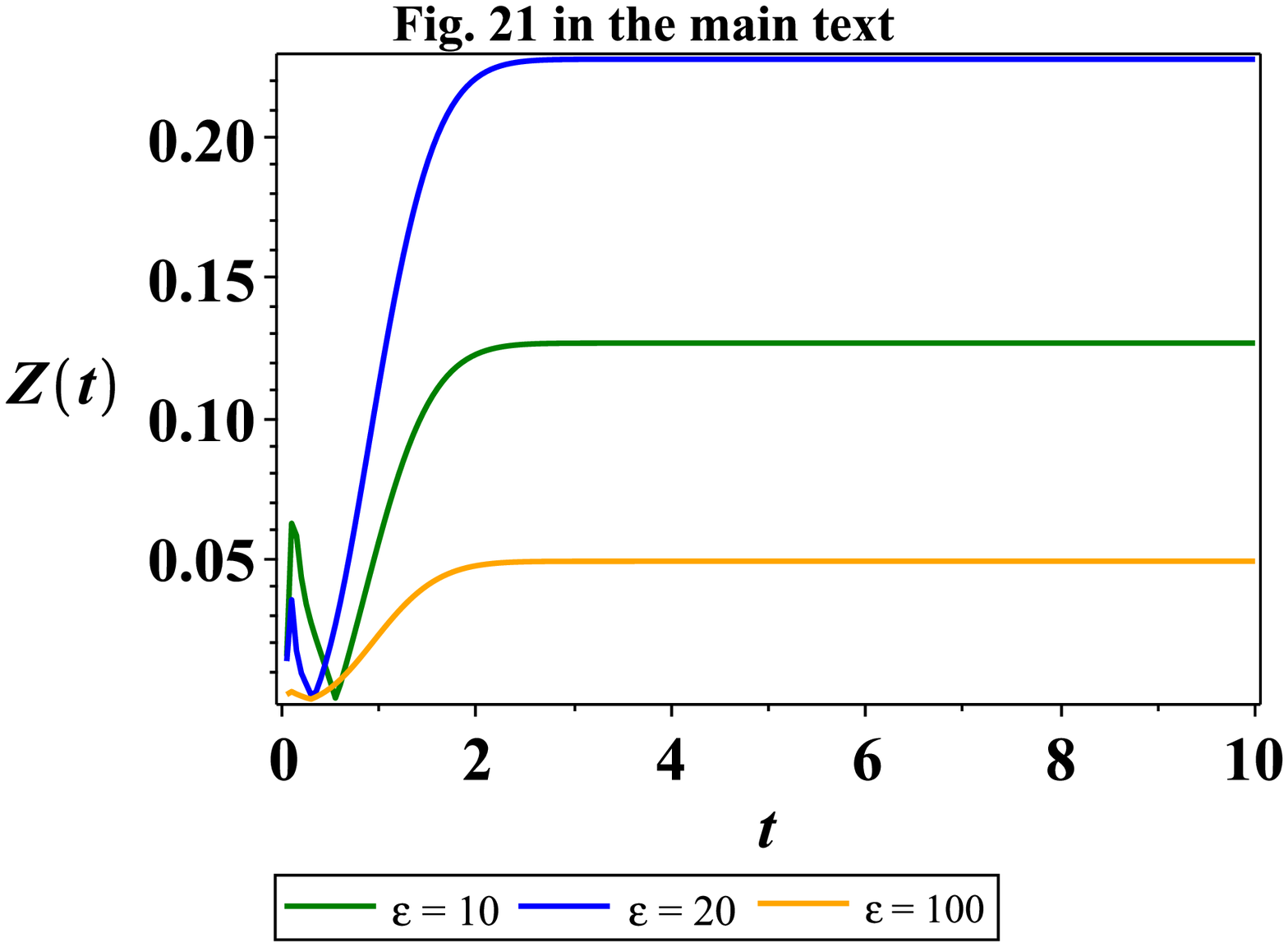}}
\scalebox{0.415}{\includegraphics{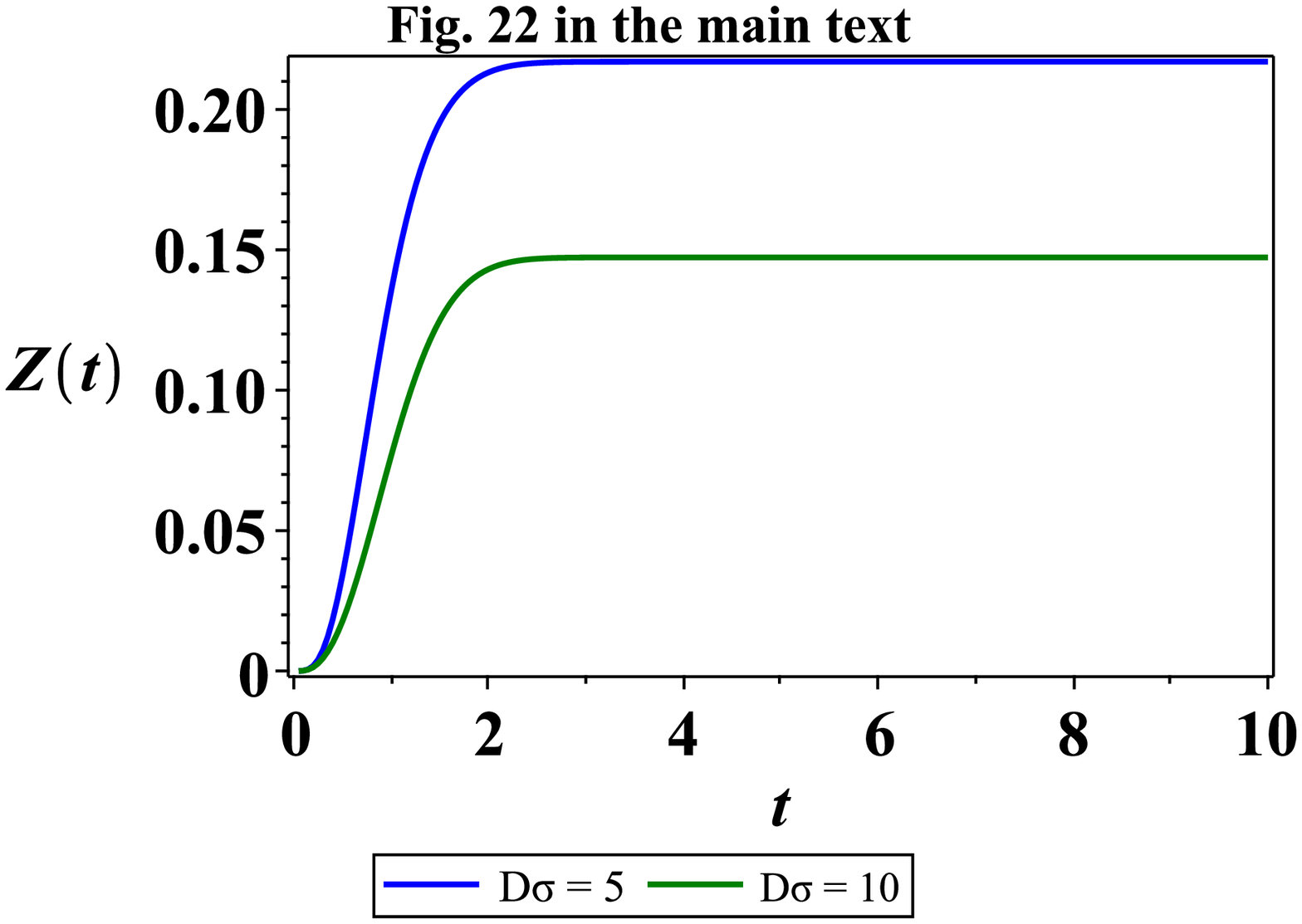}}
\scalebox{0.415}{\includegraphics{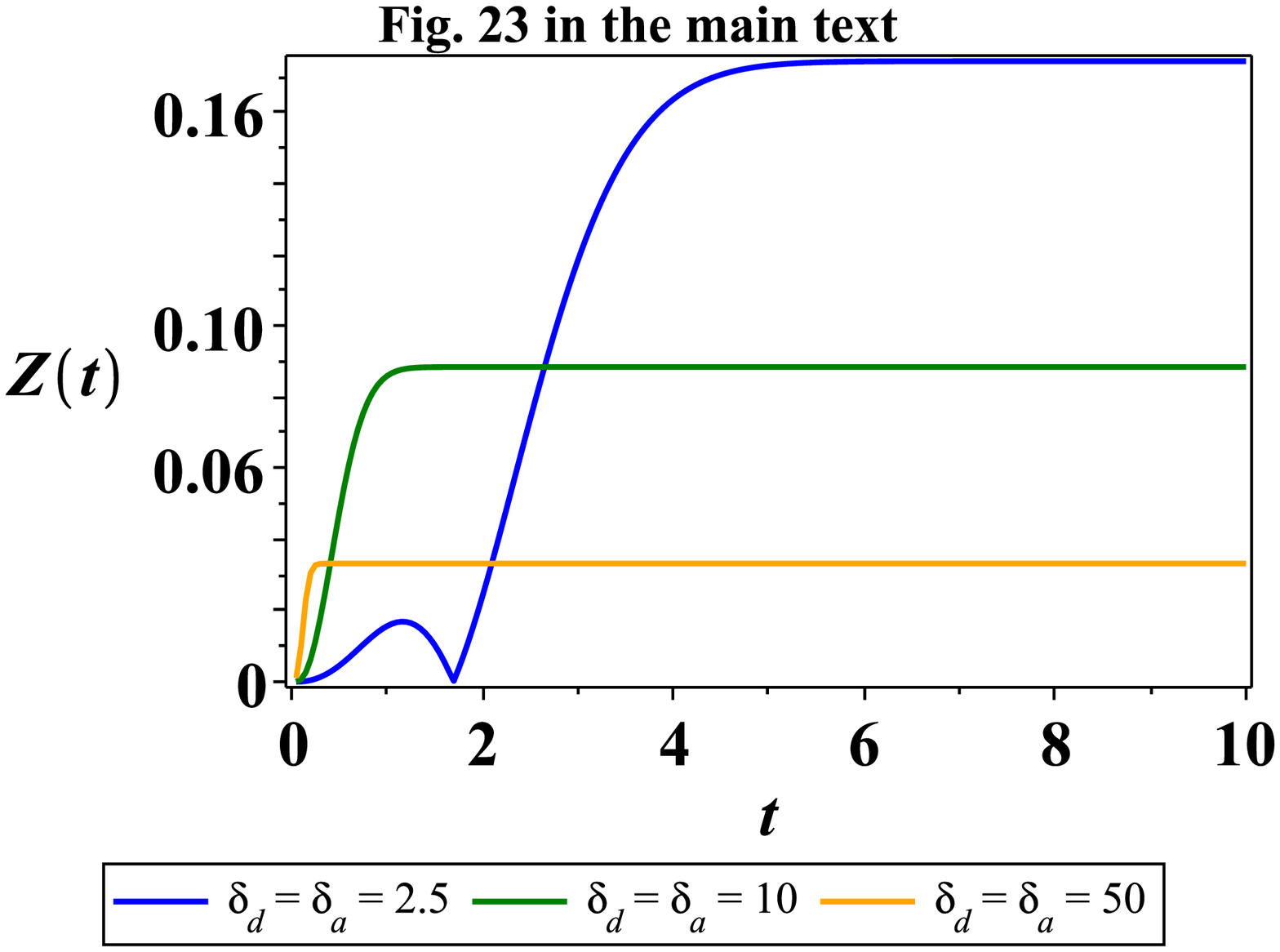}}
\scalebox{0.4}{\includegraphics{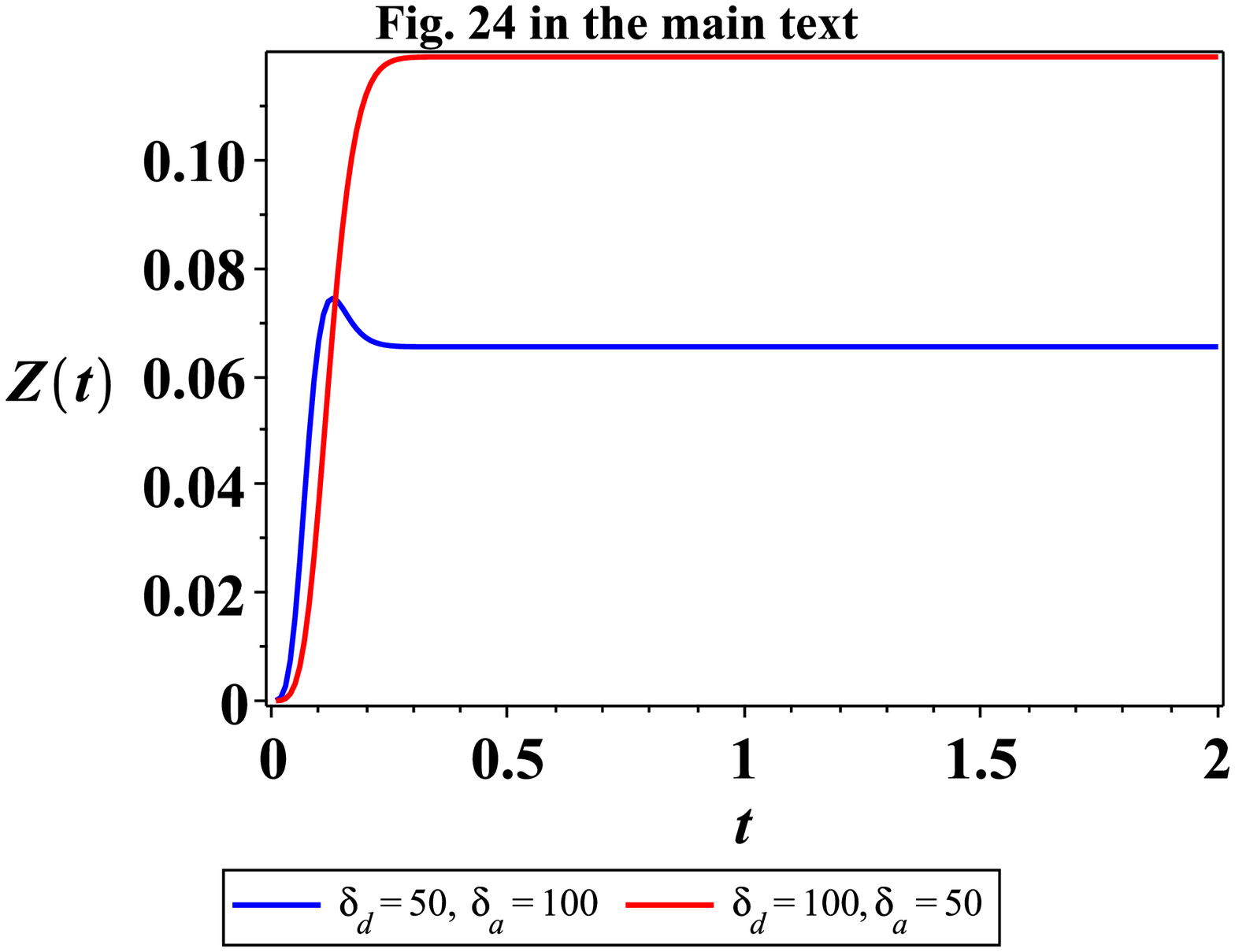}}
\scalebox{0.425}{\includegraphics{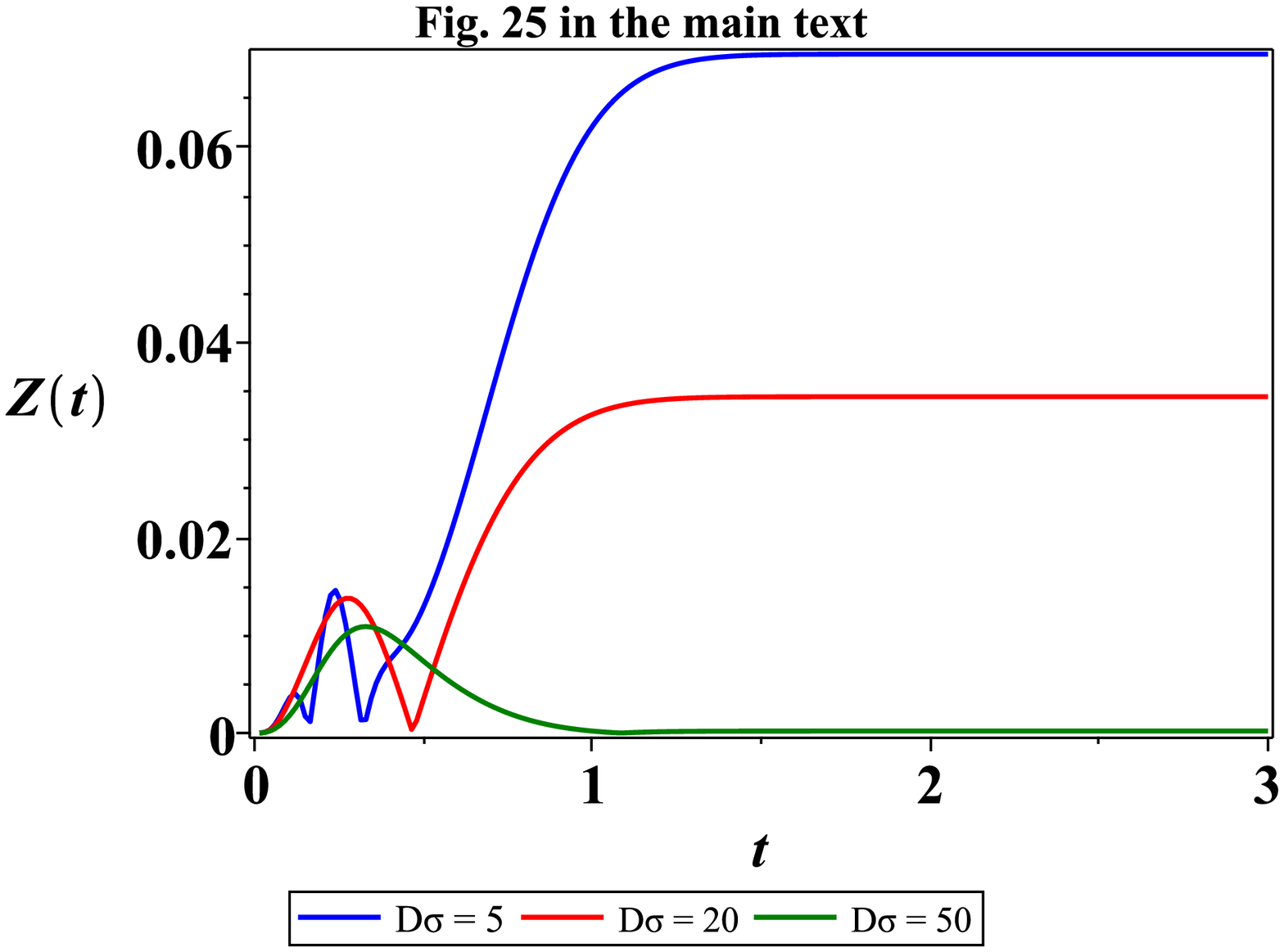}}
\scalebox{0.43}{\includegraphics{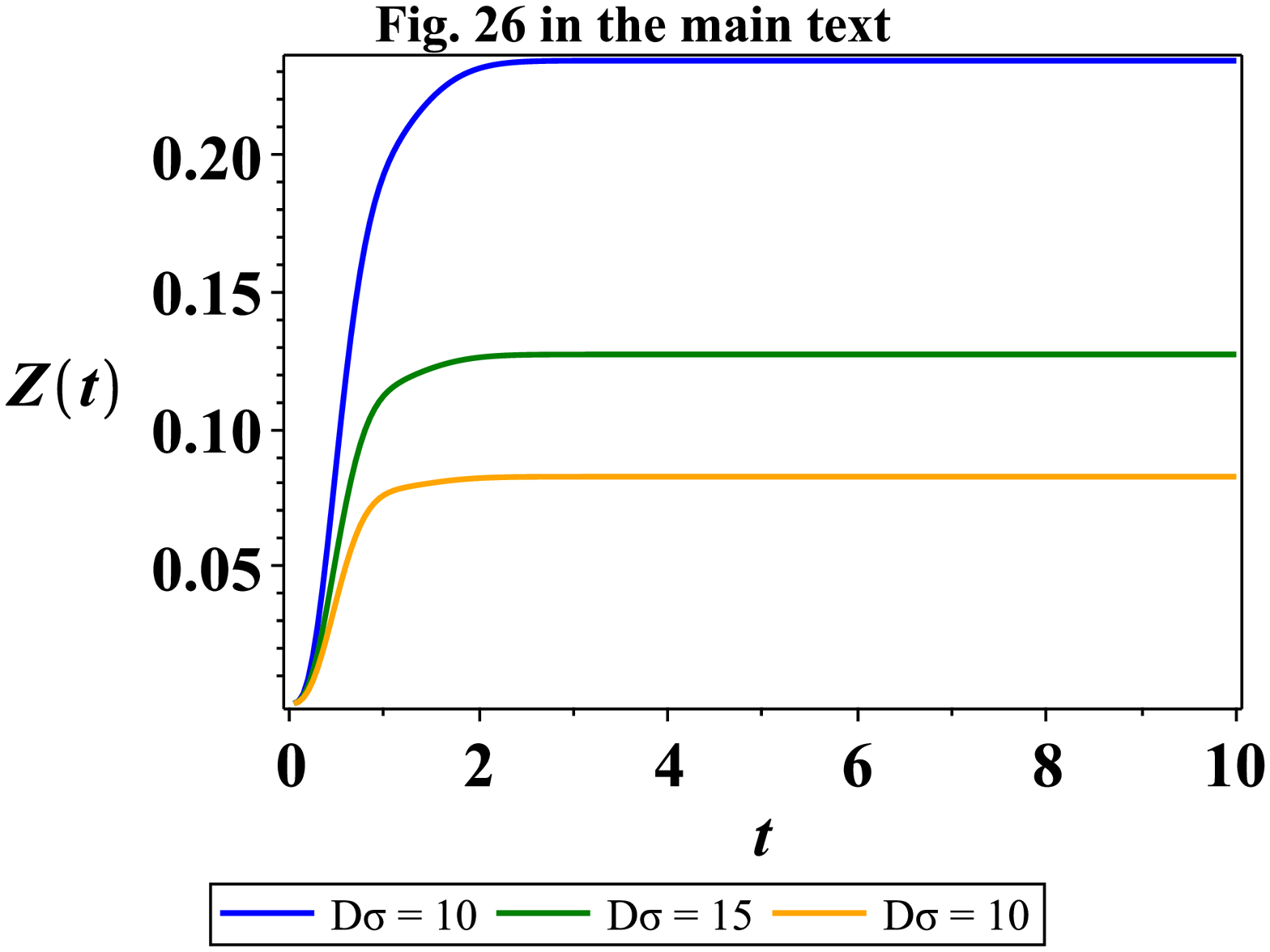}}
\end{center}
\caption{(Color online)  Dependence of $Z$ on time $t$. Estimates are made 
for the results presented in  Figs. 21 -- 26 in the main text.
\label{SM11}}
\end{figure}

\end{widetext}


\begin{thebibliography}{10}
\providecommand{\url}[1]{{#1}}
\providecommand{\urlprefix}{URL }
\expandafter\ifx\csname urlstyle\endcsname\relax
  \providecommand{\doi}[1]{DOI \discretionary{}{}{}#1}\else
  \providecommand{\doi}{DOI \discretionary{}{}{}\begingroup
  \urlstyle{rm}\Url}\fi


\bibitem{Len}
K.L.M. Lewis, F.D. Fuller, J.A. Myers, C.F. Yocum, S. Mukamel,  D. Abramavicius, and J. P. Ogilvie,  J. Phys. Chem. A, {\bf 117}, 34 (2012).

\bibitem{ECR}
G.~Engel, T.~Calhoun, E.~Read, T.~Ahn, T. Mancal, Y.~Cheng, R.~Blankenship, and
  G.~Fleming, Nature Letters \textbf{446}, 782 (2007).

\bibitem{IF}
  A. Ishizaki and G.R. Fleming, J. Chem. Phys., {\bf 130}, 234110 (2009).
  
\bibitem{IFG}
A.~Ishizaki and G.~Fleming, PNAS \textbf{106}, 17255 (2009).

\bibitem{CWWC}
E.~Collini, C.~Wong, K.~Wilk, P.~Curmi, P.~Brumer, G.~Scholes, Nature Letters
  \textbf{463}, 644 (2010).

\bibitem{PHFC}
G.~Panitchayangkoon, D.~Hayes, K.~Fransted, J.~Caram, E.~Harel, J.~Wenb,   R.~Blankenship, G.~Engel, PNAS USA \textbf{107}, 12766 (2010).

\bibitem{WD}
D.M. Wilkins and N.S. Dattani, J. Chem. Theory Comput. {\bf 11}, 3411 (2015). 

\bibitem{book1} 
M. Mohseni, Y. Omar, G. Engel, and M.B. Plenio (eds.), Quantum Effects in Biology (Cambridge University Press, 2014).

\bibitem{RMKL}
P.~Rebentrost, M.~Mohseni, I.~Kassal, S.~Lloyd, and A.~Aspuru-Guzik, New J. Phys.
  \textbf{11}(3), 033003 (2009)

\bibitem{CFMB}
G.~Celardo, F.~Borgonovi, M.~Merkli, V.~Tsifrinovich, and G.~Berman,
 {J. Phys. Chem.} \textbf{116},  {22105} ({2012}).
 
\bibitem{BAF}
D.I.G. Bennett, K. Amarnath, and G.R. Fleming, {JACS}, {\bf 135}, 9164 (2013).

\bibitem{MRSN}
 {{R.}~ Marcus}  and { {N.}~ Sutin},  {Biochimica et Biophysica Acta},
  \textbf{811}, {265} ({1985}).

\bibitem{HDR}
X.~Hu, A.~Damjanovic, T.~Ritz, and K.~Schulten, Proc. Natl. Acad. Sci. USA
  \textbf{95}, 5935 (1998).
  
  \bibitem{MBS}
  M. Merkli, G.P. Berman, R.T. Sayre, S. Gnanakaran, M.K\"onenberg, 
  A.I. Nesterov, and H. Song, J. Math. Chem., {\bf 54}, 866 (2016).


\bibitem{MSB}
M. Merkli, I.M. Sigal, and G.P. Berman,  Phys. Rev. Lett. {\bf 98}, 130401 (2007).

\bibitem{MSBMulti}
M. Merkli, H. Song, an G.P. Berman,   J. Phys. A: Math. Theor. {\bf 48}, 275304 (2015). 


\bibitem{MFL}
B.~McMahon, P.~Fenimore, and M.~LaBute, in \emph{{Fluctuations and Noise in Biological, Biophysical, and Biomedical Systems}}, \emph{Proceedings of
  SPIE}, vol. 5110, ed. by S.M. Bezrukov, H.~Frauenfelder, F.~Moss (2003), \emph{Proceedings of SPIE}, vol. 5110, pp. 10 -- 21

\bibitem{DB}
  T.G. Dewey and J.G. Bann, Biophysical Journal, {\bf 63}, 594 (1992).

  \bibitem{G}
  R. Grima, J. Chem. Phys., {\bf 132}, 185102 (2010).

  \bibitem{CBC}
  P. Carlini, A.R. Bizzarri, and S. Cannistraro, Physica D, {\bf 165}, 242 (2002).

 \bibitem{SPA}
  M.S. Samoilov, G. Price, and A.P. Arkin, Science's STKE, {\bf 2006}, re17 (2006).

\bibitem{MBR2}
M. Merkli, G.P. Berman, and A. Redondo, {J. Phys. A, Math.  Theor.}, {\bf 44}, 305306 (2011).

 \bibitem{BGA}
 J. Bergli, Y.M. Galperin, and B.L. Altshuler, New Journal of Physics, {\bf 11}, 025002 (2009).
 
 \bibitem{GABS}
 Y.M. Galperin, B.L. Altshuler, J.~Bergli, D.~Shantsev, and V.~Vinokur, Phys. Rev. B
 \textbf{76}, 064531 (2007).
 
 \bibitem{NB1}
 A.I. Nesterov and G.P. Berman, Phys. Rev. A, {\bf 85}, 052125 (2012).
 
\bibitem{Govorov}
A. Govorov, P.L.H. Mart\'{i}nez, and H.V. Demir (Eds.), Understanding and Modeling F\"{o}rster-type Resonance Energy Transfer (FRET), Introduction to FRET, Vol. 1, (Springer, 2016).

\bibitem{Andrews}
G. Juzeli\={u}nas and D.L. Andrews, Advs. in Chem. Phys., {\bf 112}, 357 (2000). 

\bibitem{Milonni}
P.W. Milonni and S.M.H. Rafsanjani, Phys. Rev. A, {\bf 92}, 062711 (2015).

  \bibitem{WW}
V.F. Weisskopf and  E.P. Wigner, Z. Physics \textbf{63}, 54  (1930).

\bibitem{SM}
S.~Mukamel, \emph{{Principles} of {Nonlinear} {Optical} {Spectroscopy}} (Oxford
  University Press, New York, 1995).
   
\bibitem{G0}
S. Gurvitz, A.I. Nesterov, and G.P. Berman, J. Phys. A: Math. Theor., {\bf 50},  365601 (2017).

 \bibitem{abr}
 M.~Abramowitz, I.A. Stegun (eds.), \emph{{Handbook of Mathematical Functions}}
 (Dover, New York, 1965).
 
\bibitem{Struve}
W.S. Struve, In: Anoxygenic Photosynthetic Bacteria, R.E. Blankenship, M.T. Madigan and C.E. Bauer (Eds.), Chapter 15, p. 297
(Kluwer Academic Publishers. Printed in The Netherlands, 1995).

\bibitem{Forster2}
T. F\"{o}rster, Ann. Phys. (Leipzig) {\bf 2}, 55 (1948).

  A. Aharony, S. Gurvitz, O. Entin-Wohlman, and S. Dattagupta, Phys. Rev. B, {\bf 82}, 245417 (2010).

\bibitem{NBSS}
A.I. Nesterov, G.P. Berman, J.M. S\'anchez M\'artinez, and R. Sayre, J. Math. Chem., {\bf 51}, 1 (2013).

 \bibitem{Muh}
 F.M\"{u}h, D. Lindorfer, M. Schmidt am Busch, and T. Renger, Phys. Chem. Chem.Phys., {\bf 16}, 11848 (2014). 

B. Elattari and S.A. Gurvitz, Phys. Rev. B, {\bf 62}, 032102 (2000).

\bibitem{SBK}
V.V. Shubin, I.N. Bezsmertnaya, and  N.V. Karapetyanpanel,   J. Photochemistry and Photobiology B: Biology, {\bf  30}, 153 (1995).

\bibitem{MHU}
M. Mimuro, K. Hirayama,  K. Uezono, H. Miyashita, and S. Miyachi, Physica Acta, {\bf  1456},  27 (2000).

\bibitem{S1}
H. Sumi,  J. Phys. Chem. B, {\bf 106},  13370 (2002).

\bibitem{S2}
H. Sumi,  J. Phys. Chem. B, {\bf 108}, 11792 (2004).

\bibitem{LLC}
P. Loughlin,  Y. Lin, and M. Chen, Photosynth Res, {\bf 116},  277 (2013).

\bibitem{AKZ}
S. I. Allakhverdiev, V. D. Kreslavski, S. K. Zharmukhamedov, R. A. Voloshin, D. V. Korolyakova, T. Tomo, and J.R. Shen, Biochemistry (Moscow),  {\bf 81}, 201-212 (2016).

\bibitem{KND}
M. Kaucikas, D. N\"urnberg, G. Dorlhiac, A.W. Rutherford, and Jasper J. van Thor, Biophysical Journal, {\bf 112},  234 (2017).

\bibitem{TKK}
L.M. Tan, J. Yu, T. Kawakami,  M. Kobayashi,  P. Wang, Z.Y.W. Otomo,  and J.P. Zhang, J Phys. Chem. Lett., {\bf 9}, 3278  (2018).


\bibitem{MA1}
S. Buckhout-White, M. Ancona, E. Oh, J.R. Deschamps, M.H. Stewart, J.B. Blanco-Canosa, P.E. Dawson, E.R. Goldman, and I.L. Medintz, ACS Nano, {\bf 6}, 1026 (2012).

\bibitem{MA2}
C.M. Spillmann, M.G. Ancona, S. Buckhout-White, W.R. Algar,
M.H. Stewart, K. Susumu, A.L. Huston, E.R. Goldman, and I.L. Medintz,
ACS Nano, {\bf 7(8)}, 7101 (2013).

\bibitem{NS}
N.C. Seeman, J. Theor. Biol., {\bf 99}, 237 ( 1982).


\end{thebibliography}

\begin{thebibliography}{10}
\providecommand{\url}[1]{{#1}}
\providecommand{\urlprefix}{URL }
\expandafter\ifx\csname urlstyle\endcsname\relax
  \providecommand{\doi}[1]{DOI \discretionary{}{}{}#1}\else
  \providecommand{\doi}{DOI \discretionary{}{}{}\begingroup
  \urlstyle{rm}\Url}\fi

  \bibitem{KV2}
V.~Klyatskin.
 {\em Dynamics of Stochastic Systems}.
 Elsevier, 2005.

\bibitem{KV3}
V.~Klyatskin.
 {\em Lectures on Dynamics of Stochastic Systems}.
 Elsevier, 2011.

 \bibitem{NB1}
  A.I. Nesterov and G.P. Berman, Phys. Rev. A, {\bf 85}, 052125 (2012).

\bibitem{NBSS}
A.I. Nesterov, G.P. Berman, J.M. S\'anchez M\'artinez, and R. Sayre, J. Math. Chem., {\bf 51}, 1 
(2013).

 \bibitem{abr}
M.~Abramowitz, I.A. Stegun (eds.), \emph{{Handbook of Mathematical Functions}}
  (Dover, New York, 1965).
 
 \bibitem{PR1}
A.P.~Prudnikov, Yu. A. Brychkov and O. I. Marchev,  \emph{{Integrals and Series Volume 1: 
Elementary Functions }}(Gordon and Breach, Amsterdam, 1998).

\bibitem{PR2}
A.P.~Prudnikov, Yu. A. Brychkov and O. I. Marchev,  \emph{{Integrals and Series Volume 2: 
Special Functions }}(Gordon and Breach, Amsterdam, 1998).
\end{thebibliography}
\end{document}